\documentclass[11pt,a4paper]{article}
\usepackage{jheppub}
\usepackage[usenames,dvipsnames,table]{xcolor}
\usepackage[T1]{fontenc} 
\usepackage[utf8]{inputenc}
\usepackage{tikz, pgf}
\usepackage{amsmath, mathtools}
\usetikzlibrary{shapes.misc}
\usetikzlibrary{shapes}
\usetikzlibrary{fit}
\usetikzlibrary{arrows}
\usepackage{array,tikz-cd}
\usepackage{subfigure}
\usepackage{braket}
\usepackage{environ}
\usepackage{float}

\def\b{\beta}
\def\o{\omega}

\def\Mc{{\bf{\widetilde{M}}}}

\def\Mhotwo{{\bf M}}

\def\Mtwo{{\bf M}}
\def\Mtildetwo{{\bf {\widetilde{M}}}}

\def\Mtildethree{{\bf \widetilde{M}}}

\def\Mtildefour{{\bf {\widetilde{M}}}}

\newcommand{\matM}[1]{{\bf M}}
\newcommand{\matMt}[1]{{\bf \widetilde{M}}}

\newcommand{\eP}[1]{e^{(#1)}_{_P}}
\newcommand{\eF}[1]{e^{(#1)}_{_F}}
\newcommand{\ebP}[1]{\bar{e}^{(#1)}_{_P}}
\newcommand{\ebF}[1]{\bar{e}^{(#1)}_{_F}}

\NewEnviron{eqn}{
\begin{align}
\begin{split}
  \BODY
\end{split}
\end{align}
}

\definecolor{rust}{rgb}{0.8,0.2,0.2}

\def\b{\beta}
\def\o{\omega}

\newcommand{\be}{\begin{equation}}
\newcommand{\ee}{\end{equation}}
\newcommand{\ba}{\begin{eqnarray}}
\newcommand{\ea}{\end{eqnarray}}
\newcommand{\bi}{\begin{itemize}}
\newcommand{\ei}{\end{itemize}}
\newcommand{\bse}{\begin{subequations}}
\newcommand{\ese}{\end{subequations}}

\newcommand{\bose}{\mathfrak{f}}

\newcommand{\Op}[1]{\mathbb{#1}}

\newcommand{\contourDDDD}[5]{
\begin{tikzpicture}[scale = 1.5]
\draw[thick,color=red](-1.2, #2) circle (0.5ex);
\draw[thick,color=red](-1.2, #2 - 0.5) circle (0.5ex);
\draw[thick,color=red](-0.4, #3) circle (0.5ex);
\draw[thick,color=red](-0.4, #3 - 0.5) circle (0.5ex);
\draw[thick,color=red](0.4, #4) circle (0.5ex);
\draw[thick,color=red](0.4, #4 - 0.5) circle (0.5ex);
\draw[thick,color=red](1.2, #5) circle (0.5ex);
\draw[thick,color=red](1.2, #5 - 0.5) circle (0.5ex);
\draw[thick, color=black]
{
 (-1.2, #2) node [above] {\large{-1}}
 (-1.2, #2 - 0.5) node [above] {\large{+1}}
 (-0.4, #3) node [above] {\large{-1}}
 (-0.4, #3 - 0.5) node [above] {\large{+1}}
 (0.4, #4) node [above] {\large{-1}}
 (0.4, #4 - 0.5) node [above] {\large{+1}}
 (1.2, #5) node [above] {\large{-1}}
 (1.2, #5 - 0.5) node [above] {\large{+1}}
};
\foreach \x in {0,...,#1}{
\pgfmathtruncatemacro\z{int(-\x + 1)}
\draw[thick,color=blue] (2 + 0.25, \x cm - 0.25cm) node[right] {\footnotesize{\z}};
\draw[thick,color=violet,->] (-2,\x cm) -- (2,\x cm) ;
\draw[thick,color=violet,->] (2, \x cm - 0.5 cm) -- (-2, \x cm - 0.5 cm) ;
\draw[thick,color=violet,->] (2, \x cm) arc (90:-90:0.25);
}
\foreach \y in {-1,...,#1}{
\draw[thick,color=violet,->] (-2,\y cm + 0.5 cm) arc (90:270:0.25);
\pgfmathtruncatemacro\w{int(-\y + 1)}
}
\end{tikzpicture}
}

\newcommand{\contourDDD}[4]{
\begin{tikzpicture}[scale = 1.5]
\draw[thick,color=red](-1, #2) circle (0.5ex);
\draw[thick,color=red](-1, #2 - 0.5) circle (0.5ex);
\draw[thick,color=red](0, #3) circle (0.5ex);
\draw[thick,color=red](0, #3 - 0.5) circle (0.5ex);
\draw[thick,color=red](1, #4) circle (0.5ex);
\draw[thick,color=red](1, #4 - 0.5) circle (0.5ex);
\draw[thick, color=black]
{
 (-1, #2) node [above] {\large{{-1}}}
 (-1, #2 - 0.5) node [above] {\large{{+1}}}
 (0, #3) node [above] {\large{{-1}}}
 (0, #3 - 0.5) node [above] {\large{{+1}}}
 (1, #4) node [above] {\large{{-1}}}
 (1, #4 - 0.5) node [above] {\large{{+1}}}
};
\foreach \x in {0,...,#1}{
\pgfmathtruncatemacro\z{int(-\x + 1)}
\draw[thick,color=violet,->] (-2,\x cm) -- (2,\x cm) ;
\draw[thick,color=violet,->] (2, \x cm - 0.5 cm) -- (-2, \x cm - 0.5 cm) ;
\draw[thick,color=violet,->] (2, \x cm) arc (90:-90:0.25);
\draw[thick,color=blue] (2 + 0.25, \x cm - 0.25cm) node[right] {\footnotesize{\z}};
}
\foreach \y in {-1,...,#1}{
\pgfmathtruncatemacro\w{int(-\y + 1)}
\draw[thick,color=violet,->] (-2,\y cm + 0.5 cm) arc (90:270:0.25);
}
\end{tikzpicture}
}

\newcommand{\contourDD}[3]{
\begin{tikzpicture}[scale = 1.5]
\draw[thick,color=red](-1, #2) circle (0.5ex);
\draw[thick,color=red](-1, #2 - 0.5) circle (0.5ex);
\draw[thick,color=red](1, #3) circle (0.5ex);
\draw[thick,color=red](1, #3 - 0.5) circle (0.5ex);
\draw[thick, color=black]
{
 (-1, #2) node [above] {\large{{-1}}}
(-1, #2 - 0.5) node [above] {\large{{+1}}}
 (1, #3) node [above] {\large{{-1}}}
 (1, #3- 0.5) node [above] {\large{{+1}}}
};
\foreach \x in {0,...,#1}{
\pgfmathtruncatemacro\z{int(-\x + 1)}
\draw[thick,color=violet,->] (-2,\x cm) -- (2,\x cm) ;
\draw[thick,color=violet,->] (2, \x cm - 0.5 cm) -- (-2, \x cm - 0.5 cm) ;
\draw[thick,color=violet,->] (2, \x cm) arc (90:-90:0.25);
\draw[thick,color=blue] (2 + 0.25, \x cm - 0.25cm) node[right] {\footnotesize{\z}};
}
\foreach \y in {-1,...,#1}{
\pgfmathtruncatemacro\w{int(-\y + 1)}
\draw[thick,color=violet,->] (-2,\y cm + 0.5 cm) arc (90:270:0.25);
}
\end{tikzpicture}
}

\newcommand{\contourSS}[3]{
\begin{tikzpicture}[scale = 1.5]
\draw[thick,color=red](-1, #2) circle (0.5ex);
\draw[thick,color=red](1, #3) circle (0.5ex);
\draw[thick, color=black]
{
 (-1, #2) node [above] {\large{{2}}}
 (1, #3) node [above] {\large{{1}}}
};
\foreach \x in {0,...,#1}{
\pgfmathtruncatemacro\z{int(-\x + 1)}
\draw[thick,color=violet,->] (-2,\x cm) -- (2,\x cm) ;
\draw[thick,color=violet,->] (2, \x cm - 0.5 cm) -- (-2, \x cm - 0.5 cm) ;
\draw[thick,color=violet,->] (2, \x cm) arc (90:-90:0.25);
\draw[thick,color=blue] (2 + 0.25, \x cm - 0.25cm) node[right] {\footnotesize{\z}};
}
\foreach \y in {-1,...,#1}{
\pgfmathtruncatemacro\w{int(-\y + 1)}
\draw[thick,color=violet,->] (-2,\y cm + 0.5 cm) arc (90:270:0.25);
\draw[thick,color=blue] (-2 - 0.25, \y cm + 0.25cm); 
}
\end{tikzpicture}
}

\title{ Spectral Representation of Thermal OTO Correlators }

\author[a]{Soumyadeep Chaudhuri}
\author[a,b]{\!, Chandramouli Chowdhury}
\author[a]{\!, R. Loganayagam}

\vspace{2cm}

\affiliation[a]{International Centre for Theoretical Sciences (ICTS-TIFR), \\
Tata Institute of Fundamental Research, \\
Shivakote, Hesaraghatta, \\
Bangalore 560089, INDIA}

\affiliation[b]{Department of Physics, \\
Indian Institute of Technology Guwahati,\\
Guwahati 781039, INDIA}
\emailAdd{soumyadeep.chaudhuri@icts.res.in}
\emailAdd{chandramouli.chowdhury@gmail.com }
\emailAdd{nayagam@icts.res.in}

\abstract{We study the spectral representation of finite temperature, out of time ordered (OTO) correlators on the multi-time-fold generalised Schwinger-Keldysh contour. We write  the contour-ordered correlators 
as a sum over time-order permutations acting on a fundamental array of Wightman correlators. We decompose this Wightman array in a basis of column vectors, which provide a natural generalisation of the 
familiar retarded-advanced  basis  in the finite temperature Schwinger-Keldysh formalism. The coefficients of this decomposition take the form of 
generalised spectral functions, which are Fourier transforms of nested and double commutators.  Our construction extends a variety of classical results on spectral functions in the SK formalism at finite temperature to the OTO case.}

\begin{document}
\maketitle


\section{Introduction}
\label{sec:intro}
The dynamics of quantum field theory at finite temperature is of fundamental interest in fields ranging from dynamical critical phenomena to cosmology, in blackhole physics/quantum gravity. Until recently, it had been conventional to assume that, in principle all the observables of  real time, finite temperature quantum field theory are encoded in its Schwinger Keldysh correlators\cite{Schwinger:1960qe,Keldysh:1964ud,Feynman:1963fq,Chou:1984es, Calzetta:2008iqa, Kamenev:2011, Kamenev:2009jj, stefanucci2013nonequilibrium, Sieberer:2015svu,Haehl:2016pec}. This statement has been upended by the advent of out of time ordered correlators (OTOCs)\cite{1969JETP...28.1200L} which fall beyond the conventional Schwinger-Keldysh formalism and the usual edifice of intuitions, approximations and computations built around it. These OTOCs 
require that we extend the standard Schwinger-Keldysh formalism to path integrals with many time-fold contours\cite{Aleiner:2016eni,Haehl:2017qfl}.

From the viewpoint of a non-equilibrium field theorist, three crucial questions could be asked regarding OTOCs :
\begin{itemize}
\item What new \emph{physics} do these OTOCs encode ? A growing literature has shown relations to notions of chaos vs ergodicity say in blackholes\cite{Shenker:2013pqa,Maldacena:2015waa} via its relation to Lodschmidt echo\footnote{See \url{http://www.scholarpedia.org/article/Loschmidt_echo} for a description.}, thermalisation vs localisation
\cite{2016arXiv160801914F,2016arXiv160802765C,Swingle:2016jdj, 2017AnP...52900318H}, quantum information measures related to joint quasi-probablibilities weak measurements \cite{Hosur:2015ylk,Roberts:2016hpo,Tsuji:2016kep,Halpern:2016zcm,Halpern:2017abm},
generalised discontinuities of the correlators  \cite{Caron-Huot:2017vep,Simmons-Duffin:2017nub,Iliesiu:2018fao} which encode useful spectral information in CFTs. The analytic structure of OTOCs in quantum thermal systems has led to bounds on chaos\cite{Maldacena:2015waa,Blake:2017ris, Haehl:2017pak,Haehl:2018izb,Basu:2018akv}, generalised FDTs\cite{Haehl:2017eob} and in generalising Eigenstate hypothesis\cite{Foini:2018sdb}. This fast growing array of ideas show the usefulness of studying OTOCs.
\item Secondly, how are they to be \emph{measured} in experiments ? The dogma that only time-ordered correlators can be measured in an experiment has yielded ground to an ingenious set of experiments/ experimental  proposals aimed at reverse time evolution/weak measurements \cite{Zhu:2016uws,Garttner:2016mqj}. Despite, this, we are far from having experimental protocols to measure OTOCs in complex systems.
\item Thirdly, What are the most efficient ways to \emph{compute} these correlators ? Any attempt at setting up a 
naive diagrammatic perturbation theory, even in the simplest of quantum field theories, runs aground
with a proliferation of fields and their Feynman vertices. This definitely calls for new computational frameworks to systematise  such calculations. 
\end{itemize} 
In this work, we will primarily address the last issue by constructing a practical framework to compute and classify OTOCs of a system at thermal equilibrium. 
Stated briefly, this can be done by recognising that the core physics of the system can be encoded in certain  \emph{spectral functions} and the structure of thermal correlators
naturally admit  \emph{spectral representations} in terms of them. This statement is a finite-temperature generalisation of the Kallen-Lehmann spectral representations in the 
zero temperature quantum field theory (See, for example, \S\S 10.7 of  \cite{Weinberg:1995mt} for a textbook discussion).

The idea of spectral representation for Schwinger-Keldysh real time correlators has a long history \cite{Evans:1990qh,Evans:1990hy,Evans:1991ky,Taylor:1993ub,Carrington:1996rx,Hou:1998yc,Wang:1998wg,Hou:1999zv}
(for a discussion in terms of discontinuities see \cite{Guerin:1993ik,Guerin:2001mx}). Such spectral representations have   been found useful  in developing efficient perturbative formalisms  \cite{Chu:1993nc,Henning:1993gh,Henning:1995sm, 2007EPJC...50..711C}.
They have found applications in transport computations at finite temperature and  in developing effective methods to truncate to kinetic theory 
descriptions (including  effective actions encoding hard thermal loops of gluons at high temperature ala Braaten-Pisarski\cite{Braaten:1991gm}). Our aim in this work is to develop a similarly useful formalism for
out of time ordered thermal perturbation theory.

We will now describe in slightly more detail, the idea of spectral functions/representations.  For example, in the above mentioned  works, it was recognised that the 2-pt and 3-pt SK correlators in the thermal state can be 
written down  in terms of thermal expectation values of fully nested time-ordered commutators (also termed fully retarded Green functions \cite{Evans:1991ky}) of the form :
\[ \langle \Theta_{12} [O(t_1),O(t_2)]+\Theta_{21} [O(t_2),O(t_1)] \rangle_\beta\ ,\qquad \langle \Theta_{123} [[O(t_1),O(t_2)] ,O(t_3)] +\text{permutations}\rangle_\beta \ .\]
Here $\Theta_{ij\ldots}$ are step functions enforcing time-ordering $t_i>t_j>\ldots$. These are the aforementioned spectral functions which are nicer objects to compute than the full real-time correlators and they are also easiest to obtain by analytic continuation
from Euclidean correlators \cite{Evans:1991ky,Baier:1993yh}. Commutators have nice causality properties in time domain which, via Kramers-Kronig type arguments, enforce
good analytic behaviour in appropriate regions of the frequency domain.  

Another key insight relevant to this work is the following :  there is a natural formalism in terms of arrays of certain column vectors which provides 
a convenient way to organise and use  such spectral representations \cite{Chu:1993nc,Henning:1993gh,Henning:1995sm,Hou:1998yc,Wang:1998wg}. This column vector basis is also naturally related
to what is termed retarded-advanced (RA) basis\cite{aurenche1992comparison,van1992finite,Baier:1993yh,vanEijck:1994rw} in the thermal SK formalism.

When we move to 4-pt correlators, the time-ordered commutators are no more sufficient to capture all thermal correlations \cite{Haehl:2017eob}, and
OTO commutators/spectral functions should be added to the set of spectral functions. The addition of OTO spectral functions into the analysis, clears up the 
complexity visible in older analysis of thermal SK correlators.  The authors of  \cite{Haehl:2017eob} showed that, by adding in the OTO spectral functions, one can indeed 
reconstruct all n-point Wightman correlators. In fact, the constraints imposed by thermal periodicity can be completely solved for an arbitrary n-pt function , and a simple
formula can be written down  expressing arbitrary Wightman correlators in terms of spectral functions \cite{Haehl:2017eob}. 

Wightman correlators, however, are not natural objects to formulate perturbation theory or to set up diagrammatics. Diagrammatics and 
path integral formalism naturally work with contour-ordered correlators on the multi-time-fold contours.  In principle, this is a simple matter of 
expressing contour correlators in terms of Wightman functions and using the relations derived in  \cite{Haehl:2017eob}. In practice, however,
combinatorics overwhelm this exercise, resulting in  complicated looking expressions which hide much of the structure. 

Inspired by the
previous work on SK correlators, in this work, we will extend the column vector/retarded-advanced formalism to generalised SK correlators. Our basis is chosen such 
that, \emph{on a time contour with $k$ timefolds, we have $k$ `retarded' combinations which can  occur within a correlator only in the causal past of some other operators and 
$k$ advanced combinations which can occur only in the causal future}. This is a natural generalisation of the usual retarded-advanced formalism with a single retarded and 
a single advanced field.  Our primary  aim here is to express the contour correlators in terms of spectral functions within such a  formalism.

The paper is organised as follows : we will begin in \S\ref{sec:2ptSK} by reviewing spectral representation of Schwinger-Keldysh two point functions in terms of column vectors. The material here is well-known
and is discussed in a variety of reviews and textbooks (see, for example \cite{Chou:1984es}).  We write down many equivalent expressions for the two point functions and note  their
underlying structure.  Our notation and  emphasis here are aimed  towards further  generalisation. The reader familiar with this material may wish to skim these sections 
and move ahead to \S\ref{sec:2ptGenSK} where we extend 
this spectral representation to two point functions in the generalised SK contour. This section brings out the main ideas behind the construction 
of these representations which is then applied to higher point functions. This is followed by  section \S\ref{sec:npt} where we quote  the results for  higher point functions within
generalised Schwinger-Keldysh formalism. We end with a discussion of future directions in  \S\ref{sec:discussion}. 

For the convenience of the reader, many of the technical details
are relegated to the appendices : in appendix~\S\ref{basis}, we summarise the basis of column vectors on which our spectral representations are based. The appendix~\ref{app:rules}
details the structure of arguments used to constrain the structure of the contour-ordered thermal correlators. In appendices~\S\ref{app2pt}, \S\ref{app3pt} and \S\ref{app4pt}, we present the 
analyses of 2 point, 3 point and 4 point functions respectively.

\section{Spectral representation of SK two point functions}
\label{sec:2ptSK}

\subsection{Example of a free scalar field}

Before going into the general contour correlators and their relations, let us begin with a simple example.
Consider the contour-ordered, thermal two point functions of a free real scalar field in SK formalism (in the mostly plus metric convention) : 
\begin{equation}\label{eq:twopt}
\begin{split}
\langle \mathcal{T}_C \phi_1(x_1) \phi_1(x_2) \rangle 
&= \langle \mathcal{T} \phi(x_1) \phi(x_2) \rangle=\int_p \rho_p(\Theta_{12}+\bose_p)e^{ip\cdot(x_1-x_2)}\ , \\
\langle \mathcal{T}_C \phi_1(x_1) \phi_2(x_2) \rangle  &= \langle \phi(x_2)  \phi(x_1)\rangle= \int_p \rho_p \bose_p e^{ip\cdot(x_1-x_2)}\ ,\\
\langle \mathcal{T}_C \phi_2(x_1) \phi_1(x_2) \rangle & =\langle \phi(x_1)  \phi(x_2)\rangle= \int_p \rho_p(1+\bose_p)e^{ip\cdot(x_1-x_2)}\ , \\
\langle \mathcal{T}_C \phi_2(x_1) \phi_2(x_2) \rangle &= \langle \mathcal{T}^\ast \phi(x_1) \phi(x_2) \rangle= \int_p \rho_p(\Theta_{21}+\bose_p)e^{ip\cdot(x_1-x_2)}\ . 
\end{split}
\end{equation}
Here, $\phi_1$ is the `ket' field with time-ordered propagator whereas $\phi_2$ is the `bra' field of Schwinger formalism with anti-time-ordered
propagators. The symbol $\mathcal{T}_C$ denotes SK contour ordering and $\Theta_{12}$ denotes Heaviside step function in time. We have 
written down the corresponding correlators in the single-copy notation  (with the time-ordering operator $\mathcal{T}$ and anti-time-ordering operator $\mathcal{T}^\ast$ ) 
for the convenience of the reader.
 
 The symbol $\rho_p$ in the above equation stands  for the \emph{spectral function} which in a free scalar theory takes the form 
\[ \rho_p \equiv 2\pi\ \text{sign}(p^0)\delta(p^2+m^2)= \frac{2\pi}{2\omega_p} [\delta(\omega-\omega_p)-\delta(\omega+\omega_p)]\ . \]
Here $\omega_p=\sqrt{\mathbf{p}^2+m^2}$. The spectral function is  also directly related to the Fourier-transform of commutators in the theory, viz.,
\[ \int_p \rho_p e^{ip\cdot(x_1-x_2)} =\langle[\phi(x_1),\phi(x_2)]\rangle\  \]
and it neatly encodes all the theory-dependent information. The factor $\bose_p$ is the Bose-Einstein factor
\[ \bose_p \equiv \frac{1}{e^{\beta p^0}-1}\  \]
which obeys $1+\bose_p+\bose_{-p} =0 $ and $\bose_p=e^{-\beta p^0}(1+\bose_p)$. These Bose-Einstein factors
are universal and the way they occur in the correlators are completely fixed by general arguments. 

Further, we have used the notation
\[ \int_p \equiv \int \frac{d^dp}{(2\pi)^d} \]
to denote the momentum integrals in $d$ spacetime dimensions. Using these relations, we get the more familiar two point correlators :
\begin{equation}
\begin{split}
\langle \mathcal{T}_C \phi_1(x_1) \phi_1(x_2) \rangle 
&= \langle \mathcal{T} \phi(x_1) \phi(x_2) \rangle\\
&= \int \frac{d^{d-1}p}{(2\pi)^{d-1}2\omega_p} \left[ (\Theta_{12}+\bose_p)e^{ip\cdot(x_1-x_2)} +(\Theta_{21}+\bose_p)e^{-ip\cdot(x_1-x_2)}\right]_{p^0=\omega_p}\ ,\\
\langle \mathcal{T}_C \phi_1(x_1) \phi_2(x_2) \rangle  
&= \langle \phi(x_2)  \phi(x_1)\rangle\\
&= \int\frac{d^{d-1}p}{(2\pi)^{d-1}2\omega_p} \left[ \bose_p e^{ip\cdot(x_1-x_2)} +(1+\bose_p)e^{-ip\cdot(x_1-x_2)}\right]_{p^0=\omega_p}\ ,\\
\langle \mathcal{T}_C \phi_2(x_1) \phi_1(x_2) \rangle 
&=\langle \phi(x_1)  \phi(x_2)\rangle\\
&= \int \frac{d^{d-1}p}{(2\pi)^{d-1}2\omega_p}  \left[ (1+\bose_p)e^{ip\cdot(x_1-x_2)} +\bose_p e^{-ip\cdot(x_1-x_2)}\right]_{p^0=\omega_p}\ ,\\
\langle \mathcal{T}_C \phi_2(x_1) \phi_2(x_2) \rangle 
&= \langle \mathcal{T}^\ast \phi(x_1) \phi(x_2) \rangle\\
&=\int\frac{d^{d-1}p}{(2\pi)^{d-1}2\omega_p}  \left[(\Theta_{21}+\bose_p)e^{ip\cdot(x_1-x_2)} +(\Theta_{12}+\bose_p)e^{-ip\cdot(x_1-x_2)}\right]_{p^0=\omega_p}\ .
\end{split}
\end{equation}

The reader can readily verify the correctness of the above expressions by starting with the free theory mode expansion
\begin{equation}
\begin{split}
\phi(x) 
= \int \frac{d^{d-1}p}{(2\pi)^{d-1}\sqrt{2\omega_p}}  \left[ a_p e^{ip\cdot x} +a^\dag_p e^{-ip\cdot x}\right]_{p^0=\omega_p}\\
\end{split}
\end{equation}
 and using the thermal expectation values $\langle a^\dag_{p_1} a_{p_2}\rangle = (2\pi)^{d-1} \delta^{d-1}(\vec{p}_1-\vec{p}_2)\bose_{p_1}$ and $\langle a_{p_1} a^\dag_{p_2}\rangle = (2\pi)^{d-1} \delta^{d-1}(\vec{p}_1-\vec{p}_2)(1+\bose_{p_1})$.


\subsection{The column vector structure}

For a general scalar operator $\Phi(x)$ instead of the free field,  the above form of two point functions in \eqref{eq:twopt} still holds
in SK formalism, just with a different spectral function still defined by 
\[ \int_p \rho_p e^{ip\cdot(x_1-x_2)} \equiv\langle[\Phi(x_1),\Phi(x_2)]\rangle\ . \]
This is the SK analog of the famous Kallen-Lehman representation in zero temperature QFT and is a direct consequence of periodicity in imaginary time of thermal 
correlators, viz., 
\[ \langle \Phi(x_1-i\beta)  \Phi(x_2)\rangle = \langle \Phi(x_2)  \Phi(x_1)\rangle\ . \]
Here $\beta^\mu$ is a time-like vector defining thermal equilibrium with its direction giving the rest frame and its magnitude (also denoted by $\beta$) giving rest frame inverse temperature. 

The statement of periodicity is also termed Kubo-Martin-Schwinger(KMS) relations and is the underlying reason behind fluctuation-dissipation theorems in QFTs. Using these relations 
along with the second equation of  \eqref{eq:twopt} (which can be taken as the definition of $\rho_p$) , the rest of \eqref{eq:twopt} follows. Thus, the \emph{four} two point functions of 
SK formalism depend eventually on only \emph{one} system-dependent spectral function and  thermality fixes the rest, as advertised.

We will find it convenient to write the above correlators as an array : 
\begin{equation}
\begin{split}
\langle \mathcal{T}_C & \Phi_i(x_1) \Phi_j(x_2) \rangle = \int_p \rho_p\left(\begin{array}{cc}\Theta_{12}+\bose_p & \bose_p \\ 1+\bose_p & \Theta_{21}+\bose_p\end{array}\right)
e^{ip\cdot(x_1-x_2)} \\
&= \Theta_{12} \int_p\rho_p \left(\begin{array}{cc}1+\bose_p & \bose_p \\ 1+\bose_p & \bose_p\end{array}\right)
e^{ip\cdot(x_1-x_2)} 
 + \Theta_{21}\int_p\rho_p \left(\begin{array}{cc} \bose_p & \bose_p \\ 1+\bose_p &1+\bose_p\end{array}\right) e^{ip\cdot(x_1-x_2)}\\
&= \Theta_{12} \int_p \rho_p   \begin{pmatrix} 1\\ 1 \end{pmatrix}  e^{ip\cdot {x_1}}\otimes  \begin{pmatrix} 1+\bose_p\\ \bose_p  \end{pmatrix} e^{-ip\cdot {x_2}}
 + \Theta_{21}\int_p\rho_p  \begin{pmatrix} \bose_p\\   1+ \bose_p  \end{pmatrix} e^{ip\cdot {x_1}}\otimes    \begin{pmatrix} 1\\ 1 \end{pmatrix}  e^{-ip\cdot {x_2}}\ ,
 \end{split}
\end{equation}
where in the last line we have re-written the answer as tensor products of certain set of column vectors for later convenience. 

Such arrays and the column vectors 
have various structural features which generalise to the case of OTOCs as well as higher point functions. 
Note that the array that appears along 
with the step function $\Theta_{21}$ can be obtained from \emph{transposing}  the array that appears  with the step function $\Theta_{12}$, followed
 by a map $p\mapsto -p$ under which $\rho_p\mapsto -\rho_p$ and $\bose_p\mapsto -(1+\bose_p)$. At the level of tensor products, 
the transpose appears as a permutation in the order of tensor products as the time-order changes.

A more symmetric representation is obtained by defining 
\[  \int_{p_1}\int_{p_2} \rho[12]\ e^{i(p_1\cdot x_1+p_2\cdot x_2)} \equiv \langle[\Phi(x_1),\Phi(x_2)]\rangle \]
in terms of which we can write a spectral representation\cite{Evans:1990hy,Carrington:1996rx,Hou:1998yc,Wang:1998wg}
\begin{equation}\label{eq:2ptSK}
\begin{split}
\langle \mathcal{T}_C \Phi_i(x_1) \Phi_j(x_2) \rangle 
 &= \int_{p_1}\int_{p_2} \Bigl\{\rho[12]\  \Theta_{12} \begin{pmatrix} -1\\ -1 \end{pmatrix}e^{ip_1\cdot x_1}\otimes   \begin{pmatrix} \bose_2\\ 1+\bose_2 \end{pmatrix}  e^{ip_2\cdot x_2}\\
&\qquad +\rho[21]\  \Theta_{21}   \begin{pmatrix} \bose_1\\    1+\bose_1 \end{pmatrix} e^{ip_1\cdot x_1}\otimes \begin{pmatrix} -1\\ -1 \end{pmatrix} e^{ip_2\cdot x_2}\Bigr\}\ ,
 \end{split}
\end{equation}
where we have used the notation $\bose_1\equiv \bose_{p_1}$, $\rho[12]\equiv \rho[p_1,p_2]$ etc. For a free scalar,
\[ \rho[12] \equiv 2\pi\ \text{sign}(p_1^0)\delta(p_1^2+m^2)\times (2\pi)^d \delta^d(p_1+p_2) = -\rho[21]\ . \]
In this presentation, the action on the array can be described as the joint permutation of the time ordering, the array indices and the momenta.

\subsection{The Wightman array}
In the end of the last subsection, we had obtained 
\begin{equation}
\begin{split}
\langle \mathcal{T}_C \Phi_i(x_1) \Phi_j(x_2) \rangle 
 &= \Theta_{12} \Mtwo(x_1,x_2) + \text{permutation}
 \end{split}
\end{equation}
where
\begin{equation}
\begin{split}
\Mtwo(x_1,x_2) 
 &\equiv 
  \int_{p_1}\int_{p_2} \rho[12]\   \begin{pmatrix} -1\\ -1 \end{pmatrix}e^{ip_1\cdot x_1}\otimes   \begin{pmatrix} \bose_2\\ 1+\bose_2 \end{pmatrix}  e^{ip_2\cdot x_2} \ .
 \end{split}
\end{equation}
The array correlator $\Mtwo$ is actually  an array of Wightman correlators :
\begin{equation}
\begin{split}
\Mtwo(x_1,x_2)  &=\left(\begin{array}{cc}\langle  \Phi_1(x_1) \Phi_2(x_2) \rangle & \langle \Phi_2(x_2) \Phi_1(x_1) \rangle \\  \langle \Phi_1(x_1) \Phi_2(x_2) \rangle &\langle \Phi_2(x_2) \Phi_1(x_1) \rangle\end{array}\right) =\left(\begin{array}{cc}\langle 12\rangle & \langle 21 \rangle \\  \langle12 \rangle &\langle 21 \rangle\end{array}\right) \ .
 \end{split}
\end{equation}
Here we have introduced a useful notation for Wightman correlators\cite{Haehl:2017eob} whereby only insertion points and their ordering are retained. The Fourier representation is 
then obtained by using KMS relations :
\begin{equation}
\begin{split}
\langle 12\rangle &=  -\int_{p_1}\int_{p_2}  \rho[12] \bose_2 e^{ip_k\cdot x_k}\ ,\\
\langle 21\rangle &=  -\int_{p_1}\int_{p_2}  \rho[12] (1+\bose_2) e^{ip_k\cdot x_k}\  .
 \end{split}
\end{equation}

Note that the array of Wightman correlators above is constructed so as to agree with the contour-ordered correlators for a particular time-ordering of insertions,viz.,
\begin{equation}
\begin{split}
 \Theta_{12}\langle \mathcal{T}_C \Phi_i(x_1) \Phi_j(x_2) \rangle 
 &= \Theta_{12} \Mtwo(x_1,x_2) \ .
 \end{split}
\end{equation}
Such an arrangement of Wightman correlators play a crucial role throughout this work and we will henceforth refer to it as the \emph{Wightman array corresponding to a time-ordering} and denote it by 
$\Mtwo$. Often, it is convenient to deal with the Fourier transform of the Wightman array which we will denote by $\Mtildetwo$ : 
\begin{equation}
\begin{split}
\Mtwo(x_1,x_2) 
 &\equiv  \int_p \Mtildetwo(p)
e^{ip\cdot(x_1-x_2)}\qquad \text{with}\quad \Mtildetwo(p) \equiv \rho_p \left(\begin{array}{cc}1+\bose_p & \bose_p \\ 1+\bose_p & \bose_p\end{array}\right) \ .
 \end{split}
\end{equation}
or
\[ \Mtwo(x_1,x_2) 
 \equiv \int_{p_1}\int_{p_2} \Mtildetwo(p_1,p_2)  e^{ip_k\cdot x_k} \]
  with
\begin{equation}
\begin{split}
\Mtildetwo(p_1,p_2) 
 &\equiv \rho[12]\  \begin{pmatrix} -1\\ -1 \end{pmatrix}\otimes   \begin{pmatrix} \bose_2\\ 1+\bose_2 \end{pmatrix} \ . 
 \end{split}
\end{equation}
This formula is the basic building block out of which spectral representations are constructed via Fourier transforms and sum over time-orderings.
We note the following features :
\begin{itemize}
\item First of all, there is a clear separation here between the theory dependent information in the spectral function(viz. the Fourier transform of the commutators) and \emph{the array structure imposed by KMS relations} captured by the column vectors . In practical
terms, it is always easier to compute $ \rho[12]$ and use the above representation than computing each of these thermal correlators in turn.
\item Next one notes the \emph{causal structure} of these correlators made manifest via step-functions in time. We note that the correlators here are written as a sum over 
various time-orderings. Within each time-ordering, specific spectral functions appear in conjunction with a particular tensor product of column vectors.
\item  As we permute across time-orderings, the arguments of spectral functions  get permuted along with a permutation in the order in which the tensor products are taken.
\end{itemize}
As we will see later on, all these features directly generalise to spectral representations of higher point thermal correlators (whether time-ordered or out of time-ordered).

Before we move to the generalisation of these results, let us focus on an example of how causal structure is encoded in these column vectors : consider taking either the first field to be a SK difference field $\Phi_d\equiv \Phi_1-\Phi_2$. This is equivalent to contracting the first vector of the product with a row vector $(1\ -1)$ resulting in 
\begin{equation}
\begin{split}
\langle \mathcal{T}_C (\Phi_1(x_1)-\Phi_2(x_1)) \Phi_j(x_2) \rangle 
 &= \Theta_{21}  \int_{p_1}\int_{p_2} \rho[21]\    \bose_1 e^{ip_1\cdot x_1}\ \begin{pmatrix} 1\\ 1 \end{pmatrix} e^{ip_2\cdot x_2}\ .
 \end{split}
\end{equation}
We note that this vanishes unless the difference operator at $x_1$ is actually in the past of $x_2$. What we have shown is the \emph{largest time equation}
for difference operators : \emph{any correlator with  the future-most operator being the  difference operator, vanishes}\cite{Chou:1984es}.

The structure of the two-point thermal correlators that we just reviewed raises a variety of questions : how much of these structures could be generalised to higher point functions ?
What is the systematic way to derive similar results ? Could one systematically understand the structure of the column vectors whose tensor-products appear in such formulae ?
How do we generalise these results in the context of out of time order correlators beyond the usual SK formalism ?

\section{Spectral representation of generalised SK two point functions}
\label{sec:2ptGenSK}
\subsection{Structure of generalised SK two point functions}
We would now like to generalise the column vector representation in Eqn\eqref{eq:2ptSK} to two point functions on a generalised SK contour like the one shown below :
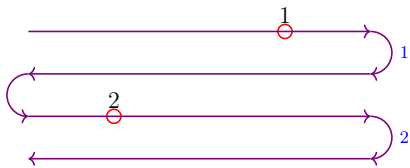
\begin{center}
\scalebox{0.75}{\contourSS{-1}{-1}{0}}
\end{center}
Each of the of the two insertions can lie on any of the four legs of the  contour thus resulting in a $4\times 4$ array of contour ordered two point functions. This is a simple enough
correlator that all the contour-ordering can be explicitly worked out. We obtain
\begin{equation}
\begin{split}
\langle \mathcal{T}_C \Phi_i(x_1) \Phi_j(x_2) \rangle 
 &= \Theta_{12} \Mtwo(x_1,x_2) + \text{permutation}
 \end{split}
\end{equation}
with the Wightman array
 \be
\begin{split}
\Mhotwo= 
  \begin{pmatrix}
    \langle 12\rangle & \langle  21\rangle & \langle  21\rangle & \langle  21\rangle \\
     \langle 12\rangle & \langle  21\rangle & \langle  21\rangle &  \langle  21\rangle \\
     \langle 12\rangle &  \langle 12\rangle &  \langle 12\rangle & \langle  21\rangle \\
       \langle 12\rangle & \langle 12\rangle &  \langle 12\rangle & \langle  21\rangle \\
  \end{pmatrix}\ .
  \end{split}
\ee
In Fourier domain, we have
\[ \Mtwo(x_1,x_2) 
 \equiv \int_{p_1}\int_{p_2} \Mtildetwo(p_1,p_2)  e^{ip_k\cdot x_k} \]
with
 \be \label{Eq:Mtildetwo}
\begin{split}
\Mtildetwo= -\rho[12]
  \begin{pmatrix}
    \bose_2 & 1+ \bose_2 & 1 + \bose_2 & 1+ \bose_2 \\
      \bose_2 & 1+ \bose_2 & 1 + \bose_2 & 1+ \bose_2 \\
     \bose_2 & \bose_2 & \bose_2 & 1+ \bose_2  \\     
     \bose_2 & \bose_2 & \bose_2 & 1+ \bose_2  \\  
  \end{pmatrix}\ .
  \end{split}
\ee
We want to now choose a judicious basis of column vectors which make the causal structure of this array manifest. To see why a good basis
is required, note that, a generic $4\times 4$ array decomposed in a general basis is a sum of $16$ tensor products. In the context of perturbation theory,
using the above two point function as the propogator, this is the statement that, naively we seem to have 4 fields in the path-integral 
which all get changed into each other during time-evolution, thus producing $16$ propagators. The concomitant proliferation of diagrams arising from this fact
 seems to be intimidating to  
all but profligate diagrammars.

This way of proceeding is, however, excessively inefficient for the array under question. For example, here is a column vector decomposition which does much better (with only two tensor products) :
\be
\begin{split}
\Mtildetwo= \rho[12]
\left\{\left(
\begin{array}{c}
-1\\
-1\\
0 \\ 
0
\end{array}
\right)
 \otimes \left(
\begin{array}{c}
\bose_2\\
1+ \bose_2\\
1 + \bose_2 \\ 
1+ \bose_2
\end{array}
\right)
 +    
\left(
\begin{array}{c}
0\\
0\\
-1 \\ 
-1
\end{array}
 \right)
 \otimes 
\left(
\begin{array}{c}
\bose_2\\
\bose_2\\
 \bose_2 \\ 
1+ \bose_2
\end{array}
\right)
\right\}\ .
  \end{split}
  \label{eq:MtildeTwo}
\ee
This is the OTO analogue of the familiar statement in real time SK perturbation theory: by a judicious choice of basis which exploits the causal/KMS structure, 
the number of propagators/diagrams can be reduced drastically.

\subsection{A basis of row vectors from causality and KMS}
Let us pause to examine why such a simplification is made possible. By analysing  the array in Eq.\eqref{Eq:Mtildetwo}, we note that $\Mtildetwo$ is annihilated by the following 
row vectors contracted to its first index (i.e., the index corresponding to its future-most operator) :
\be
\begin{split}
&\eF{1}(\omega_1)\equiv (-1 \ , 1 \ , 0\ , 0 )\ ,\\
&\eF{2}(\omega_1)\equiv (0 \  ,0 \ , -1\  ,1 )\ .
\end{split}
\label{eq:eFDef}
\ee
Here the subscript $F$ is to remind us that these vectors annihilate the future-most index. This is the multi time-fold analogue of the largest time equation, whereby 
if the future-most operator is set to be a difference operator, the correlator vanishes.

In pictures, the annihilation by the row vector $\eF{1}$ is the statement that the following combination of correlators vanish (irrespective of the position of the operator insertion $2$, provided it is in the past of insertion $1$) :
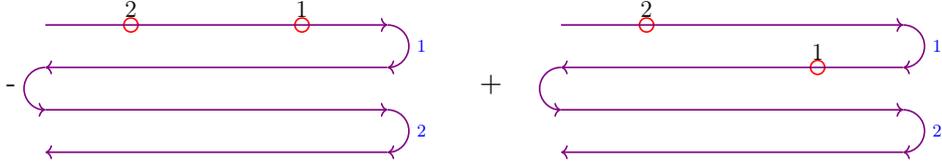
\begin{center}
-\raisebox{-0.9 cm}{\scalebox{0.75}{\contourSS{-1}{-0}{-0}}} \quad $+$\quad \raisebox{-0.9 cm}{\scalebox{0.75}{\contourSS{-1}{-0}{-0.5}}}
\end{center}
Thus, we say row vector $\eF{1}$ encodes the sliding of operators against the first future turning point.  A similar picture of the row vector $\eF{2}$ describing the sliding across
the \emph{second} future turning point is :
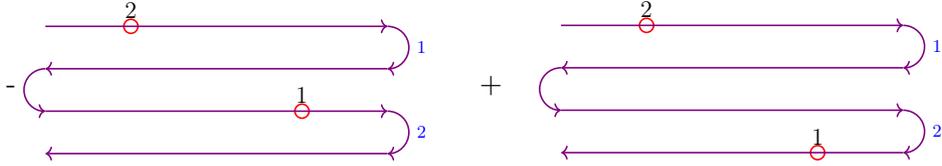
\begin{center}
-\raisebox{-0.9 cm}{\scalebox{0.75}{\contourSS{-1}{-0}{-1.0}}} \quad $+$\quad \raisebox{-0.9 cm}{\scalebox{0.75}{\contourSS{-1}{-0}{-1.5}}}
\end{center}
These statements immediately generalise to any number of time-fold contours. In the case of $k$ time-folds, the row-vectors that annihilate the future most index
are $2k$ dimensional and they are $k$ in number :
 \begin{equation}
\begin{split}
\eF{1}(\omega)&\equiv (-1,1,0,...,0),\\
\eF{2}(\omega) &\equiv  (0,0,-1,1,0,...,0),\\  ....,\\\
\eF{j}(\omega) &\equiv   (0,0,0,
\ldots,-1_{2j-1},1_{2j},0,...,0),\\  ....,\\
\eF{k}(\omega) &\equiv (0,0,....,-1,1)\ .
\end{split}
\label{eq:eFDefk}
\end{equation}
Here, the row vector $\eF{j}$ describes the sliding across the  $j$'th future turning point.

The array $\Mtildetwo$ is also annihilated by the following row
vectors contracted to its second index (i.e., the index corresponding to its past-most operator) :
\be
\begin{split}
\eP{1}(\omega_2) &\equiv (1 \ , 0 \ , 0\  ,-e^{-\beta \omega_2} )\ ,\\
\eP{2}(\omega_2) &\equiv (0 \ , -1 \ , 1\  , 0 )\ ,
\end{split}
\label{eq:ePDef}
\ee
where we have used $\bose_2 =e^{-\beta \omega_2}(1+\bose_2)$.  Here the subscript $P$ is to remind us that these vectors annihilate the past-most index.

In pictures, the annihilation by the row vector $\eP{1}$ is the statement that the following combination of correlators vanish (irrespective of the position of the operator insertion $1$, provided it is in the future of insertion $2$) :
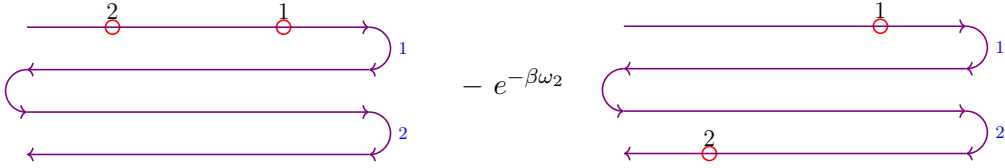
\begin{center}
\raisebox{-0.9 cm}{\scalebox{0.75}{\contourSS{-1}{-0}{-0}}} \quad $-\ e^{-\beta \omega_2}$\quad   \raisebox{-0.9 cm}{\scalebox{0.75}{\contourSS{-1}{-1.5}{-0}}}
\end{center}
This is a frequency domain version of the KMS relation
\be
\langle \Op{O}_1(t_1) \Op{O}_2(t_2) \rangle=\langle \Op{O}_2(t_2-i\beta) \Op{O}_1(t_1) \rangle\ .
\label{2pt:18}
\ee
The readers familiar with the thermal SK formalisms will recognise the above as the combination which occurs in the retarded-advanced (RA) formalism  for SK correlators\cite{aurenche1992comparison,van1992finite,Baier:1993yh,vanEijck:1994rw}. Thus, 
the row vector $\eP{1}$ describes the sliding across the thermal density matrix, which, by convention, is treated as the first past turning point.

 A similar picture of the row vector $\eP{2}$ describing the sliding across
the \emph{second} past turning point is :
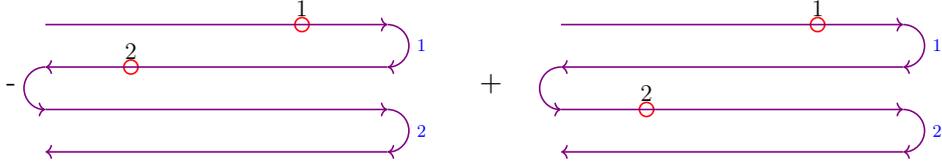
\begin{center}
-\raisebox{-0.9 cm}{\scalebox{0.75}{\contourSS{-1}{-0.5}{-0.0}}} \quad $+$\quad \raisebox{-0.9 cm}{\scalebox{0.75}{\contourSS{-1}{-1.0}{-0.0}}}
\end{center}
These statements again generalise to any number of time-fold contours. In the case of $k$ time-folds, the row-vectors that annihilate the past most index
are $2k$ dimensional and they are $k$ in number :
\begin{equation}
\begin{split}
\eP{1}(\omega) &\equiv (1,0,0,0...,0,-e^{-\b\o}),\\
\eP{2}(\omega) &\equiv  (0,-1,1,0,....,0),\\
\eP{3}(\omega) &\equiv   (0,0,0,-1,1,0...),\\  ....,\\
\eP{j}(\omega) &\equiv   (0,0,0,
\ldots,-1_{2j-2},1_{2j-1},0,...,0)\ .
\end{split}
\label{eq:ePDefk}
\end{equation}
Here, the row vector $\eP{j}$ describes the sliding across the  $j^{\text{th}}$ past turning point. Thus causality and KMS conditions naturally choose a basis\footnote{More precisely, one obtains 
a basis at non-zero frequencies $\omega\neq 0$. Throughout this work, we will stay away from the special points where any  one or more of the external frequencies go to zero. The expressions we 
write down, in general, receive contact term corrections at these special loci. } of $2k$
row vectors $\{\eF{j},\eP{j}\}$ which annihilate the future-most and past-most indices, thus implementing largest and smallest time equations.

Returning back to the case of $k=2$ time-folds, we conclude that, due to causality and KMS conditions,  $\Mtildetwo$  contracted with the following $12$ of the $16$ basis tensors vanish :
 \begin{equation}
\begin{split}
\eF{1}\otimes \eP{1}\ ,&\qquad \eF{1}\otimes \eP{2}\ ,\qquad \eF{1}\otimes \eF{1}\ ,\qquad \eF{1}\otimes \eF{2}\ ,\\
\eF{2}\otimes \eP{1}\ ,&\qquad \eF{2}\otimes \eP{2}\ ,\qquad \eF{2}\otimes \eF{1}\ ,\qquad \eF{2}\otimes \eF{2}\ ,\\
\eP{1}\otimes \eP{1}\ ,&\qquad \eP{1}\otimes \eP{2}\ ,\\
\eP{2}\otimes \eP{1}\ ,&\qquad \eP{2}\otimes \eP{2}\  .
\end{split}
\end{equation}
We will call such tensor structure orthogonal to $\Mtildetwo$ as \emph{orthogonal tensors}. We will also introduce the following notation to denote the array 
$\Mtildetwo$ contracted against these row vectors 
\begin{equation}
\begin{split}
\Mtildetwo_{AB}^{rs}&\equiv \Mtildetwo_{ij}\ (e^{(r)}_A)^i\  (e^{(s)}_B)^j\equiv \Mtildetwo \cdot  e^{rs}_{AB} \  ,
\end{split}
\end{equation}
where $A,B\in\{P,F\}$ and $i,j\in \{1,2\}$ (or more generally $i,j\in\{1,2,\ldots,k\}$). We have also introduced the convenient notation $ e^{rs}_{AB} \equiv e^{(r)}_A\otimes e^{(s)}_B$.

For example, the orthogonal tensors listed above imply that the following contractions of  the array 
$\Mtildetwo$ are 
zero :
\begin{equation}
\begin{split}
\Mtildetwo_{FP}^{11}\ , &\qquad  \Mtildetwo_{FP}^{12}\ , \qquad  \Mtildetwo_{FF}^{11}\ ,\qquad  \Mtildetwo_{FF}^{12}\ ,\\
\Mtildetwo_{FP}^{21}\ , &\qquad  \Mtildetwo_{FP}^{22}\ , \qquad  \Mtildetwo_{FF}^{21}\ ,\qquad  \Mtildetwo_{FF}^{22}\ ,\\
\Mtildetwo_{PP}^{11}\ , &\qquad  \Mtildetwo_{PP}^{12}\ ,\\
\Mtildetwo_{PP}^{21}\ , &\qquad  \Mtildetwo_{PP}^{22}\  .
\end{split}
\end{equation}

\subsection{Dual basis of Column vectors}
We concluded the previous subsection with the result that contractions of  the array 
$\Mtildetwo$ with many of the basis tensors vanish. By elementary linear algebra,
these contractions are essentially components of the array in the \emph{dual} basis.

To see this, let us begin by computing the basis of column vectors which is dual to the basis of 4-dimensional row vectors mentioned above. We have
\be
\begin{split}
&\ebP{1}(\omega)\equiv (1+\bose(\omega) \ ,\ 1+\bose(\omega) \ ,\ 1+\bose(\omega)\  ,1+\bose(\omega) )^T\ ,\\
&\ebP{2}(\omega)\equiv (\bose(\omega) \ ,\ \bose(\omega) \ ,\ 1+\bose(\omega)\  ,1+\bose(\omega) )^T\ ,\\
&\ebF{1}(\omega)\equiv (\bose(\omega) \ ,\ 1+\bose(\omega) \ ,\ 1+\bose(\omega)\  ,1+\bose(\omega) )^T\ ,\\
&\ebF{2}(\omega)\equiv (\bose(\omega) \ ,\ \bose(\omega) \ ,\ \bose(\omega)\  ,1+\bose(\omega) )^T\ ,
\end{split}
\label{2pt:25}
\ee
for any frequency $\omega\neq 0$, where $\bose(\omega) \equiv \frac{1}{e^{\beta \omega}-1}$ is the Bose-Einstein distribution.  These dual column vectors satisfy the following:
\be
\begin{split}
&\eP{i}(\omega)\cdot\ebF{j}(\omega)=\eF{i}(\omega)\cdot\ebP{j}(\omega)=0\ ,\\
&\eP{i}(\omega)\cdot\ebP{j}(\omega)=\eF{i}(\omega)\cdot\ebF{j}(\omega)=\delta^{ij}\ ,
\end{split}
\label{2pt:26}
\ee
for $i,j\in\{1,2\}$. A dual basis for $k$ time-folds can also be constructed and  takes the form :
\begin{equation}
\begin{split}
 \bar{e}_{_P}^{(j)}(\omega) &\equiv   \{ \underbrace{\bose(\omega),\bose(\omega),\ldots,\bose(\omega)}_{2 j -2 \mbox{ times} },\underbrace{1+\bose(\omega),1+\bose(\omega),\ldots,1+\bose(\omega)}_{2 k-2j+2 \mbox{  times} }   \}^T\ ,   \\
 \bar{e}_{_F}^{(j)}(\omega) &\equiv   \{ \underbrace{\bose(\omega),\bose(\omega),\ldots,\bose(\omega)}_{2 j -1 \mbox{ times} },\underbrace{1+\bose(\omega),1+\bose(\omega),\ldots,1+\bose(\omega)}_{2 k-2j+1 \mbox{  times} }   \}^T    \ .
\end{split}
\end{equation}

One can then expand the array $ \Mtildetwo$ in the basis of tensor products of these column vectors defined at $\omega_1$ and $\omega_2$ :
\be
\begin{split}
\Mtildetwo&= \Mc_{PP}^{ij}\ \ebP{i}(\omega_1)\otimes \ebP{j}(\omega_2)+ \Mc_{FP}^{ij}\ \ebF{i}(\omega_1)\otimes \ebP{j}(\omega_2)\\
&\qquad+ \Mc_{FF}^{ij}\ \ebF{i}(\omega_1)\otimes \ebF{j}(\omega_2)+ \Mc_{PF}^{ij}\ \ebP{i}(\omega_1)\otimes \ebF{j}(\omega_2)\\
&= \Mc_{PF}^{ij}\ \ebP{i}(\omega_1)\otimes \ebF{j}(\omega_2)\ .
\end{split}
\label{2pt:27}
\ee
Here $\Mtildetwo_{AB}^{rs}$ denotes the array contracted against the basis row vectors and in the last step, we have used the fact that many of these contractions vanish.
So, we have to fix the 4 coefficients - $\Mc_{PF}^{11},\Mc_{PF}^{12},\Mc_{PF}^{21}$ and $\Mc_{PF}^{22}$.

Let us look at the expansion of these coefficients in terms of the elements of the array $\Mtildetwo$ (i.e., the usual contour ordered correlators):
\be
\begin{split}
& \Mc_{PF}^{11}=-\Mtildetwo_{11}+\Mtildetwo_{12}+e^{-\beta \omega_1}\Mtildetwo_{41}-e^{-\beta \omega_1}\Mtildetwo_{42}= -\rho[12]\ , \\ 
& \Mc_{PF}^{12}=-\Mtildetwo_{13}+\Mtildetwo_{14}+e^{-\beta \omega_1}\Mtildetwo_{43}-e^{-\beta \omega_1}\Mtildetwo_{44}= e^{-\beta\omega_1} \rho[12]\ , \\  
&  \Mc_{PF}^{21}=\Mtildetwo_{21}-\Mtildetwo_{22}-\Mtildetwo_{31}+\Mtildetwo_{32} = \rho[12]\ ,\\ 
 & \Mc_{PF}^{22}=\Mtildetwo_{23}-\Mtildetwo_{24}-\Mtildetwo_{33}+\Mtildetwo_{34}= -\rho[12]\ . 
\end{split}
\label{2pt:31}
\ee
Here, we have computed the correlators directly term by term. We note a few salient aspects of this result : first, many of the contractions with these basis vectors naturally 
evaluate to the spectral function $ \rho[12]$. Second, since all the components are proportional to $ \rho[12]$, we can deduce additional linear combinations which vanish :
\be
\begin{split}
 \Mc_{PF}^{11}+\Mc_{PF}^{21} \ ,\quad \Mc_{PF}^{22}+e^{\beta\omega_1}\Mc_{PF}^{12}\ ,\quad \Mc_{PF}^{22}-\Mc_{PF}^{11} . 
\end{split}
\ee
This is equivalent to the statement that there are three additional, non-trivial orthogonal tensors to $\Mtildetwo$ : 
\be
\begin{split}
 \eP{1}\otimes \eF{1}+  \eP{2}\otimes \eF{1}\ ,\quad \eP{2}\otimes \eF{2}+e^{\beta\omega_1}\eP{1}\otimes \eF{2}\ ,\quad \eP{2}\otimes \eF{2}- \eP{1}\otimes \eF{1} . 
\end{split}
\ee
If we could somehow deduce the complete set of orthogonal tensors independently, then the only explicit computation needed is that of $\Mc_{PF}^{11}$. We will develop a method to do so
in the appendix~\S\ref{app:rules}, using which we systematically tabulate all the orthogonal tensors for $2,3$ and $4$ point functions in the appendices~\S\ref{app2pt}, \S\ref{app3pt} and \S\ref{app4pt} respectively.

Returning to  the array $\Mtildetwo$, it can be expressed as
\be
\begin{split}
\Mtildetwo=\rho[12] & \Big (\ebP{2}(\omega_1)\otimes \ebF{1}(\omega_2)-\ebP{1}(\omega_1)\otimes \ebF{1}(\omega_2)\\&+e^{-\beta \omega_1}\ebP{1}(\omega_1)\otimes \ebF{2}(\omega_2)-\ebP{2}(\omega_1)\otimes \ebF{2}(\omega_2) \Big )\ .
\end{split}
\label{2pt:48}
\ee
Let us define 
\be
\begin{split}
\eP{3}(\omega) &\equiv (e^{\b\o},0,0,-1) = e^{\b\o}  \eP{1}(\omega)\ ,   \\
\ebP{3}(\omega) &\equiv e^{-\beta \omega} \ebP{1}(\omega)=\frac{\bose(\omega)}{1+\bose(\omega)}\ebP{1}(\omega)=(\bose(\omega),\ \bose(\omega),\ \bose(\omega),\ \bose(\omega))^T\ .
\end{split}
\label{2pt:49}
\ee
Then, the final expression of the array $\Mtildetwo$ in terms of tensor products of the column vectors is as follows:
\be
\begin{split}
\Mtildetwo&=\rho[12]  \sum_{i=1}^2 \Big (\ebP{i+1}(\omega_1)\otimes \ebF{i}(\omega_2)-\ebP{i}(\omega_1)\otimes \ebF{i}(\omega_2) \Big ) \\ &=\rho[12]  \sum_{i=1}^2 \Big (\ebP{i+1}(\omega_1)-\ebP{i}(\omega_1)\Big )\otimes \ebF{i}(\omega_2)
\end{split}
\label{2pt:50}
\ee
One can check that these two terms correspond exactly to the terms encountered in Eq.\eqref{eq:MtildeTwo}:
\[
\underbrace{\left( 
\begin{array}{c}
-1\\
-1\\
0 \\ 
0
\end{array}
\right)}_{\ebP{2} - \ebP{1}}
 \otimes 
\underbrace{\left(
\begin{array}{c}
\bose_2\\
1+ \bose_2\\
1 + \bose_2 \\ 
1+ \bose_2
\end{array}
\right)}_{\ebF{1}}
 +    
\underbrace{\left(
\begin{array}{c}
0\\
0\\
-1 \\ 
-1
\end{array}
 \right)}_{\ebP{3} - \ebP{2}}
 \otimes 
\underbrace{\left(
\begin{array}{c}
\bose_2\\
\bose_2\\
 \bose_2 \\ 
1+ \bose_2
\end{array}
\right)}_{\ebF{2}}\ .
 \]

As we will show  in appendix~\S\ref{app2pt}, for $k$ time folds, the above result simply generalises to 
\be
\boxed{\begin{split}
\matMt{k,2}(\text{2-Pt})&=\rho[12]\ \sum_{r=1}^k  \Big(\ebP{r+1}-\ebP{r}\Big)\otimes \ebF{r}\ .
\end{split}}
\ee
Here, as in $k=2$, we have defined 
\begin{equation}
\begin{split}
\eP{k+1}(\omega) &\equiv (e^{\b\o},0,0,0...,0,-1) = e^{\b\o}  \eP{1}(\omega)\ ,   \\
\ebP{k+1}(\omega)  &\equiv   \{ \bose(\omega),\bose(\omega),\ldots,\bose(\omega) \}^T = e^{-\b\o}  \ebP{1}(\omega)   \ .
\end{split}
\end{equation}
Instead of $4k^2$ tensor products, only $2k$ tensor products appear, illustrating how much of simplification a judicious choice of basis can achieve.

 
 \section{Spectral representation of Higher point  correlators} 
 \label{sec:npt}
 We will now move on to the question of how we get a spectral representation for the higher point functions. Given the description in the previous sections, the basic logic
 on how to proceed is clear.

 First, we systematically construct all the orthogonal tensors of the Wightman array in Fourier domain. This constrains the form of the array
 to a great extent with a few undetermined coefficients.  In fact, we can count the number of orthogonal tensors  using the following  fact derived in \cite{Haehl:2017eob} :
 after KMS conditions are imposed, for sufficiently large $k$,  there are  $(n-1)!$ independent $n$ point correlators. Thus, among $(2k)^n$ tensors, $(2k)^n-(n-1)!$ linear combinations would be 
 orthogonal to the Wightman array. While the logic is straightforward, one needs to proceed systematically and algorithmically. We will describe precisely such  a systematic method to list the orthogonal 
 tensors in  the appendix~\S\ref{app:rules}, where the curious reader can find the details behind our results. 
 
 Once the orthogonal tensors have been  enumerated and the most general form with $(n-1)!$ undetermined coefficients is written down, we can then 
 fix the undetermined coefficients by $(n-1)!$ calculations. 
 
 \subsection{Spectral representation of  three point functions}
 Implementing the above logic for three point functions, we find a simple expression in 
 the column vector basis : 
  \be \boxed{
\begin{split}
\matMt{k,3}(\text{3-Pt})&=\rho[321]  \sum_{r=1}^{k} ( \ebP{r+1}\otimes \ebP{r+1}-\ebP{r} \otimes \ebP{r})\otimes \ebF{r}\\
&\quad-\rho[123]  \sum_{r=1}^{k} ( \ebP{r+1}-\ebP{r})\otimes \ebF{r}\otimes \ebF{r}\ .  \end{split}}
\ee
 Here, the spectral functions that appear in the Wightman array are defined by 
\begin{equation}
\begin{split}
 \int_{p_1}\int_{p_2}\int_{p_3} \rho[123]\ e^{i(p_1\cdot x_1+p_2\cdot x_2+p_3\cdot x_3)} &\equiv \langle[[\Phi(x_1),\Phi(x_2)],\Phi(x_3)]\rangle \ ,\\
  \int_{p_1}\int_{p_2}\int_{p_3} \rho[321]\ e^{i(p_1\cdot x_1+p_2\cdot x_2+p_3\cdot x_3)} &\equiv \langle[[\Phi(x_3),\Phi(x_2)],\Phi(x_1)]\rangle \ .
 \end{split}
\end{equation}
The contour ordered correlators are then given by
\begin{equation}
\begin{split}
\langle \mathcal{T}_C \Phi_i(x_1) \Phi_j(x_2) \Phi_k(x_3)\rangle &\equiv  \int_{p_1}\int_{p_2} \int_{p_3} \Theta_{123}\ \Mtildethree(\text{3-Pt})\  e^{i(p_1\cdot x_1+p_2\cdot x_2+p_3\cdot x_3)} \\
&\qquad +\text{(Rest of the $3!$ permutations)}\ .
 \end{split}
\end{equation}
 In the above expression, when the time-orderings are permuted, the tensors should also be  permuted as before. We note again the drastic reduction in number of tensor products
 due to the choice of the basis : we go from $8k^3$ possible terms to only $4k$ non-zero terms. Among the $8$ possible set of permutations of $\{P,F\}$ that can occur, causality
 forbids all combinations except two : $PPF$ and $PFF$. And when we contract $\Mtildethree$ with row vectors in these two sectors, we naturally obtain the two independent spectral functions that characterise $3$-pt. functions.
 
 Let us look at some examples. Consider the example of $k=1$ (Schwinger-Keldysh). We get   :
 \begin{equation}
\begin{split}
 \Mtildethree(\text{3-Pt})
 &= -\rho[123]\   \begin{pmatrix} -1\\ -1 \end{pmatrix}\otimes  \begin{pmatrix} \bose_2\\ 1+\bose_2 \end{pmatrix}
 \otimes    \begin{pmatrix} \bose_3\\ 1+\bose_3 \end{pmatrix} \\
&\ +\rho[321]\   \left[ \begin{pmatrix} \bose_1\\    \bose_1 \end{pmatrix} \otimes  \begin{pmatrix} \bose_2\\    \bose_2 \end{pmatrix} \otimes \begin{pmatrix} \bose_3\\ 1+\bose_3 \end{pmatrix} -
 \begin{pmatrix} 1+\bose_1\\    1+\bose_1 \end{pmatrix}\otimes  \begin{pmatrix} 1+\bose_2\\    1+\bose_2 \end{pmatrix} \otimes \begin{pmatrix} \bose_3\\ 1+\bose_3 \end{pmatrix} \right]\ .
 \end{split}
\end{equation}
This structure, when multiplied by the step function $\Theta_{123}$ and then summed over all its permutations, yields the contour-ordered 3-point functions of the Keldysh contour.
For example, here is the term obtained by $1\leftrightarrow3$ exchange (i.e., the combination that multiplies $\Theta_{321}$) :  
\begin{equation}
\begin{split}
&-\rho[321]\   \begin{pmatrix} \bose_1\\ 1+\bose_1 \end{pmatrix}\otimes   \begin{pmatrix} \bose_2\\ 1+\bose_2 \end{pmatrix}  
 \otimes   \begin{pmatrix} -1\\ -1 \end{pmatrix}  \\
&\qquad +\rho[123]\  \left[  \begin{pmatrix} \bose_1\\   1+ \bose_1 \end{pmatrix}\otimes  \begin{pmatrix} \bose_2\\    \bose_2 \end{pmatrix} \otimes \begin{pmatrix} \bose_3\\ \bose_3 \end{pmatrix} -  
\begin{pmatrix} \bose_1\\    1+\bose_1 \end{pmatrix} \otimes  \begin{pmatrix} 1+\bose_2\\    1+\bose_2 \end{pmatrix} \otimes \begin{pmatrix} 1+\bose_3\\ 1+\bose_3 \end{pmatrix} \right]\ .
 \end{split}
\end{equation}

The Wightman array for the $k=2$ contour is  given similarly by 
\begin{equation}
\begin{split}
& \Mtildethree(\text{3-Pt})=-\rho[123]\ 
\left[ 
\underbrace{\left( 
\begin{array}{c}
-1\\
-1\\
0 \\ 
0
\end{array}
\right)}_{\ebP{2} - \ebP{1}}
 \otimes 
\underbrace{\left(
\begin{array}{c}
\bose_2\\
1+ \bose_2\\
1 + \bose_2 \\ 
1+ \bose_2
\end{array}
\right)}_{\ebF{1}}
\otimes 
\underbrace{\left(
\begin{array}{c}
\bose_3\\
1+ \bose_3\\
1 + \bose_3 \\ 
1+ \bose_3
\end{array}
\right)}_{\ebF{1}}
 +    
\underbrace{\left(
\begin{array}{c}
0\\
0\\
-1 \\ 
-1
\end{array}
 \right)}_{\ebP{3} - \ebP{2}}
 \otimes 
\underbrace{\left(
\begin{array}{c}
\bose_2\\
\bose_2\\
 \bose_2 \\ 
1+ \bose_2
\end{array}
\right)}_{\ebF{2}}
\otimes 
\underbrace{\left(
\begin{array}{c}
\bose_3\\
\bose_3\\
 \bose_3 \\ 
1+ \bose_3
\end{array}
\right)}_{\ebF{2}}
\right] \\
&\ +\rho[321]\  \left[  
\underbrace{\left(  
\begin{array}{c}
\bose_1\\
\bose_1\\
1 + \bose_1 \\ 
1+ \bose_1
\end{array}
 \right)}_{\ebP{2}}
 \otimes 
\underbrace{\left(
\begin{array}{c}
\bose_2\\
\bose_2\\
1 + \bose_2 \\ 
1+ \bose_2
\end{array}
 \right)}_{\ebP{2}}
\otimes 
\underbrace{ \left(
\begin{array}{c}
\bose_3\\
1+ \bose_3\\
1 + \bose_3\\ 
1+ \bose_3
\end{array}
\right)}_{\ebF{1}}  - 
\underbrace{  \left( 
\begin{array}{c}
1+ \bose_1\\
1+ \bose_1\\
1 + \bose_1 \\ 
1+ \bose_1
\end{array}
\right)}_{\ebP{1}}
 \otimes 
\underbrace{\left(
\begin{array}{c}
1+ \bose_2\\
1+ \bose_2\\
1 + \bose_2 \\ 
1+ \bose_2
\end{array}
 \right)}_{\ebP{1}}
\otimes 
\underbrace{ \left(
\begin{array}{c}
\bose_3\\
1+ \bose_3\\
1 + \bose_3\\ 
1+ \bose_3
\end{array}
\right)}_{\ebF{1}}\ \right.\\
&\ \left.+\underbrace{\left( 
\begin{array}{c}
\bose_1\\
\bose_1\\
\bose_1 \\ 
\bose_1
\end{array}
\right)}_{\ebP{3}}
 \otimes 
\underbrace{\left( 
\begin{array}{c}
\bose_2\\
\bose_2\\
\bose_2 \\ 
\bose_2
\end{array}
 \right)}_{\ebP{3}}
\otimes
\underbrace{ \left(
\begin{array}{c}
\bose_3\\
\bose_3\\
\bose_3 \\ 
1+ \bose_3
\end{array}
\right)}_{\ebF{2}} - 
\underbrace{\left( 
\begin{array}{c}
\bose_1\\
\bose_1\\
1 + \bose_1 \\ 
1+ \bose_1
\end{array}
\right)}_{\ebP{2}}
 \otimes 
\underbrace{\left(
\begin{array}{c}
\bose_2\\
\bose_2\\
1 + \bose_2 \\ 
1+ \bose_2
\end{array}
 \right)}_{\ebP{2}} 
\otimes 
\underbrace{ \left(
\begin{array}{c}
\bose_3\\
\bose_3\\
\bose_3 \\ 
1+ \bose_3
\end{array}
\right)}_{\ebF{2}}
 \right]\ .
 \end{split}
\end{equation}
As for the $k=1$ case, this structure when multiplied by the step function $\Theta_{123}$, summed over the permutations and Fourier transformed gives the contour ordered correlators.

 \subsection{Spectral representation of  four point functions}
 We now turn to the 4-point correlators. Classifying the orthogonal tensors and fixing the remaining coefficients, we can write the 
 column vector representation for $4$pt functions. There are $16$ combinations which are a priori possible, but most of them are forbidden by 
 causality. Only $4$ combinations $PPPF,PFFF,PFPF,PPFF$ are allowed. We can thus write
\be
\boxed{\begin{split}
\matMt{k,4}(\text{4-Pt})&=\matMt{k,4}_{PPPF}+\matMt{k,4}_{PFFF}+\matMt{k,4}_{PFPF}+\matMt{k,4}_{PPFF}
\end{split}}
\ee

where we obtain

\be
\boxed{\begin{split}
\matMt{k,4}_{PPPF}&=-\rho[4321]
\sum_{r=1}^k \Big (\ebP{r+1} \otimes \ebP{r+1}\otimes \ebP{r+1} \otimes \ebF{r}-\ebP{r} \otimes \ebP{r} \otimes \ebP{r} \otimes \ebF{r}\Big )\ , \\
\matMt{k,4}_{PFFF}&=\rho[1234]
\sum_{r=1}^k \Big (\ebP{r+1}-\ebP{r}\Big ) \otimes \ebF{r} \otimes \ebF{r}\otimes \ebF{r}\ ,\\
\matMt{k,4}_{PFPF}&= \sum_{r,s=1}^k  \Big(\theta_{r>s}\ \rho[12][34]+\theta_{r \leq s}\ \rho[34][12]\Big)\Big (\ebP{r+1}-\ebP{r}\Big ) \otimes \ebF{r} \otimes \ebP{s+1} \otimes \ebF{s}\\
& -\sum_{r,s=1}^k   \Big(\theta_{r\geq s} \rho[12][34] +\theta_{r < s}\ \rho[34][12]\Big)  \Big(\ebP{r+1}-\ebP{r}\Big) \otimes \ebF{r}\otimes \ebP{s} \otimes \ebF{s}\ ,\\
\matMt{k,4}_{PPFF}= \sum_{r,s=1}^k &  \Big(\theta_{r>s}\ \rho[13][24]
+\theta_{r \leq s}\ \rho[24][13] \Big)\Big (\ebP{r+1}-\ebP{r}\Big ) \otimes \ebP{s+1} \otimes \ebF{r} \otimes \ebF{s}\\
& -\sum_{r,s=1}^k \Big(\theta_{r\geq s}\ \rho[13][24]+\theta_{r < s}\ \rho[24][13] \Big)  \Big (\ebP{r+1}-\ebP{r}\Big ) \otimes \ebP{s} \otimes \ebF{r} \otimes \ebF{s}\\
& +\sum_{r,s=1}^k \Big(\theta_{r\geq s}\ \rho[14][23] +\theta_{r < s}\ \rho[23][14] \Big)  \ebP{r+1}\otimes \Big (\ebP{s+1}-\ebP{s}\Big )\otimes \ebF{s} \otimes \ebF{r}\\
&-\sum_{r,s=1}^k   \Big(\theta_{r> s}\ \rho[14][23]+\theta_{r \leq s}\ \rho[23][14] \Big)  \ebP{r}\otimes \Big (\ebP{s+1}-\ebP{s}\Big )\otimes \ebF{s} \otimes \ebF{r}\\
&+\rho[2314]\ \sum_{r=1}^k \Big (\ebP{r+1}\otimes \ebP{r+1}-\ebP{r}\otimes \ebP{r}\Big) \otimes \ebF{r} \otimes \ebF{r}\ .
\end{split}}
\ee
Here we define the spectral functions via 
\[  \int_{p_1}\int_{p_2}\int_{p_3} \int_{p_4} \rho[1234]\ e^{i(p_1\cdot x_1+p_2\cdot x_2+p_3\cdot x_3+p_4\cdot x_4)} \equiv \langle[[[\Phi(x_1),\Phi(x_2)],\Phi(x_3)],\Phi(x_4)]\rangle\ , \]
\[  \int_{p_1}\int_{p_2}\int_{p_3} \int_{p_4} \rho[12][34]\ e^{i(p_1\cdot x_1+p_2\cdot x_2+p_3\cdot x_3+p_4\cdot x_4)} \equiv \langle[\Phi(x_1),\Phi(x_2)][\Phi(x_3),\Phi(x_4)]\rangle\ . \]
Our previous comments about the reduction of number of terms extend to the four point functions : the column vector basis reduces the $16k^4$ terms that can potentially appear
in the 4-pt vertices of a $k$ time-fold contour to $12k^2+6k=6k(2k+1)$ number of terms.

For example, in the Schwinger Keldysh case, the above formulae evaluates to 
\begin{equation}
\begin{split}
 \Mtildefour(\text{4-Pt})_{\text{Nested}} &= \rho[1234]\   \begin{pmatrix} -1\\ -1 \end{pmatrix}\otimes   \begin{pmatrix} \bose_2\\ 1+\bose_2 \end{pmatrix}  
 \otimes   \begin{pmatrix} \bose_3\\ 1+\bose_3 \end{pmatrix}  \otimes   \begin{pmatrix} \bose_4\\ 1+\bose_4 \end{pmatrix} \\
&\qquad +\rho[4321]\    \begin{pmatrix} 1+\bose_1\\    1+\bose_1 \end{pmatrix}  \otimes  \begin{pmatrix} 1+\bose_2\\    1+\bose_2 \end{pmatrix}  \otimes \begin{pmatrix} 1+\bose_3\\    1+\bose_3 \end{pmatrix}  \otimes\begin{pmatrix} \bose_4\\ 1+\bose_4 \end{pmatrix}  \\
&\qquad -\rho[4321]\    \begin{pmatrix} \bose_1\\    \bose_1 \end{pmatrix}  \otimes  \begin{pmatrix} \bose_2\\    \bose_2 \end{pmatrix}  \otimes  \begin{pmatrix} \bose_3\\    \bose_3 \end{pmatrix}  \otimes \begin{pmatrix} \bose_4\\ 1+\bose_4 \end{pmatrix}  \\
&\qquad -\rho[2314]\    \begin{pmatrix} 1+\bose_1\\    1+\bose_1 \end{pmatrix}  \otimes  \begin{pmatrix} 1+\bose_2\\    1+\bose_2 \end{pmatrix}  \otimes \begin{pmatrix} \bose_3\\    1+\bose_3 \end{pmatrix}  \otimes\begin{pmatrix} \bose_4\\ 1+\bose_4 \end{pmatrix}  \\
&\qquad +\rho[2314]\    \begin{pmatrix} \bose_1\\    \bose_1 \end{pmatrix}  \otimes  \begin{pmatrix} \bose_2\\    \bose_2 \end{pmatrix}  \otimes  \begin{pmatrix} \bose_3\\    1+\bose_3 \end{pmatrix}  \otimes \begin{pmatrix} \bose_4\\ 1+\bose_4 \end{pmatrix}  \ ,
 \end{split}
\end{equation}
and
\begin{equation}
\begin{split}
 \Mtildefour(\text{4-Pt})_{\text{Double}}&=\rho[12][34]\   \begin{pmatrix} 1\\ 1 \end{pmatrix} \otimes   \begin{pmatrix} \bose_2\\ 1+\bose_2 \end{pmatrix}   
 \otimes   \begin{pmatrix} 1+\bose_3\\ 1+\bose_3 \end{pmatrix}    \otimes   \begin{pmatrix} \bose_4\\ 1+\bose_4 \end{pmatrix}   \\
&\qquad - \rho[34][12]\   \begin{pmatrix} 1\\ 1 \end{pmatrix} \otimes   \begin{pmatrix} \bose_2\\ 1+\bose_2 \end{pmatrix}   
 \otimes   \begin{pmatrix} \bose_3\\ \bose_3 \end{pmatrix}    \otimes   \begin{pmatrix} \bose_4\\ 1+\bose_4 \end{pmatrix}   \\
&\qquad +\rho[13][24]\   \begin{pmatrix} 1\\ 1 \end{pmatrix} \otimes   \begin{pmatrix} 1+\bose_2\\ 1+\bose_2 \end{pmatrix}   
 \otimes   \begin{pmatrix} \bose_3\\ 1+\bose_3 \end{pmatrix}    \otimes   \begin{pmatrix} \bose_4\\ 1+\bose_4 \end{pmatrix}   \\
&\qquad - \rho[24][13]\   \begin{pmatrix} 1\\ 1 \end{pmatrix} \otimes   \begin{pmatrix} \bose_2\\ \bose_2 \end{pmatrix}   
 \otimes   \begin{pmatrix} \bose_3\\ 1+\bose_3 \end{pmatrix}    \otimes   \begin{pmatrix} \bose_4\\ 1+\bose_4 \end{pmatrix}   \\
&\qquad -\rho[14][23]\ \begin{pmatrix} \bose_1\\ \bose_1 \end{pmatrix}  \otimes     \begin{pmatrix} 1\\ 1 \end{pmatrix}   
 \otimes   \begin{pmatrix} \bose_3\\ 1+\bose_3 \end{pmatrix}    \otimes   \begin{pmatrix} \bose_4\\ 1+\bose_4 \end{pmatrix}   \\
&\qquad + \rho[23][14]\   \begin{pmatrix} 1+\bose_1\\ 1+\bose_1 \end{pmatrix} \otimes   \begin{pmatrix} 1\\ 1 \end{pmatrix}   
 \otimes   \begin{pmatrix} \bose_3\\ 1+\bose_3 \end{pmatrix}    \otimes   \begin{pmatrix} \bose_4\\ 1+\bose_4 \end{pmatrix}  \ .
 \end{split}
\end{equation}

The boxed equations of this and the previous subsections are the main results of this work. On, one hand they give an efficient way to parametrise the contour-ordered correlators
in terms of spectral functions which are easier to compute. On the other hand, they give a basis in which the diagrammatics is simplified and the number of vertices/propagators are 
reduced drastically. The applications of this formalism will be described elsewhere.


\section{Conclusions and Discussions}
\label{sec:discussion}
In this work, we have set up the basic formalism of spectral representations of thermal out of time order correlators. We have also explicitly worked out case
of $n=2,3,4$ point functions which, in an appropriate basis,  take a nice and useful form that automatically encodes the causality and KMS conditions. This 
opens up a  way to simplify OTO perturbation theory, Feynman rules and diagrammatics at finite temperature. A particular application would be to develop a 
full-fledged OTO kinetic theory and hydrodynamics from a consistent truncation of OTO Schwinger Dyson equations. We hope our formalism can play the 
role RA formalism has played in traditional Schwinger-Keldysh applications. 

While our final results for the spectral representations take a simple and compact form, their derivation as we sketch in our appendices is somewhat elaborate due to the combinatorics
involved. Perhaps, a simpler and shorter derivation of the spectral representations that appear in this work, would provide more insight into the physics behind the simplifications we see in our final results.

A set of interesting questions for future research would be to derive cutting rules for OTOCs. Ideally, we would like to predict the OTO imaginary parts and give an `on-shell' picture 
of the physics behind them. Such a work should extend the classic work of Kobes and Semenoff \cite{Kobes:1985kc,Kobes:1986za, kobes1991retarded}
in the SK formalism (See also \cite{Guerin:1993ik}). It should also automatically incorporate the emerging understanding of the physics behind OTOCs
via operator spreading/`infection' models or OTO combustion waves \cite{Sekino:2008he,2012arXiv1212.0331O,2013PhRvL.111l7205K, Stanford:2015owe,Aleiner:2016eni}
as well as reveal the physical picture behind the OPE inversion formula and double discontinuities in CFTs \cite{Caron-Huot:2017vep,Simmons-Duffin:2017nub,Iliesiu:2018fao}.
From the viewpoint of thermal field theory, it would be interesting to extend the existing intuitions regarding hard thermal loops(HTL) \cite{Braaten:1991gm, CaronHuot:2007nw}
to OTOCs and for example, enquire whether thermal OTOCs of QCD leave a signature in the  heavy ion collision experiments.

Another set of interesting questions revolve around holography and black holes in AdS. It would be nice to have a derivation of the OTOC spectral representations in this work,
from gravity, say along the lines of \cite{Herzog:2002pc,Skenderis:2008dh}. Such a framework should allow us to compute OTO spectral functions of energy momentum tensor and
currents in strongly coupled gauge theories study their  low frequency, high temperature behaviour that gives rise to OTO hydrodynamics. 

Finally, the spectral representations we derive in this work are valid modulo contact terms in the frequency domain (since the basis row vectors we use,
become linearly dependent when any one of the external frequencies of a correlator is taken to vanish.) This is a limitation already  in the
RA formalism of SK correlators, which the extended formalism inherits. This is usually addressed by shifting to a Keldysh basis, in which only causality 
conditions are implemented (and not the KMS conditions). Consequently, the column vector type representations in such a basis contain more terms,
but have the merit of being applicable in non-equilibrum situations. A Keldysh type basis for OTOCs was described by \cite{Haehl:2017qfl} and it would 
be interesting to see how our results can be extended away from equilibrium using similar basis \cite{simpOTOdiag}.


\acknowledgments
We would like to thank Anish Kulkarni, Bibhas Ranjan Majhi, Gautam Mandal,  Shiraz Minwalla, Prithvi Narayan, Amin A. Nizami, Mukund Rangamani, Victor Ivan Giraldo Rivera,  Suvrat Raju and Spenta Wadia, 
for a variety of discussions pertaining to this work. SC would like to thank the organisers of  Bangalore Area String Meeting (BASM 2017) at ICTS-TIFR and the Gong show at 
Kavli Asian Winter School at ICTS-TIFR, where this work was presented. CC would like to thank  ICTS-TIFR for supporting him as a part of the S.N. Bhatt Memorial Fellowship  2017, when this work was done.
CC would also like to thank IIT Guwahati for academic support during the time this work was under progress.  RL would like to thank the organisers of  Saha Theory Workshop on Modern Aspects of String 
Theory at SINP, Kolkata (Feb 2018) for the opportunity to present this work. We 
 would also like to acknowledge our debt to the people of India for their steady and generous support to research in the basic sciences.


\appendix
\section{Basis of column vectors}
\label{basis}
In the series of  appendices that follow, we will sketch the computations behind the spectral representations quoted in the main text. For simplicity, we 
will suppress the spatial dependence and the dependence on the spatial momenta in the all the correlators that appear in these appendices. We will 
only retain the time/frequency dependencies. 

We will begin by describing the basis vectors that we use on the $k$-fold time contour. We expand the array  ${\bf \widetilde{M}}(\omega_1,\cdots,\omega_n)$ in the basis of tensor products of the following column vectors:

\begin{equation}
\begin{split}
 \bar{e}_{_P}^{(j)}(\omega) &\equiv   \{ \underbrace{\bose(\omega),\bose(\omega),\ldots,\bose(\omega)}_{2 j -2 \mbox{ times} },\underbrace{1+\bose(\omega),1+\bose(\omega),\ldots,1+\bose(\omega)}_{2 k-2j+2 \mbox{  times} }   \}^T  \ , \\
 \bar{e}_{_F}^{(j)}(\omega) &\equiv   \{ \underbrace{\bose(\omega),\bose(\omega),\ldots,\bose(\omega)}_{2 j -1 \mbox{ times} },\underbrace{1+\bose(\omega),1+\bose(\omega),\ldots,1+\bose(\omega)}_{2 k-2j+1 \mbox{  times} }   \}^T\ ,    \\
\end{split}
\end{equation}
where $\bose(\omega)\equiv \frac{1}{e^{\b\o}-1}$ is the Bose-Einstein distribution and $j=1,2,\ldots,k$.

The dual basis of row vectors consists of the following vectors:
\begin{equation}
\begin{split}
\eP{1}(\omega) &\equiv (1,0,0,0...,0,-e^{-\b\o}),\\
\eP{2}(\omega) &\equiv  (0,-1,1,0,....,0),\\
\eP{3}(\omega) &\equiv   (0,0,0,-1,1,0...),\\  ....,\\
\eP{j}(\omega) &\equiv   (0,0,0,
\ldots,-1_{2j-2},1_{2j-1},0,...,0),\\  ....,\\
\eP{k}(\omega) &\equiv (0,0,....,0,-1,1,0)\ ,\\ \\
\eF{1}(\omega)&\equiv (-1,1,0,...,0),\\
\eF{2}(\omega) &\equiv  (0,0,-1,1,0,...,0),\\  ....,\\\
\eF{j}(\omega) &\equiv   (0,0,0,
\ldots,-1_{2j-1},1_{2j},0,...,0),\\  ....,\\
\eF{k}(\omega) &\equiv (0,0,....,-1,1) .
\end{split}
\end{equation}

We then have 
\begin{equation}
\begin{split}
\eP{i}(\omega)  \cdot  \ebP{j}(\omega)  = \eF{i}(\omega)  \cdot \ebF{j}(\omega)  &= \delta^{ij}\ ,  \\
\eP{i}(\omega)  \cdot \ebF{j}(\omega)   = \eF{i}(\omega)  \cdot  \ebP{j}(\omega)  &=0 \ .
\end{split}
\end{equation}

Here $\eP{j}$corresponds to the difference field across $j^{th}$ past turning point (with density matrix being counted as the first past turning point) counted from the ket to the bra, whereas  $\eF{j}$corresponds to the difference field across $j^{th}$ future turning point counted from the ket to the bra. For notational convenience, we will extend the definition of these vectors to all integers  via Bloch-Floquet periodicity :
\begin{equation}
\begin{split}
\eP{j+k}(\omega)  &\equiv e^{\b\o}  \eP{j}(\omega)   \ , \qquad \ebP{j+k} (\omega)  \equiv e^{-\b\o}  \ebP{j}(\omega)\ , \\
\eF{j+k}(\omega)  &\equiv e^{\b\o}  \eF{j}(\omega)   \ , \qquad \ebF{j+k} (\omega) \equiv e^{-\b\o}  \ebF{j}(\omega) \ ,
\end{split}
\end{equation}
with
\begin{equation}
\begin{split}
\eP{i}(\omega)  \cdot  \ebP{j} (\omega)  = \eF{i}(\omega)  \cdot  \ebF{j}(\omega)  &= e^{\b\o(i-j)}\delta^{i-j\ \text{mod}\ k,0}\ ,  \\
\eP{i}(\omega)  \cdot  \ebF{j}(\omega)   = \eF{i}(\omega)  \cdot  \ebP{j}(\omega)  &=0 \ .
\end{split}
\end{equation}
In particular, we have
\begin{equation}
\begin{split}
\eP{k+1}(\omega) &\equiv (e^{\b\o},0,0,0...,0,-1) = e^{\b\o}  \eP{1}(\omega)\ ,   \\
\ebP{k+1}(\omega)  &\equiv   \{ \bose(\omega),\bose(\omega),\ldots,\bose(\omega) \}^T = e^{-\b\o}  \ebP{1}(\omega) \ ,  \\
\eF{0}(\omega)  &\equiv   \{ 0,0,\ldots,0,-e^{-\b\o}, e^{-\b\o}  \}^T = e^{-\b\o}  \eF{k}(\omega)\ ,  \\
\ebF{0}(\omega)  &\equiv   \{ 1+\bose(\omega),1+\bose(\omega),\ldots,\bose(\omega), e^{\b\o} (1+\bose(\omega))  \}^T = e^{\b\o} \ebF{k}(\omega)\ .   \\
\end{split}
\end{equation}


\section{Rules of contraction for general k}
\label{app:rules}
\subsection{Summary of the rules}
\label{rulesummary}
The contractions of the array $\matMt{k,n}(\omega_1,\cdots,\omega_n)$ with the tensor products of the row vectors introduced above give the components of the array in the basis of tensor products of the dual column vectors. For instance, 
\be
\Mc_{PPF}^{rsu}=\sum_{a=1}^4 \sum_{b=1}^4 \sum_{c=1}^4\ \eP{r}(\omega_1)_a\ \eP{s}(\omega_2)_b\ \eF{u}(\omega_3)_c \ \matMt{2,3}(\text{3-Pt})_{abc}\ .
\ee
In this example the indices of $\eF{u}(\omega_3)$ and $\eP{s}(\omega_2)$ contract with the third and the second indices of $ \matMt{2,3}(\text{3-Pt})$ respectively. These 2 indices correspond to the positions of past-most insertion and the next insertion to its future in the array $\matM{2,3}(t_1,t_2,t_3)$ which is obtained by taking the inverse Fourier transform of  $ \matMt{2,3}(\text{3-Pt})$ and multiplying by a theta function $\Theta_{123}$. In what follows, for such a contraction, we would loosely say that $\eF{u}(\omega_3)$ lies to the past of $\eP{s}(\omega_2)$. In a similar sense, we would say that  $\eP{r}(\omega_1)$ lies to the future of $\eP{s}(\omega_2)$.

As we noted in the main text, the components obtained from these contractions with tensor products of the row vectors are not all independent. Here, we enumerate a set of rules that such contractions satisfy:
\begin{enumerate}
\item \textbf{F-collapse(Largest time Eqn) :} The contraction is zero if there is an $\eF{r}$ and if there is no  $\eP{s}$ to its future such that  $s\in\{r,r+1\}$. 
\item \textbf{P-collapse(Smallest time Eqn) :} The contraction is zero if there is an $\eP{r}$ and if there is no  $\eF{s}$ to its past such that  $s\in\{r,r-1\}$. 
\item \textbf{F-sliding :} One can replace $\eF{r-1}$ by $-\eF{r}$ without changing the value of the contraction,
 \begin{enumerate}
 \item if there is an $\eP{r}$ to its future (Anchor condition) and
 \item if  there is no other  $\eF{r}$ or $\eF{r-1}$ to the past of $\eP{r}$ (Eclipse condition).
 \end{enumerate}
\item \textbf{P-sliding :} One can replace $\eP{r+1}$ by $-\eP{r}$ without changing the value of the contraction,, 
 \begin{enumerate}
 \item if there is an $\eF{r}$ to its past (Anchor condition) and
 \item if there is no other  $\eP{r}$ or $\eP{r+1}$ to the future of $\eF{j}$ (Eclipse condition).
 \end{enumerate}
\item \textbf{C-shift : } One can do a global contour translation, viz., shift all the indices by a given number i.e. do the following replacement :
\begin{equation}
\begin{split}
 \eF{r} &\mapsto \eF{r+m} \ ,\\
 \eP{s} &\mapsto \eP{s+m} 
 \end{split}
\end{equation}
for all $\eF{r} $ and $\eP{s} $ in the contraction and any integer m, without changing the value contraction.
\item \textbf{F-fragmentation :}  For a given $r$, one can do the following replacements together without changing the value of the contraction:
\begin{equation*}
\begin{split}
 \eF{r} &\mapsto \eF{r}+\eF{r+1}\ , \\
 \eF{r+m} & \mapsto \eF{r+1+m} \ \forall \ m>0\ ,\\
 \eP{r+m} &\mapsto \eP{r+1+m}\ \forall  \ m>0\ .
\end{split}
\end{equation*}
\item \textbf{P-fragmentation :} For a given $r$, one can do the following replacements  together without changing the value of the contraction:
\begin{equation*}
\begin{split}
 \eP{r} &\mapsto \eP{r}+\eP{r+1}\ , \\
 \eP{r+m} & \mapsto \eP{r+1+m} \ \forall  \ m>0\ ,\\
 \eF{r-1+m} &\mapsto \eF{r+m}\ \forall  \ m>0\ .
\end{split}
\end{equation*}

 \end{enumerate}

We are going to introduce a diagrammatic scheme to represent the contractions and then show some cases where the above rules can be applied to demonstrate why they are valid. 

\subsection{A diagrammatic scheme for the contractions:}
As we have seen earlier, unitarity of the theory allows one to slide insertions in a correlator along the contour without changing the value of the correlator as long as it does not encounter another insertion. To impose these conditions between correlators in the time domain, we can assume that similar relations hold between the corresponding elements of the array $ \matMt{k,n}(\omega_1,\cdots,\omega_n)$. So we can represent the array elements by contour diagrams with insertions as is demonstrated by the following example:
\be
\begin{split}
\matMt{2,2}(\text{2-Pt})_{42}=\raisebox{-0.9 cm}{\scalebox{0.75}{\contourSS{-1.5}{-0.5}{-1.5}}}
\end{split}
\label{appdiascheme:1}
\ee

Note that the array $\matMt{2,2}(\text{2-Pt})$ is constructed to reproduce the contour ordered correlators in the domain $t_1>t_2$. So,we put the first insertion to the right of the second insertion in the above diagram. We emphasize that this is just a digrammatic way to represent the array elements of $\matMt{k,n}(\omega_1,\cdots,\omega_n)$ and the exact horizontal position of any insertion is not important. One should just make sure that the relative horizontal positions of the insertions are in the correct order. Now, notice that each of the dual vectors $e_P^{(r)}(\omega_i)$ and $e_F^{(r)}(\omega_i)$ have just 2 nonzero elements. We can represent contraction of the array  $\matMt{k,n}(\omega_1,\cdots,\omega_n)$  with any of these vectors by drawing 2 insertions at the same horizontal position but on 2 different legs where the components of the vector are nonzero. With each insertion, we associate the corresponding element of the vector. For instance, one can represent the contraction of $\matMt{2,2}(\text{2-Pt})$  with $e_P^{(2)}(\omega_1)\otimes e_F^{(1)}(\omega_2)$ by the following diagram:

\be
\begin{split}
\matMt{2,2}(\text{2-Pt}) \cdot \Big(e_P^{(2)}(\omega_1)\otimes e_F^{(1)}(\omega_2) \Big)=\raisebox{-0.9 cm}{\scalebox{0.75}{\contourDD{-1.5}{-0}{-0.5}}}
\end{split}
\label{appdiascheme:2}
\ee
Here we remind the reader that $e_P^{(2)}(\omega_1)=(0,-1,1,0)$ and $e_F^{(1)}(\omega_2)=(-1,1,0,0)$.
The rule for obtaining the contraction from such a diagram with multiple insertions on the same horizontal position is to choose one insertion for each horizontal position, calculate the  corresponding correlator, multiply by the product of the factors associated with each chosen insertion, and then take a sum over all such possible choices. For instance, in the example given in \eqref{appdiascheme:2}, we have
\be
\begin{split}
\Mtildetwo_{PF}^{21}&=\raisebox{-0.9 cm}{\scalebox{0.75}{\contourDD{-1.5}{-0}{-0.5}}}\\
&= (-1)(-1)\raisebox{-0.9 cm}{\scalebox{0.75}{\contourSS{-1.5}{-0}{-0.5}}}+(1)(-1)\raisebox{-0.9 cm}{\scalebox{0.75}{\contourSS{-1.5}{-0}{-1}}}\\ \\
&\quad+(-1)(1)\raisebox{-0.9 cm}{\scalebox{0.75}{\contourSS{-1.5}{-0.5}{-0.5}}}+(1)(1)\raisebox{-0.9 cm}{\scalebox{0.75}{\contourSS{-1.5}{-0.5}{-1}}}\\ \\
&=\matMt{2,2}(\text{2-Pt})_{21}-\matMt{2,2}(\text{2-Pt})_{22}-\matMt{2,2}(\text{2-Pt})_{31}+\matMt{2,2}(\text{2-Pt})_{32}\ .
\end{split}
\label{appdiascheme:3}
\ee

The numbers on top of the insertions in the diagram in the first line of the above equations are the factors which are to be multiplied to get the contraction. On the other hand, the numbers on top of the insertions in the diagrams in the second and the third lines indicate horizontal positions of the insertions. In such diagrams, one can slide any insertion down or up a leg without changing the value of the contraction as long as such a sliding is not obstructed by another insertion. If one slides any insertion from the bottom-most leg to the top-most one, then one picks up an additional factor of $e^{\beta \omega}$ because of the KMS relations.

Now, let us look at some examples of contractions which are related to each other by the rules mentioned in appendix~\S\ref{rulesummary} using such diagrams. We will work with the k=2 case for the collapse rules, the sliding rules and the C-shift. For the fragmentation rules, we will give examples in the $k=4$ case.

\subsection{Some examples demonstrating the rules of contraction}
\label{examplerules}

\subsection*{F-collapse :}
Consider the contraction of $\matMt{2,3}(\text{3-Pt})$ with the tensor $\Big(e_F^{(2)}(\omega_1)\otimes e_P^{(2)}(\omega_2)\otimes e_F^{(2)}(\omega_3)\Big)$. The corresponding diagram is 

\be
\begin{split}
&\matMt{2,3}(\text{3-Pt}) \cdot \Big(e_F^{(2)}(\omega_1)\otimes e_P^{(2)}(\omega_2)\otimes e_F^{(2)}(\omega_3)\Big)\\
&=\raisebox{-0.9 cm}{\scalebox{0.75}{\contourDDD{-1.5}{-1.0}{-0.5}{-1.0}}}\\
\end{split}
\label{appfcollapse:1}
\ee

Notice that $\eF{2}(\omega_1)$ is the future-most insertion and there is no $\eP{2}$ or $\eP{1}$ to block it from collapsing. One can slide down the future-most insertion on the third leg down to the fourth leg without changing the value of the contraction as there is no other insertion to obstruct this sliding. But this leads to the pair of future-most insertions with opposite signs lying on exactly the same position. Consequently they cancel each other's contribution and the value of the contraction is $0$.

\subsection*{P-collapse :}
Consider the contraction of $\matMt{2,3}(\text{3-Pt})$ with the tensor $\Big(e_P^{(2)}(\omega_1)\otimes e_F^{(1)}(\omega_2)\otimes e_P^{(2)}(\omega_3)\Big)$. The corresponding diagram is 

\be
\begin{split}
&\matMt{2,3}(\text{3-Pt}) \cdot \Big(e_P^{(2)}(\omega_1)\otimes e_F^{(1)}(\omega_2)\otimes e_P^{(2)}(\omega_3)\Big)\\
&=\raisebox{-0.9 cm}{\scalebox{0.75}{\contourDDD{-1.5}{-0.5}{0}{-0.5}}}\\
\end{split}
\label{apppcollapse:1}
\ee

Notice that $\eP{2}(\omega_3)$ is the past-most insertion and there is no $\eF{2}$ or $\eF{1}$ to block it from collapsing. One can slide down the past-most insertion on the second leg down to the third leg without changing the value of the contraction as there is no other insertion to obstruct this sliding. But this leads to the pair of past-most insertions with opposite signs lying on exactly the same position. Consequently they cancel each other's contribution and the value of the contraction is 0.

Similarly, consider the contraction of $\matMt{2,3}(\text{3-Pt})$  with the tensor $\Big(e_P^{(2)}(\omega_1)\otimes e_F^{(1)}(\omega_2)\otimes e_P^{(1)}(\omega_3)\Big)$. The corresponding diagram is 

\be
\begin{split}
&\matMt{2,3}(\text{3-Pt}) \cdot \Big(e_P^{(2)}(\omega_1)\otimes e_F^{(1)}(\omega_2)\otimes e_P^{(1)}(\omega_3)\Big)\\
&=\raisebox{-0.9 cm}{\scalebox{0.75}{\begin{tikzpicture}[scale = 1.5]
\draw[thick,color=red](-1, 0) circle (0.5ex);
\draw[thick,color=red](-1, -1.5) circle (0.5ex);
\draw[thick,color=red](0, 0) circle (0.5ex);
\draw[thick,color=red](0, - 0.5) circle (0.5ex);
\draw[thick,color=red](1, -0.5) circle (0.5ex);
\draw[thick,color=red](1, -1.0) circle (0.5ex);
\draw[thick, color=black]
{
 (-1, 0) node [above] {\large{{1}}}
 (-1,  - 1.5) node [above] {\large{{-$e^{-\beta \omega_3}$}}}
 (0, 0) node [above] {\large{{-1}}}
 (0, - 0.5) node [above] {\large{{+1}}}
 (1, -0.5) node [above] {\large{{-1}}}
 (1, - 1.0) node [above] {\large{{+1}}}
};
\foreach \x in {0,...,-1.5}{
\pgfmathtruncatemacro\z{int(-\x + 1)}
\draw[thick,color=violet,->] (-2,\x cm) -- (2,\x cm) ;
\draw[thick,color=violet,->] (2, \x cm - 0.5 cm) -- (-2, \x cm - 0.5 cm) ;
\draw[thick,color=violet,->] (2, \x cm) arc (90:-90:0.25);
\draw[thick,color=blue] (2 + 0.25, \x cm - 0.25cm) node[right] {\footnotesize{\z}};
}
\foreach \y in {-1,...,-.5}{
\pgfmathtruncatemacro\w{int(-\y + 1)}
\draw[thick,color=violet,->] (-2,\y cm + 0.5 cm) arc (90:270:0.25);
}
\end{tikzpicture}
}}\\
\end{split}
\label{apppcollapse:2}
\ee

Again, notice that $\eP{1}(\omega_3)$ is the past-most insertion and there is no $\eF{1}$ or $\eF{2}$ to block it from collapsing. One can slide the past-most insertion on the fourth leg up to the first leg  picking up a factor $e^{\beta \omega_3}$ because of the KMS relations . Again, this leads to the pair of past-most insertions with opposite signs lying on exactly the same position. Consequently,as before, they cancel each other's contribution and the value of the contraction is 0.

\subsection*{F-sliding :}
Consider the contraction of $\matMt{2,3}(\text{3-Pt})$ with the tensor $\Big(e_P^{(2)}(\omega_1)\otimes e_P^{(2)}(\omega_2)\otimes e_F^{(1)}(\omega_3)\Big)$. The corresponding diagram is 

\be
\begin{split}
&\matMt{2,3}(\text{3-Pt}) \cdot \Big(e_P^{(2)}(\omega_1)\otimes e_P^{(2)}(\omega_2)\otimes e_F^{(1)}(\omega_3)\Big)\\
&=\raisebox{-0.9 cm}{\scalebox{0.75}{\contourDDD{-1.5}{0}{-0.5}{-0.5}}}\\
\end{split}
\label{appfsliding:1}
\ee

Notice that $\eF{1}(\omega_3)$ is the past-most insertion. There is an $\eP{2}(\omega_2)$  to its future. But, apart from $\eF{1}(\omega_3)$,there is no other $\eF{1}$ or $\eF{2}$ to the past of $\eP{2}(\omega_2)$. If we choose the past-most insertion with the the factor (-1) on the first leg, then there is no insertion to block the 2 insertions corresponding to $\eP{2}(\omega_2)$ from collapsing on to each other i.e.  one can slide either of those insertions to the position of the other and their contributions would exactly cancel each other. Therefore, we have 
\be
\begin{split}
&\matMt{2,3}(\text{3-Pt}) \cdot \Big(e_P^{(2)}(\omega_1)\otimes e_P^{(2)}(\omega_2)\otimes e_F^{(1)}(\omega_3)\Big)\\
&=\raisebox{-0.9 cm}{\scalebox{0.75}{\begin{tikzpicture}[scale = 1.5]
\draw[thick,color=red](-1, -0.5) circle (0.5ex);
\draw[thick,color=red](0, -0.5) circle (0.5ex);
\draw[thick,color=red](0, - 1.0) circle (0.5ex);
\draw[thick,color=red](1, -0.5) circle (0.5ex);
\draw[thick,color=red](1, -1.0) circle (0.5ex);
\draw[thick, color=black]
{
 (-1,  - 0.5) node [above] {\large{{+1}}}
 (0, -0.5) node [above] {\large{{-1}}}
 (0, - 1.0) node [above] {\large{{+1}}}
 (1, -0.5) node [above] {\large{{-1}}}
 (1, - 1.0) node [above] {\large{{+1}}}
};
\foreach \x in {0,...,-1.5}{
\pgfmathtruncatemacro\z{int(-\x + 1)}
\draw[thick,color=violet,->] (-2,\x cm) -- (2,\x cm) ;
\draw[thick,color=violet,->] (2, \x cm - 0.5 cm) -- (-2, \x cm - 0.5 cm) ;
\draw[thick,color=violet,->] (2, \x cm) arc (90:-90:0.25);
\draw[thick,color=blue] (2 + 0.25, \x cm - 0.25cm) node[right] {\footnotesize{\z}};
}
\foreach \y in {-1,...,-.5}{
\pgfmathtruncatemacro\w{int(-\y + 1)}
\draw[thick,color=violet,->] (-2,\y cm + 0.5 cm) arc (90:270:0.25);
}
\end{tikzpicture}
}}\\
\end{split}
\label{appfsliding:2}
\ee
One can slide the past-most insertion on the second leg down to the third leg without changing the value of the contraction and obtain
\be
\begin{split}
&\matMt{2,3}(\text{3-Pt}) \cdot \Big(e_P^{(2)}(\omega_1)\otimes e_P^{(2)}(\omega_2)\otimes e_F^{(1)}(\omega_3)\Big)\\
&=\raisebox{-0.9 cm}{\scalebox{0.75}{\begin{tikzpicture}[scale = 1.5]
\draw[thick,color=red](-1, -1.0) circle (0.5ex);
\draw[thick,color=red](0, -0.5) circle (0.5ex);
\draw[thick,color=red](0, - 1.0) circle (0.5ex);
\draw[thick,color=red](1, -0.5) circle (0.5ex);
\draw[thick,color=red](1, -1.0) circle (0.5ex);
\draw[thick, color=black]
{
 (-1,  - 1.0) node [above] {\large{{+1}}}
 (0, -0.5) node [above] {\large{{-1}}}
 (0, - 1.0) node [above] {\large{{+1}}}
 (1, -0.5) node [above] {\large{{-1}}}
 (1, - 1.0) node [above] {\large{{+1}}}
};
\foreach \x in {0,...,-1.5}{
\pgfmathtruncatemacro\z{int(-\x + 1)}
\draw[thick,color=violet,->] (-2,\x cm) -- (2,\x cm) ;
\draw[thick,color=violet,->] (2, \x cm - 0.5 cm) -- (-2, \x cm - 0.5 cm) ;
\draw[thick,color=violet,->] (2, \x cm) arc (90:-90:0.25);
\draw[thick,color=blue] (2 + 0.25, \x cm - 0.25cm) node[right] {\footnotesize{\z}};
}
\foreach \y in {-1,...,-.5}{
\pgfmathtruncatemacro\w{int(-\y + 1)}
\draw[thick,color=violet,->] (-2,\y cm + 0.5 cm) arc (90:270:0.25);
}
\end{tikzpicture}
}}\\
\end{split}
\label{appfsliding:3}
\ee
Now, one can add another past-most insertion with a factor (-1) on the fourth leg without changing the value of the contraction because if we choose this new insertion, then again there is no insertion to block the pair of points of $\eP{2}(\omega_2)$ from collapsing onto each other. Therefore, we have 
\be
\begin{split}
&\matMt{2,3}(\text{3-Pt}) \cdot \Big(e_P^{(2)}(\omega_1)\otimes e_P^{(2)}(\omega_2)\otimes e_F^{(1)}(\omega_3)\Big)\\
&=\raisebox{-0.9 cm}{\scalebox{0.75}{\begin{tikzpicture}[scale = 1.5]
\draw[thick,color=red](-1, -1.0) circle (0.5ex);
\draw[thick,color=red](-1, -1.5) circle (0.5ex);
\draw[thick,color=red](0, -0.5) circle (0.5ex);
\draw[thick,color=red](0, - 1.0) circle (0.5ex);
\draw[thick,color=red](1, -0.5) circle (0.5ex);
\draw[thick,color=red](1, -1.0) circle (0.5ex);
\draw[thick, color=black]
{
 (-1,  - 1.0) node [above] {\large{{+1}}}
  (-1,  -1.5) node [above] {\large{{-1}}}
 (0, -0.5) node [above] {\large{{-1}}}
 (0, - 1.0) node [above] {\large{{+1}}}
 (1, -0.5) node [above] {\large{{-1}}}
 (1, - 1.0) node [above] {\large{{+1}}}
};
\foreach \x in {0,...,-1.5}{
\pgfmathtruncatemacro\z{int(-\x + 1)}
\draw[thick,color=violet,->] (-2,\x cm) -- (2,\x cm) ;
\draw[thick,color=violet,->] (2, \x cm - 0.5 cm) -- (-2, \x cm - 0.5 cm) ;
\draw[thick,color=violet,->] (2, \x cm) arc (90:-90:0.25);
\draw[thick,color=blue] (2 + 0.25, \x cm - 0.25cm) node[right] {\footnotesize{\z}};
}
\foreach \y in {-1,...,-.5}{
\pgfmathtruncatemacro\w{int(-\y + 1)}
\draw[thick,color=violet,->] (-2,\y cm + 0.5 cm) arc (90:270:0.25);
}
\end{tikzpicture}
}}\\ \\
&=-\raisebox{-0.9 cm}{\scalebox{0.75}{\begin{tikzpicture}[scale = 1.5]
\draw[thick,color=red](-1, -1.0) circle (0.5ex);
\draw[thick,color=red](-1, -1.5) circle (0.5ex);
\draw[thick,color=red](0, -0.5) circle (0.5ex);
\draw[thick,color=red](0, - 1.0) circle (0.5ex);
\draw[thick,color=red](1, -0.5) circle (0.5ex);
\draw[thick,color=red](1, -1.0) circle (0.5ex);
\draw[thick, color=black]
{
 (-1,  - 1.0) node [above] {\large{{-1}}}
  (-1,  - 1.5) node [above] {\large{{+1}}}
 (0, -0.5) node [above] {\large{{-1}}}
 (0, - 1.0) node [above] {\large{{+1}}}
 (1, -0.5) node [above] {\large{{-1}}}
 (1, - 1.0) node [above] {\large{{+1}}}
};
\foreach \x in {0,...,-1.5}{
\pgfmathtruncatemacro\z{int(-\x + 1)}
\draw[thick,color=violet,->] (-2,\x cm) -- (2,\x cm) ;
\draw[thick,color=violet,->] (2, \x cm - 0.5 cm) -- (-2, \x cm - 0.5 cm) ;
\draw[thick,color=violet,->] (2, \x cm) arc (90:-90:0.25);
\draw[thick,color=blue] (2 + 0.25, \x cm - 0.25cm) node[right] {\footnotesize{\z}};
}
\foreach \y in {-1,...,-.5}{
\pgfmathtruncatemacro\w{int(-\y + 1)}
\draw[thick,color=violet,->] (-2,\y cm + 0.5 cm) arc (90:270:0.25);
}
\end{tikzpicture}
}}\\
&=-\matMt{2,3}(\text{3-Pt}) \cdot \Big(e_P^{(2)}(\omega_1)\otimes e_P^{(2)}(\omega_2)\otimes e_F^{(2)}(\omega_3)\Big)\ .
\end{split}
\label{appfsliding:4}
\ee
So, we see that, in this case the transformation $e_F^{(1)}(\omega_3)\mapsto -e_F^{(2)}(\omega_3)$ keeps the value of the contraction unchanged.


\subsection*{P-sliding :}
Consider the contraction of $\matMt{2,3}(\text{3-Pt})$ with the tensor $\Big(e_P^{(2)}(\omega_1)\otimes e_F^{(1)}(\omega_2)\otimes e_F^{(1)}(\omega_3)\Big)$. The corresponding diagram is 

\be
\begin{split}
&\matMt{2,3}(\text{3-Pt}) \cdot \Big(e_P^{(2)}(\omega_1)\otimes e_F^{(1)}(\omega_2)\otimes e_F^{(1)}(\omega_3)\Big)\\
&=\raisebox{-0.9 cm}{\scalebox{0.75}{\contourDDD{-1.5}{0}{-0}{-0.5}}}\\
\end{split}
\label{apppsliding:1}
\ee

Notice that $\eP{2}(\omega_1)$ is the future-most insertion. There is an $\eF{1}(\omega_2)$  to its past. But, apart from $\eP{2}(\omega_1)$,there is no other $\eP{1}$ or $\eP{2}$ to the future of $\eF{1}(\omega_2)$. If we choose, the future-most insertion with the the factor (+1) on the third leg, then there is no insertion to block the 2 insertions corresponding to $\eF{1}(\omega_2)$ from collapsing on to each other i.e. one can slide either of those insertions to the position of the other and their contributions would exactly cancel each other. Therefore, we have 
\be
\begin{split}
&\matMt{2,3}(\text{3-Pt}) \cdot \Big(e_P^{(2)}(\omega_1)\otimes e_F^{(1)}(\omega_2)\otimes e_F^{(1)}(\omega_3)\Big)\\
&=\raisebox{-0.9 cm}{\scalebox{0.75}{\begin{tikzpicture}[scale = 1.5]
\draw[thick,color=red](-1, -0) circle (0.5ex);
\draw[thick,color=red](-1, -0.5) circle (0.5ex);
\draw[thick,color=red](0, - 0) circle (0.5ex);
\draw[thick,color=red](0, -0.5) circle (0.5ex);
\draw[thick,color=red](1, -0.5) circle (0.5ex);
\draw[thick, color=black]
{
 (-1,  - 0) node [above] {\large{{-1}}}
 (-1, -0.5) node [above] {\large{{+1}}}
 (0, - 0) node [above] {\large{{-1}}}
 (0, -0.5) node [above] {\large{{+1}}}
 (1, - 0.5) node [above] {\large{{-1}}}
};
\foreach \x in {0,...,-1.5}{
\pgfmathtruncatemacro\z{int(-\x + 1)}
\draw[thick,color=violet,->] (-2,\x cm) -- (2,\x cm) ;
\draw[thick,color=violet,->] (2, \x cm - 0.5 cm) -- (-2, \x cm - 0.5 cm) ;
\draw[thick,color=violet,->] (2, \x cm) arc (90:-90:0.25);
\draw[thick,color=blue] (2 + 0.25, \x cm - 0.25cm) node[right] {\footnotesize{\z}};
}
\foreach \y in {-1,...,-.5}{
\pgfmathtruncatemacro\w{int(-\y + 1)}
\draw[thick,color=violet,->] (-2,\y cm + 0.5 cm) arc (90:270:0.25);
}
\end{tikzpicture}
}}\\
\end{split}
\label{apppsliding:2}
\ee
One can slide the future-most insertion on the second leg up to the first leg without changing the value of the contraction and obtain
\be
\begin{split}
&\matMt{2,3}(\text{3-Pt}) \cdot \Big(e_P^{(2)}(\omega_1)\otimes e_F^{(1)}(\omega_2)\otimes e_F^{(1)}(\omega_3)\Big)\\
&=\raisebox{-0.9 cm}{\scalebox{0.75}{\begin{tikzpicture}[scale = 1.5]
\draw[thick,color=red](-1, -0) circle (0.5ex);
\draw[thick,color=red](-1, -0.5) circle (0.5ex);
\draw[thick,color=red](0, - 0) circle (0.5ex);
\draw[thick,color=red](0, -0.5) circle (0.5ex);
\draw[thick,color=red](1, -0) circle (0.5ex);
\draw[thick, color=black]
{
 (-1,  - 0) node [above] {\large{{-1}}}
 (-1, -0.5) node [above] {\large{{+1}}}
 (0, - 0) node [above] {\large{{-1}}}
 (0, -0.5) node [above] {\large{{+1}}}
 (1, - 0) node [above] {\large{{-1}}}
};
\foreach \x in {0,...,-1.5}{
\pgfmathtruncatemacro\z{int(-\x + 1)}
\draw[thick,color=violet,->] (-2,\x cm) -- (2,\x cm) ;
\draw[thick,color=violet,->] (2, \x cm - 0.5 cm) -- (-2, \x cm - 0.5 cm) ;
\draw[thick,color=violet,->] (2, \x cm) arc (90:-90:0.25);
\draw[thick,color=blue] (2 + 0.25, \x cm - 0.25cm) node[right] {\footnotesize{\z}};
}
\foreach \y in {-1,...,-.5}{
\pgfmathtruncatemacro\w{int(-\y + 1)}
\draw[thick,color=violet,->] (-2,\y cm + 0.5 cm) arc (90:270:0.25);
}
\end{tikzpicture}
}}\\
\end{split}
\label{apppsliding:3}
\ee
Now, one can add another future-most insertion with a factor ($e^{-\beta \omega_1}$) on the fourth leg without changing the value of the contraction because if we choose this new insertion, then again there is no insertion to block the pair of points of $\eF{1}(\omega_2)$ from collapsing onto each other. Therefore, we have 
\be
\begin{split}
&\matMt{2,3}(\text{3-Pt}) \cdot \Big(e_P^{(2)}(\omega_1)\otimes e_P^{(2)}(\omega_2)\otimes e_F^{(1)}(\omega_3)\Big)\\
&=\raisebox{-0.9 cm}{\scalebox{0.75}{\begin{tikzpicture}[scale = 1.5]
\draw[thick,color=red](-1, -0) circle (0.5ex);
\draw[thick,color=red](-1, -0.5) circle (0.5ex);
\draw[thick,color=red](0, - 0) circle (0.5ex);
\draw[thick,color=red](0, -0.5) circle (0.5ex);
\draw[thick,color=red](1, -0) circle (0.5ex);
\draw[thick,color=red](1, -1.5) circle (0.5ex);
\draw[thick, color=black]
{
 (-1,  - 0) node [above] {\large{{-1}}}
 (-1, -0.5) node [above] {\large{{+1}}}
 (0, - 0) node [above] {\large{{-1}}}
 (0, -0.5) node [above] {\large{{+1}}}
 (1, - 0) node [above] {\large{{-1}}}
  (1, - 1.5) node [above] {\large{{$e^{-\beta \omega_1}$}}}
};
\foreach \x in {0,...,-1.5}{
\pgfmathtruncatemacro\z{int(-\x + 1)}
\draw[thick,color=violet,->] (-2,\x cm) -- (2,\x cm) ;
\draw[thick,color=violet,->] (2, \x cm - 0.5 cm) -- (-2, \x cm - 0.5 cm) ;
\draw[thick,color=violet,->] (2, \x cm) arc (90:-90:0.25);
\draw[thick,color=blue] (2 + 0.25, \x cm - 0.25cm) node[right] {\footnotesize{\z}};
}
\foreach \y in {-1,...,-.5}{
\pgfmathtruncatemacro\w{int(-\y + 1)}
\draw[thick,color=violet,->] (-2,\y cm + 0.5 cm) arc (90:270:0.25);
}
\end{tikzpicture}
}}\\ \\
&=-\raisebox{-0.9 cm}{\scalebox{0.75}{\begin{tikzpicture}[scale = 1.5]
\draw[thick,color=red](-1, -0) circle (0.5ex);
\draw[thick,color=red](-1, -0.5) circle (0.5ex);
\draw[thick,color=red](0, - 0) circle (0.5ex);
\draw[thick,color=red](0, -0.5) circle (0.5ex);
\draw[thick,color=red](1, -0) circle (0.5ex);
\draw[thick,color=red](1, -1.5) circle (0.5ex);
\draw[thick, color=black]
{
 (-1,  - 0) node [above] {\large{{-1}}}
 (-1, -0.5) node [above] {\large{{+1}}}
 (0, - 0) node [above] {\large{{-1}}}
 (0, -0.5) node [above] {\large{{+1}}}
 (1, - 0) node [above] {\large{{+1}}}
  (1, - 1.5) node [above] {\large{{-$e^{-\beta \omega_1}$}}}
};
\foreach \x in {0,...,-1.5}{
\pgfmathtruncatemacro\z{int(-\x + 1)}
\draw[thick,color=violet,->] (-2,\x cm) -- (2,\x cm) ;
\draw[thick,color=violet,->] (2, \x cm - 0.5 cm) -- (-2, \x cm - 0.5 cm) ;
\draw[thick,color=violet,->] (2, \x cm) arc (90:-90:0.25);
\draw[thick,color=blue] (2 + 0.25, \x cm - 0.25cm) node[right] {\footnotesize{\z}};
}
\foreach \y in {-1,...,-.5}{
\pgfmathtruncatemacro\w{int(-\y + 1)}
\draw[thick,color=violet,->] (-2,\y cm + 0.5 cm) arc (90:270:0.25);
}
\end{tikzpicture}
}}\\
&=-\matMt{2,3}(\text{3-Pt}) \cdot \Big(e_P^{(1)}(\omega_1)\otimes e_F^{(1)}(\omega_2)\otimes e_F^{(1)}(\omega_3)\Big)\ .
\end{split}
\label{apppsliding:4}
\ee
So, we see that, in this case the transformation $e_P^{(2)}(\omega_1)\mapsto -e_P^{(1)}(\omega_1)$ keeps the value of the contraction unchanged.


\subsection*{C-shift:}
Consider the contraction of $\matMt{2,2}(\text{2-Pt})$ with the tensor $\Big(e_P^{(1)}(\omega_1)\otimes e_F^{(1)}(\omega_2)\Big )$. The corresponding diagram is 

\be
\begin{split}
&\matMt{2,2}(\text{2-Pt}) \cdot \Big(e_P^{(1)}(\omega_1)\otimes e_F^{(1)}(\omega_2)\Big )=\raisebox{-0.9 cm}{\scalebox{0.75}{\begin{tikzpicture}[scale = 1.5]
\draw[thick,color=red](-1, -0) circle (0.5ex);
\draw[thick,color=red](-1, -0.5) circle (0.5ex);
\draw[thick,color=red](1, -0) circle (0.5ex);
\draw[thick,color=red](1, -1.5) circle (0.5ex);
\draw[thick, color=black]
{
 (-1,  - 0) node [above] {\large{{-1}}}
 (-1, -0.5) node [above] {\large{{+1}}}
 (1, - 0) node [above] {\large{{+1}}}
  (1, - 1.5) node [above] {\large{{-$e^{-\beta \omega_1}$}}}
};
\foreach \x in {0,...,-1.5}{
\pgfmathtruncatemacro\z{int(-\x + 1)}
\draw[thick,color=violet,->] (-2,\x cm) -- (2,\x cm) ;
\draw[thick,color=violet,->] (2, \x cm - 0.5 cm) -- (-2, \x cm - 0.5 cm) ;
\draw[thick,color=violet,->] (2, \x cm) arc (90:-90:0.25);
\draw[thick,color=blue] (2 + 0.25, \x cm - 0.25cm) node[right] {\footnotesize{\z}};
}
\foreach \y in {-1,...,-.5}{
\pgfmathtruncatemacro\w{int(-\y + 1)}
\draw[thick,color=violet,->] (-2,\y cm + 0.5 cm) arc (90:270:0.25);
}
\end{tikzpicture}
}}\\
\end{split}
\label{appcshift:1}
\ee
One can slide the insertions one after the other without changing the value of the contraction as shown below:

\be
\begin{split}
&\matMt{2,2}(\text{2-Pt}) \cdot \Big(e_P^{(1)}(\omega_1)\otimes e_F^{(1)}(\omega_2)\Big )=\raisebox{-0.9 cm}{\scalebox{0.75}{\begin{tikzpicture}[scale = 1.5]
\draw[thick,color=red](-1, -0) circle (0.5ex);
\draw[thick,color=red](-1, -0.5) circle (0.5ex);
\draw[thick,color=red](1, -0) circle (0.5ex);
\draw[thick,color=red](1, -1.5) circle (0.5ex);
\draw[thick, color=black]
{
 (-1,  - 0) node [above] {\large{{-1}}}
 (-1, -0.5) node [above] {\large{{+1}}}
 (1, - 0) node [above] {\large{{+1}}}
  (1, - 1.5) node [above] {\large{{-$e^{-\beta \omega_1}$}}}
};
\foreach \x in {0,...,-1.5}{
\pgfmathtruncatemacro\z{int(-\x + 1)}
\draw[thick,color=violet,->] (-2,\x cm) -- (2,\x cm) ;
\draw[thick,color=violet,->] (2, \x cm - 0.5 cm) -- (-2, \x cm - 0.5 cm) ;
\draw[thick,color=violet,->] (2, \x cm) arc (90:-90:0.25);
\draw[thick,color=blue] (2 + 0.25, \x cm - 0.25cm) node[right] {\footnotesize{\z}};
}
\foreach \y in {-1,...,-.5}{
\pgfmathtruncatemacro\w{int(-\y + 1)}
\draw[thick,color=violet,->] (-2,\y cm + 0.5 cm) arc (90:270:0.25);
}
\end{tikzpicture}
}}\\ \\
&=\raisebox{-0.9 cm}{\scalebox{0.75}{\begin{tikzpicture}[scale = 1.5]
\draw[thick,color=red](-1, -0) circle (0.5ex);
\draw[thick,color=red](1, -0) circle (0.5ex);
\draw[thick, color=black]
{
 (-1,  - 0) node [above] {\large{{-1}}}
  (1, - 0) node [above] {\large{{+1}}}
};
\foreach \x in {0,...,-1.5}{
\pgfmathtruncatemacro\z{int(-\x + 1)}
\draw[thick,color=violet,->] (-2,\x cm) -- (2,\x cm) ;
\draw[thick,color=violet,->] (2, \x cm - 0.5 cm) -- (-2, \x cm - 0.5 cm) ;
\draw[thick,color=violet,->] (2, \x cm) arc (90:-90:0.25);
\draw[thick,color=blue] (2 + 0.25, \x cm - 0.25cm) node[right] {\footnotesize{\z}};
}
\foreach \y in {-1,...,-.5}{
\pgfmathtruncatemacro\w{int(-\y + 1)}
\draw[thick,color=violet,->] (-2,\y cm + 0.5 cm) arc (90:270:0.25);
}
\end{tikzpicture}
}}+
\raisebox{-0.9 cm}{\scalebox{0.75}{\begin{tikzpicture}[scale = 1.5]
\draw[thick,color=red](-1, -0.5) circle (0.5ex);
\draw[thick,color=red](1, -0) circle (0.5ex);
\draw[thick, color=black]
{
 (-1,  - 0.5) node [above] {\large{{+1}}}
  (1, - 0) node [above] {\large{{+1}}}
};
\foreach \x in {0,...,-1.5}{
\pgfmathtruncatemacro\z{int(-\x + 1)}
\draw[thick,color=violet,->] (-2,\x cm) -- (2,\x cm) ;
\draw[thick,color=violet,->] (2, \x cm - 0.5 cm) -- (-2, \x cm - 0.5 cm) ;
\draw[thick,color=violet,->] (2, \x cm) arc (90:-90:0.25);
\draw[thick,color=blue] (2 + 0.25, \x cm - 0.25cm) node[right] {\footnotesize{\z}};
}
\foreach \y in {-1,...,-.5}{
\pgfmathtruncatemacro\w{int(-\y + 1)}
\draw[thick,color=violet,->] (-2,\y cm + 0.5 cm) arc (90:270:0.25);
}
\end{tikzpicture}
}}\\&+
\raisebox{-0.9 cm}{\scalebox{0.75}{\begin{tikzpicture}[scale = 1.5]
\draw[thick,color=red](-1, -0) circle (0.5ex);
\draw[thick,color=red](1, -1.5) circle (0.5ex);
\draw[thick, color=black]
{
 (-1,  - 0) node [above] {\large{{-1}}}
  (1, - 1.5) node [above] {\large{{$-e^{-\beta \omega_1}$}}}
};
\foreach \x in {0,...,-1.5}{
\pgfmathtruncatemacro\z{int(-\x + 1)}
\draw[thick,color=violet,->] (-2,\x cm) -- (2,\x cm) ;
\draw[thick,color=violet,->] (2, \x cm - 0.5 cm) -- (-2, \x cm - 0.5 cm) ;
\draw[thick,color=violet,->] (2, \x cm) arc (90:-90:0.25);
\draw[thick,color=blue] (2 + 0.25, \x cm - 0.25cm) node[right] {\footnotesize{\z}};
}
\foreach \y in {-1,...,-.5}{
\pgfmathtruncatemacro\w{int(-\y + 1)}
\draw[thick,color=violet,->] (-2,\y cm + 0.5 cm) arc (90:270:0.25);
}
\end{tikzpicture}
}}+
\raisebox{-0.9 cm}{\scalebox{0.75}{\begin{tikzpicture}[scale = 1.5]
\draw[thick,color=red](-1, -0.5) circle (0.5ex);
\draw[thick,color=red](1, -1.5) circle (0.5ex);
\draw[thick, color=black]
{
 (-1,  - 0.5) node [above] {\large{{+1}}}
  (1, - 1.5) node [above] {\large{{$-e^{-\beta \omega_1}$}}}
};
\foreach \x in {0,...,-1.5}{
\pgfmathtruncatemacro\z{int(-\x + 1)}
\draw[thick,color=violet,->] (-2,\x cm) -- (2,\x cm) ;
\draw[thick,color=violet,->] (2, \x cm - 0.5 cm) -- (-2, \x cm - 0.5 cm) ;
\draw[thick,color=violet,->] (2, \x cm) arc (90:-90:0.25);
\draw[thick,color=blue] (2 + 0.25, \x cm - 0.25cm) node[right] {\footnotesize{\z}};
}
\foreach \y in {-1,...,-.5}{
\pgfmathtruncatemacro\w{int(-\y + 1)}
\draw[thick,color=violet,->] (-2,\y cm + 0.5 cm) arc (90:270:0.25);
}
\end{tikzpicture}
}}
\\ \\
&=\raisebox{-0.9 cm}{\scalebox{0.75}{\begin{tikzpicture}[scale = 1.5]
\draw[thick,color=red](-1, -1.0) circle (0.5ex);
\draw[thick,color=red](1, -1.0) circle (0.5ex);
\draw[thick, color=black]
{
 (-1,  - 1.0) node [above] {\large{{-1}}}
  (1, - 1.0) node [above] {\large{{+1}}}
};
\foreach \x in {0,...,-1.5}{
\pgfmathtruncatemacro\z{int(-\x + 1)}
\draw[thick,color=violet,->] (-2,\x cm) -- (2,\x cm) ;
\draw[thick,color=violet,->] (2, \x cm - 0.5 cm) -- (-2, \x cm - 0.5 cm) ;
\draw[thick,color=violet,->] (2, \x cm) arc (90:-90:0.25);
\draw[thick,color=blue] (2 + 0.25, \x cm - 0.25cm) node[right] {\footnotesize{\z}};
}
\foreach \y in {-1,...,-.5}{
\pgfmathtruncatemacro\w{int(-\y + 1)}
\draw[thick,color=violet,->] (-2,\y cm + 0.5 cm) arc (90:270:0.25);
}
\end{tikzpicture}
}}+
\raisebox{-0.9 cm}{\scalebox{0.75}{\begin{tikzpicture}[scale = 1.5]
\draw[thick,color=red](-1, -1.5) circle (0.5ex);
\draw[thick,color=red](1, -1.0) circle (0.5ex);
\draw[thick, color=black]
{
 (-1,  - 1.5) node [above] {\large{{+1}}}
  (1, - 1.0) node [above] {\large{{+1}}}
};
\foreach \x in {0,...,-1.5}{
\pgfmathtruncatemacro\z{int(-\x + 1)}
\draw[thick,color=violet,->] (-2,\x cm) -- (2,\x cm) ;
\draw[thick,color=violet,->] (2, \x cm - 0.5 cm) -- (-2, \x cm - 0.5 cm) ;
\draw[thick,color=violet,->] (2, \x cm) arc (90:-90:0.25);
\draw[thick,color=blue] (2 + 0.25, \x cm - 0.25cm) node[right] {\footnotesize{\z}};
}
\foreach \y in {-1,...,-.5}{
\pgfmathtruncatemacro\w{int(-\y + 1)}
\draw[thick,color=violet,->] (-2,\y cm + 0.5 cm) arc (90:270:0.25);
}
\end{tikzpicture}
}}\\&+
\raisebox{-0.9 cm}{\scalebox{0.75}{\begin{tikzpicture}[scale = 1.5]
\draw[thick,color=red](-1, -1.0) circle (0.5ex);
\draw[thick,color=red](1, -0.5) circle (0.5ex);
\draw[thick, color=black]
{
 (-1,  - 1.0) node [above] {\large{{-1}}}
  (1, - 0.5) node [above] {\large{{$-1$}}}
};
\foreach \x in {0,...,-1.5}{
\pgfmathtruncatemacro\z{int(-\x + 1)}
\draw[thick,color=violet,->] (-2,\x cm) -- (2,\x cm) ;
\draw[thick,color=violet,->] (2, \x cm - 0.5 cm) -- (-2, \x cm - 0.5 cm) ;
\draw[thick,color=violet,->] (2, \x cm) arc (90:-90:0.25);
\draw[thick,color=blue] (2 + 0.25, \x cm - 0.25cm) node[right] {\footnotesize{\z}};
}
\foreach \y in {-1,...,-.5}{
\pgfmathtruncatemacro\w{int(-\y + 1)}
\draw[thick,color=violet,->] (-2,\y cm + 0.5 cm) arc (90:270:0.25);
}
\end{tikzpicture}
}}+
\raisebox{-0.9 cm}{\scalebox{0.75}{\begin{tikzpicture}[scale = 1.5]
\draw[thick,color=red](-1, -1.5) circle (0.5ex);
\draw[thick,color=red](1, -0.5) circle (0.5ex);
\draw[thick, color=black]
{
 (-1,  - 1.5) node [above] {\large{{+1}}}
  (1, - 0.5) node [above] {\large{{$-1$}}}
};
\foreach \x in {0,...,-1.5}{
\pgfmathtruncatemacro\z{int(-\x + 1)}
\draw[thick,color=violet,->] (-2,\x cm) -- (2,\x cm) ;
\draw[thick,color=violet,->] (2, \x cm - 0.5 cm) -- (-2, \x cm - 0.5 cm) ;
\draw[thick,color=violet,->] (2, \x cm) arc (90:-90:0.25);
\draw[thick,color=blue] (2 + 0.25, \x cm - 0.25cm) node[right] {\footnotesize{\z}};
}
\foreach \y in {-1,...,-.5}{
\pgfmathtruncatemacro\w{int(-\y + 1)}
\draw[thick,color=violet,->] (-2,\y cm + 0.5 cm) arc (90:270:0.25);
}
\end{tikzpicture}
}}\\&
=\raisebox{-0.9 cm}{\scalebox{0.75}{\contourDD{-1.5}{-1.0}{-0.5}}}
=\matMt{2,2}(\text{2-Pt}) \cdot \Big(e_P^{(2)}(\omega_1)\otimes e_F^{(2)}(\omega_2)\Big )\ .
\end{split}
\label{appcshift:2}
\ee

So, we see that, in this case the transformation $e_P^{(1)}(\omega_1)\mapsto e_P^{(2)}(\omega_1), e_F^{(1)}(\omega_2)\mapsto e_F^{(2)}(\omega_2)$ keeps the value of the contraction unchanged.


\subsection*{F-fragmentation :}
Consider the contraction of $\matMt{4,4}(\text{4-Pt})$ with the tensor $\Big(e_P^{(2)}(\omega_1)\otimes e_P^{(3)}(\omega_2)\otimes e_F^{(2)}(\omega_3)\otimes e_F^{(2)}(\omega_4)\Big )$. The corresponding diagram is 

\be
\begin{split}
&\matMt{4,4}(\text{4-Pt})\cdot \Big(e_P^{(2)}(\omega_1)\otimes e_P^{(3)}(\omega_2)\otimes e_F^{(2)}(\omega_3)\otimes e_F^{(2)}(\omega_4)\Big )\\
&=\raisebox{-1.8 cm}{\scalebox{0.75}{\contourDDDD{-3.5}{-1.0}{-1.0}{-1.5}{-0.5}}}\\
\end{split}
\label{appffragmen:1}
\ee
One can slide all the insertions that lie on the $4^{th}$ leg or below down by two legs to obtain the following diagram:

\be
\begin{split}
&\matMt{4,4}(\text{4-Pt}) \cdot \Big(e_P^{(2)}(\omega_1)\otimes e_P^{(3)}(\omega_2)\otimes e_F^{(2)}(\omega_3)\otimes e_F^{(2)}(\omega_4)\Big )\\
&=\raisebox{-1.8 cm}{\scalebox{0.75}{\begin{tikzpicture}[scale = 1.5]
\draw[thick,color=red](-1.2, -1.0) circle (0.5ex);
\draw[thick,color=red](-1.2, -2.5) circle (0.5ex);
\draw[thick,color=red](-0.4, -1.0) circle (0.5ex);
\draw[thick,color=red](-0.4, -2.5) circle (0.5ex);
\draw[thick,color=red](0.4, -2.5) circle (0.5ex);
\draw[thick,color=red](0.4, -3.0) circle (0.5ex);
\draw[thick,color=red](1.2, -0.5) circle (0.5ex);
\draw[thick,color=red](1.2, -1.0) circle (0.5ex);
\draw[thick, color=black]
{
 (-1.2,  - 1.0) node [above] {\large{{-1}}}
 (-1.2, -2.5) node [above] {\large{{+1}}}
(-0.4,  - 1.0) node [above] {\large{{-1}}}
 (-0.4, -2.5) node [above] {\large{{+1}}}
(0.4,  - 2.5) node [above] {\large{{-1}}}
 (0.4, -3.0) node [above] {\large{{+1}}}
 (1.2, - 0.5) node [above] {\large{{-1}}}
  (1.2, - 1.0) node [above] {\large{{+1}}}
};
\foreach \x in {0,...,-3.5}{
\pgfmathtruncatemacro\z{int(-\x + 1)}
\draw[thick,color=violet,->] (-2,\x cm) -- (2,\x cm) ;
\draw[thick,color=violet,->] (2, \x cm - 0.5 cm) -- (-2, \x cm - 0.5 cm) ;
\draw[thick,color=violet,->] (2, \x cm) arc (90:-90:0.25);
\draw[thick,color=blue] (2 + 0.25, \x cm - 0.25cm) node[right] {\footnotesize{\z}};
}
\foreach \y in {-1,...,-3.5}{
\pgfmathtruncatemacro\w{int(-\y + 1)}
\draw[thick,color=violet,->] (-2,\y cm + 0.5 cm) arc (90:270:0.25);
}
\end{tikzpicture}
}}\\
\end{split}
\label{appffragmen:2}
\ee

We can add a pair of points with opposite signs on the $4^{th}$ and the $5^{th}$ legs at positions corresponding to the frequency $\omega_3$ without changing the value of the contraction as shown below:

\be
\begin{split}
&\matMt{4,4}(\text{4-Pt}) \cdot \Big(e_P^{(2)}(\omega_1)\otimes e_P^{(3)}(\omega_2)\otimes e_F^{(2)}(\omega_3)\otimes e_F^{(2)}(\omega_4)\Big )\\
&=\raisebox{-1.8 cm}{\scalebox{0.75}{\begin{tikzpicture}[scale = 1.5]
\draw[thick,color=red](-1.2, -1.0) circle (0.5ex);
\draw[thick,color=red](-1.2, -2.5) circle (0.5ex);
\draw[thick,color=red](-0.4, -1.0) circle (0.5ex);
\draw[thick,color=red](-0.4, -2.5) circle (0.5ex);
\draw[thick,color=red](-0.4, -1.5) circle (0.5ex);
\draw[thick,color=red](-0.4, -2.0) circle (0.5ex);
\draw[thick,color=red](0.4, -2.5) circle (0.5ex);
\draw[thick,color=red](0.4, -3.0) circle (0.5ex);
\draw[thick,color=red](1.2, -0.5) circle (0.5ex);
\draw[thick,color=red](1.2, -1.0) circle (0.5ex);
\draw[thick, color=black]
{
 (-1.2,  - 1.0) node [above] {\large{{-1}}}
 (-1.2, -2.5) node [above] {\large{{+1}}}
(-0.4,  - 1.0) node [above] {\large{{-1}}}
 (-0.4, -2.5) node [above] {\large{{+1}}}
(-0.4,  - 1.5) node [above] {\large{{+1}}}
 (-0.4, -2.0) node [above] {\large{{-1}}}
(0.4,  - 2.5) node [above] {\large{{-1}}}
 (0.4, -3.0) node [above] {\large{{+1}}}
 (1.2, - 0.5) node [above] {\large{{-1}}}
  (1.2, - 1.0) node [above] {\large{{+1}}}
};
\foreach \x in {0,...,-3.5}{
\pgfmathtruncatemacro\z{int(-\x + 1)}
\draw[thick,color=violet,->] (-2,\x cm) -- (2,\x cm) ;
\draw[thick,color=violet,->] (2, \x cm - 0.5 cm) -- (-2, \x cm - 0.5 cm) ;
\draw[thick,color=violet,->] (2, \x cm) arc (90:-90:0.25);
\draw[thick,color=blue] (2 + 0.25, \x cm - 0.25cm) node[right] {\footnotesize{\z}};
}
\foreach \y in {-1,...,-3.5}{
\pgfmathtruncatemacro\w{int(-\y + 1)}
\draw[thick,color=violet,->] (-2,\y cm + 0.5 cm) arc (90:270:0.25);
}
\end{tikzpicture}
}}\\
\end{split}
\label{appffragmen:3}
\ee
There is no insertion to the past of these new insertions which can block them from collapsing onto one another. Now, one can add another pair of insertions with opposite signs on the $4^{th}$ and the $5^{th}$ legs at positions corresponding to the frequency $\omega_4$ without changing the value of the contraction as shown below:
\be
\begin{split}
&\matMt{4,4}(\text{4-Pt}) \cdot \Big(e_P^{(2)}(\omega_1)\otimes e_P^{(3)}(\omega_2)\otimes e_F^{(2)}(\omega_3)\otimes e_F^{(2)}(\omega_4)\Big )\\
&=\raisebox{-1.8 cm}{\scalebox{0.75}{\begin{tikzpicture}[scale = 1.5]
\draw[thick,color=red](-1.2, -1.0) circle (0.5ex);
\draw[thick,color=red](-1.2, -2.5) circle (0.5ex);
\draw[thick,color=red](-1.2, -1.5) circle (0.5ex);
\draw[thick,color=red](-1.2, -2.0) circle (0.5ex);
\draw[thick,color=red](-0.4, -1.0) circle (0.5ex);
\draw[thick,color=red](-0.4, -2.5) circle (0.5ex);
\draw[thick,color=red](-0.4, -1.5) circle (0.5ex);
\draw[thick,color=red](-0.4, -2.0) circle (0.5ex);
\draw[thick,color=red](0.4, -2.5) circle (0.5ex);
\draw[thick,color=red](0.4, -3.0) circle (0.5ex);
\draw[thick,color=red](1.2, -0.5) circle (0.5ex);
\draw[thick,color=red](1.2, -1.0) circle (0.5ex);
\draw[thick, color=black]
{
 (-1.2,  - 1.0) node [above] {\large{{-1}}}
 (-1.2, -2.5) node [above] {\large{{+1}}}
  (-1.2,  - 1.5) node [above] {\large{{+1}}}
 (-1.2, -2.0) node [above] {\large{{-1}}}
(-0.4,  - 1.0) node [above] {\large{{-1}}}
 (-0.4, -2.5) node [above] {\large{{+1}}}
(-0.4,  - 1.5) node [above] {\large{{+1}}}
 (-0.4, -2.0) node [above] {\large{{-1}}}
(0.4,  - 2.5) node [above] {\large{{-1}}}
 (0.4, -3.0) node [above] {\large{{+1}}}
 (1.2, - 0.5) node [above] {\large{{-1}}}
  (1.2, - 1.0) node [above] {\large{{+1}}}
};
\foreach \x in {0,...,-3.5}{
\pgfmathtruncatemacro\z{int(-\x + 1)}
\draw[thick,color=violet,->] (-2,\x cm) -- (2,\x cm) ;
\draw[thick,color=violet,->] (2, \x cm - 0.5 cm) -- (-2, \x cm - 0.5 cm) ;
\draw[thick,color=violet,->] (2, \x cm) arc (90:-90:0.25);
\draw[thick,color=blue] (2 + 0.25, \x cm - 0.25cm) node[right] {\footnotesize{\z}};
}
\foreach \y in {-1,...,-3.5}{
\pgfmathtruncatemacro\w{int(-\y + 1)}
\draw[thick,color=violet,->] (-2,\y cm + 0.5 cm) arc (90:270:0.25);
}
\end{tikzpicture}
}}\\
\end{split}
\label{appffragmen:4}
\ee

As before, there is no insertion to the past of these new insertions which can block them from collapsing onto one another. But this is exactly the diagram for the contraction of  $\matMt{4,4}(\text{4-Pt})$ with  
\[\Big(e_P^{(2)}(\omega_1)\otimes e_P^{(4)}(\omega_2)\otimes (e_F^{(2)}(\omega_3)+e_F^{(3)}(\omega_3))\otimes (e_F^{(2)}(\omega_4)+e_F^{(3)}(\omega_4))\Big )\ .\]
 Therefore, we have 

\be
\begin{split}
\matMt{4,4}(\text{4-Pt})\cdot &\Big(e_P^{(2)}(\omega_1)\otimes e_P^{(3)}(\omega_2)\otimes e_F^{(2)}(\omega_3)\otimes e_F^{(2)}(\omega_4)\Big )\\=\matMt{4,4}(\text{4-Pt}) \cdot &\Big(e_P^{(2)}(\omega_1)\otimes e_P^{(4)}(\omega_2)\otimes (e_F^{(2)}(\omega_3)+e_F^{(3)}(\omega_3))\\
&\otimes (e_F^{(2)}(\omega_4)+e_F^{(3)}(\omega_4))\Big )\ .
\end{split}
\label{appffragmen:5}
\ee


\subsection*{P-fragmentation :}
Consider the contraction of $\matMt{4,4}(\text{4-Pt})$ with the tensor $\Big(e_P^{(2)}(\omega_1)\otimes e_P^{(2)}(\omega_2)\otimes e_F^{(1)}(\omega_3)\otimes e_F^{(2)}(\omega_4)\Big )$. The corresponding diagram is 

\be
\begin{split}
&\matMt{4,4}(\text{4-Pt})\cdot \Big(e_P^{(2)}(\omega_1)\otimes e_P^{(2)}(\omega_2)\otimes e_F^{(1)}(\omega_3)\otimes e_F^{(2)}(\omega_4)\Big )\\
&=\raisebox{-1.8 cm}{\scalebox{0.75}{\contourDDDD{-3.5}{-1.0}{-0}{-0.5}{-0.5}}}\\
\end{split}
\label{apppfragmen:1}
\ee
One can slide all the insertions that lie on the $3^{rd}$ leg or below down by two legs to obtain the following diagram:

\be
\begin{split}
&\matMt{4,4}(\text{4-Pt}) \cdot \Big(e_P^{(2)}(\omega_1)\otimes e_P^{(2)}(\omega_2)\otimes e_F^{(1)}(\omega_3)\otimes e_F^{(2)}(\omega_4)\Big )\\
&=\raisebox{-1.8 cm}{\scalebox{0.75}{\begin{tikzpicture}[scale = 1.5]
\draw[thick,color=red](-1.2, -2.0) circle (0.5ex);
\draw[thick,color=red](-1.2, -2.5) circle (0.5ex);
\draw[thick,color=red](-0.4, -0) circle (0.5ex);
\draw[thick,color=red](-0.4, -0.5) circle (0.5ex);
\draw[thick,color=red](0.4, -0.5) circle (0.5ex);
\draw[thick,color=red](0.4, -2.0) circle (0.5ex);
\draw[thick,color=red](1.2, -0.5) circle (0.5ex);
\draw[thick,color=red](1.2, -2.0) circle (0.5ex);
\draw[thick, color=black]
{
 (-1.2,  - 2.0) node [above] {\large{{-1}}}
 (-1.2, -2.5) node [above] {\large{{+1}}}
(-0.4,  - 0) node [above] {\large{{-1}}}
 (-0.4, -0.5) node [above] {\large{{+1}}}
(0.4,  - 0.5) node [above] {\large{{-1}}}
 (0.4, -2.0) node [above] {\large{{+1}}}
 (1.2, - 0.5) node [above] {\large{{-1}}}
  (1.2, - 2.0) node [above] {\large{{+1}}}
};
\foreach \x in {0,...,-3.5}{
\pgfmathtruncatemacro\z{int(-\x + 1)}
\draw[thick,color=violet,->] (-2,\x cm) -- (2,\x cm) ;
\draw[thick,color=violet,->] (2, \x cm - 0.5 cm) -- (-2, \x cm - 0.5 cm) ;
\draw[thick,color=violet,->] (2, \x cm) arc (90:-90:0.25);
\draw[thick,color=blue] (2 + 0.25, \x cm - 0.25cm) node[right] {\footnotesize{\z}};
}
\foreach \y in {-1,...,-3.5}{
\pgfmathtruncatemacro\w{int(-\y + 1)}
\draw[thick,color=violet,->] (-2,\y cm + 0.5 cm) arc (90:270:0.25);
}
\end{tikzpicture}
}}\\
\end{split}
\label{apppfragmen:2}
\ee

We can add a pair of points with opposite signs on the $3^{rd}$ and the $4^{th}$ legs at positions corresponding to the frequency $\omega_2$ without changing the value of the contraction as shown below:

\be
\begin{split}
&\matMt{4,4}(\text{4-Pt}) \cdot \Big(e_P^{(2)}(\omega_1)\otimes e_P^{(2)}(\omega_2)\otimes e_F^{(1)}(\omega_3)\otimes e_F^{(2)}(\omega_4)\Big )\\
&=\raisebox{-1.8 cm}{\scalebox{0.75}{\begin{tikzpicture}[scale = 1.5]
\draw[thick,color=red](-1.2, -2.0) circle (0.5ex);
\draw[thick,color=red](-1.2, -2.5) circle (0.5ex);
\draw[thick,color=red](-0.4, -0) circle (0.5ex);
\draw[thick,color=red](-0.4, -0.5) circle (0.5ex);
\draw[thick,color=red](0.4, -0.5) circle (0.5ex);
\draw[thick,color=red](0.4, -2.0) circle (0.5ex);
\draw[thick,color=red](0.4, -1.0) circle (0.5ex);
\draw[thick,color=red](0.4, -1.5) circle (0.5ex);
\draw[thick,color=red](1.2, -0.5) circle (0.5ex);
\draw[thick,color=red](1.2, -2.0) circle (0.5ex);
\draw[thick, color=black]
{
 (-1.2,  - 2.0) node [above] {\large{{-1}}}
 (-1.2, -2.5) node [above] {\large{{+1}}}
(-0.4,  - 0) node [above] {\large{{-1}}}
 (-0.4, -0.5) node [above] {\large{{+1}}}
(0.4,  - 0.5) node [above] {\large{{-1}}}
 (0.4, -2.0) node [above] {\large{{+1}}}
(0.4,  - 1.0) node [above] {\large{{+1}}}
 (0.4, -1.5) node [above] {\large{{-1}}}
 (1.2, - 0.5) node [above] {\large{{-1}}}
  (1.2, - 2.0) node [above] {\large{{+1}}}
};
\foreach \x in {0,...,-3.5}{
\pgfmathtruncatemacro\z{int(-\x + 1)}
\draw[thick,color=violet,->] (-2,\x cm) -- (2,\x cm) ;
\draw[thick,color=violet,->] (2, \x cm - 0.5 cm) -- (-2, \x cm - 0.5 cm) ;
\draw[thick,color=violet,->] (2, \x cm) arc (90:-90:0.25);
\draw[thick,color=blue] (2 + 0.25, \x cm - 0.25cm) node[right] {\footnotesize{\z}};
}
\foreach \y in {-1,...,-3.5}{
\pgfmathtruncatemacro\w{int(-\y + 1)}
\draw[thick,color=violet,->] (-2,\y cm + 0.5 cm) arc (90:270:0.25);
}
\end{tikzpicture}
}}\\
\end{split}
\label{apppfragmen:3}
\ee
There is no insertion to the future of these new insertions which can block them from collapsing onto one another. Now, one can add another pair of insertions with opposite signs on the $3^{rd}$ and the $4^{th}$ legs at positions corresponding to the frequency $\omega_1$ without changing the value of the contraction as shown below:
\be
\begin{split}
&\matMt{4,4}(\text{4-Pt}) \cdot \Big(e_P^{(2)}(\omega_1)\otimes e_P^{(2)}(\omega_2)\otimes e_F^{(1)}(\omega_3)\otimes e_F^{(2)}(\omega_4)\Big )\\
&=\raisebox{-1.8 cm}{\scalebox{0.75}{\begin{tikzpicture}[scale = 1.5]
\draw[thick,color=red](-1.2, -2.0) circle (0.5ex);
\draw[thick,color=red](-1.2, -2.5) circle (0.5ex);
\draw[thick,color=red](-0.4, -0) circle (0.5ex);
\draw[thick,color=red](-0.4, -0.5) circle (0.5ex);
\draw[thick,color=red](0.4, -0.5) circle (0.5ex);
\draw[thick,color=red](0.4, -2.0) circle (0.5ex);
\draw[thick,color=red](0.4, -1.0) circle (0.5ex);
\draw[thick,color=red](0.4, -1.5) circle (0.5ex);
\draw[thick,color=red](1.2, -0.5) circle (0.5ex);
\draw[thick,color=red](1.2, -2.0) circle (0.5ex);
\draw[thick,color=red](1.2, -1.0) circle (0.5ex);
\draw[thick,color=red](1.2, -1.5) circle (0.5ex);
\draw[thick, color=black]
{
 (-1.2,  - 2.0) node [above] {\large{{-1}}}
 (-1.2, -2.5) node [above] {\large{{+1}}}
(-0.4,  - 0) node [above] {\large{{-1}}}
 (-0.4, -0.5) node [above] {\large{{+1}}}
(0.4,  - 0.5) node [above] {\large{{-1}}}
 (0.4, -2.0) node [above] {\large{{+1}}}
(0.4,  - 1.0) node [above] {\large{{+1}}}
 (0.4, -1.5) node [above] {\large{{-1}}}
 (1.2, - 0.5) node [above] {\large{{-1}}}
  (1.2, - 2.0) node [above] {\large{{+1}}}
 (1.2, - 1.0) node [above] {\large{{+1}}}
  (1.2, - 1.5) node [above] {\large{{-1}}}
};
\foreach \x in {0,...,-3.5}{
\pgfmathtruncatemacro\z{int(-\x + 1)}
\draw[thick,color=violet,->] (-2,\x cm) -- (2,\x cm) ;
\draw[thick,color=violet,->] (2, \x cm - 0.5 cm) -- (-2, \x cm - 0.5 cm) ;
\draw[thick,color=violet,->] (2, \x cm) arc (90:-90:0.25);
\draw[thick,color=blue] (2 + 0.25, \x cm - 0.25cm) node[right] {\footnotesize{\z}};
}
\foreach \y in {-1,...,-3.5}{
\pgfmathtruncatemacro\w{int(-\y + 1)}
\draw[thick,color=violet,->] (-2,\y cm + 0.5 cm) arc (90:270:0.25);
}
\end{tikzpicture}
}}\\
\end{split}
\label{apppfragmen:4}
\ee

As before, there is no insertion to the future of these new insertions which can block them from collapsing onto one another. But this is exactly the diagram for the contraction of  $\matMt{4,4}(\text{4-Pt})$ with  $\Big((e_P^{(2)}(\omega_1)+e_P^{(3)}(\omega_1))\otimes (e_P^{(2)}(\omega_2)+e_P^{(3)}(\omega_2))\otimes e_F^{(1)}(\omega_3)\otimes e_F^{(3)}(\omega_4)\Big )$. Therefore, we have 

\be
\begin{split}
\matMt{4,4}(\text{4-Pt}) \cdot &\Big(e_P^{(2)}(\omega_1)\otimes e_P^{(2)}(\omega_2)\otimes e_F^{(1)}(\omega_3)\otimes e_F^{(2)}(\omega_4)\Big )\\=\matMt{4,4}(\text{4-Pt}) \cdot &\Big((e_P^{(2)}(\omega_1)+e_P^{(3)}(\omega_1))\otimes (e_P^{(2)}(\omega_2)+e_P^{(2)}(\omega_2))\\&\otimes e_F^{(1)}(\omega_3)e_F^{(3)}(\omega_4)\Big )\ .
\end{split}
\label{apppfragmen:5}
\ee

\section{Orthogonal tensors and Column Vector Representation }
\subsection{Column Vector Representation for $2$ pt. functions }
\label{app2pt}
In this section, we are going to discuss the  column vector representation of the array of two point  correlators on a k-fold contour. We take  $\matMt{k,4}(\text{2-Pt})$
to be the Wightman array in the  Fourier domain. Its components in column vector basis  satisfy the rules mentioned in \ref{rulesummary}. Using these rules, one can show that
\be
\boxed{\begin{split}
\matMt{k,2}(\text{2-Pt})&=\rho[12]\ \sum_{r=1}^k  \Big(\ebP{r+1}-\ebP{r}\Big)\otimes \ebF{r}\ .
\end{split}}
\label{app2pt:4}
\ee

\subsection*{Orthogonal tensors for two pt. functions} 
The expression in \eqref{app2pt:4}  can be proved by demonstrating that the array of two pt. functions should be orthogonal to the following tensors :
\begin{equation}
\begin{split}
 3k^2\ \text{orthogonal tensors} :&\  e_{_{PP}}^{rs}\ ,\  e_{_{FP}}^{rs}\ ,\  e_{_{FF}}^{rs}\ , \\
 k(k-2)\ \text{orthogonal tensors} :&\ e_{_{PF}}^{rs}\ \text{for}\ r\neq s,s+1 \ ,\\
  k\ \text{orthogonal tensors} :&\  e_{_{PF}}^{rr}+ e_{_{PF}}^{(r+1)r}\ ,\\
    k-1\ \text{orthogonal tensors} :&\  e_{_{PF}}^{rr}- e_{_{PF}}^{11} \ .\\
 \end{split}
 \label{app2pt:5}
\end{equation}
Here we have used a short-hand notation for the tensor products of the row vectors. For example, we have written $\eP{r}(\omega_1)\otimes\eF{s}(\omega_2)$ as $e_{_{PF}}^{rs}$ and so on.

 The total number of elements in this array is $2k \times 2k=4k^2$. The total number of tensors that are orthogonal to the array $\matMt{k,2}(\text{2-Pt})$ is $(4k^2-1)$. These are enumerated in \eqref{app2pt:5} .Therefore, the orthogonal tensors completely fix the array upto a single function which has to be determined.

The arguments for the tensors mentioned in \eqref{app2pt:5} being orthogonal to the array of contour correlators are based on rules of contraction enumerated in appendix~\ref{rulesummary} .These arguments are given in table~\ref{tab:2pt-ortho}.

\begin{table}[H]
\centering
\begin{tabular}{|c|c|c|}
\hline
 Orthogonal  & Total no. & Argument  \\   
 tensor	&  of tensors  & \\
\hline
$e_{_{PP}}^{rs}$ & $k^2$ & $e_{_{P \textcolor{red}{P}}}^{^{r\textcolor{red}{s}}} \xrightarrow {\text{P-collapse}} 0$ \\ & & \\
\hline
$e_{_{FP}}^{rs}$ & $k^2$ & $e_{_{\textcolor{red}{F}P}}^{^{\textcolor{red}{r}s}} \xrightarrow {\text{F-collapse}} 0$ \\& & \\
\hline
$e_{_{FF}}^{rs}$ & $k^2$ & $e_{_{ \textcolor{red}{F} F} }^{^{\textcolor{red}{r} s}} \xrightarrow {\text{F-collapse}} 0$ \\ & &\\
\hline
$e_{_{PF}}^{rs}$ & $k(k-2)$ & $e_{_{P \textcolor{red}{F}} }^{^{r \textcolor{red}{s}}} \xrightarrow {\text{F-collapse}} 0$ \\ \text{for} $\ r\neq s,s+1$ & &\\
\hline
$e_{_{PF}}^{rr}+ e_{_{PF}}^{(r+1)r}$ & $k$ & $e_{_{ \textcolor{blue}{P} \textcolor{brown}{F}} }^{^{\textcolor{blue}{(r+1)}\textcolor{brown}{r} }} \xrightarrow {\text{P-sliding}} -e_{_{ P F} }^{r r} $ \\ & &\\
\hline
$e_{_{PF}}^{rr}- e_{_{PF}}^{11}$ & $k-1$ & $e_{_{PF}}^{rr}\xrightarrow {\text{C-shift}} e_{_{ PF} }^{11} $ \\ & &\\
\hline
\end{tabular}
\caption{Arguments for the tensors orthogonal to the array of 2 point contour correlators \label{tab:2pt-ortho}} 
\end{table}
In this table, we have indicated the $e_{_P}$($e_{_F}$) which leads to a P(F)-collapse by red colour, and in case of a P/F-sliding we have indicated the $e_{_P}/e_{_F}$ that slides by blue colour and the corresponding $e_{_F}/e_{_P}$ that acts as the anchor by brown colour.
The array that is orthogonal to all the tensors mentioned above must have the form

\begin{equation}
\begin{split}
&\matMt{k,2}(\text{2-Pt})=  \alpha^{(2)} \sum_{r=1}^{k} ( \ebP{r+1} (\omega_1)-\ebP{r}(\omega_1))\otimes \ebF{r}(\omega_2)  \ .
\end{split}
 \label{app2pt:6}
\end{equation}

Now, the coefficient $ \alpha^{(2)}$ is given by the contraction with the tensor $e_{PF}^{(r+1)r}$ , an example of which is the case $r=1$. Then this coefficient is given by 

\begin{equation}
\begin{split}
 \alpha^{(2)}&=\matMt{k,2}(\text{2-Pt})\cdot e_{PF}^{\ 21} =\rho[12]\ .
\end{split}
 \label{app2pt:11}
\end{equation}

Substituting the value of $\alpha^{(2)}$ that was obtained in \eqref{app2pt:11} into the equation   \eqref{app2pt:6} we get the expansion that was mentioned in \eqref{app2pt:4}.

\subsection{Column Vector Representation  for $3$ pt. functions  }
\label{app3pt}
Now, let us discuss the  column vector representation of the array of 3-point correlators on a k-fold contour. We take  $\matMt{k,4}(\text{3-Pt})$
to be the Wightman array in the Fourier domain. Its components in column vector basis  satisfy the rules mentioned in \ref{rulesummary}. Using these rules, one can show that
\be \boxed{
\begin{split}
\matMt{k,3}(\text{3-Pt})&=\rho[321]  \sum_{r=1}^{k} ( \ebP{r+1}\otimes \ebP{r+1}-\ebP{r} \otimes \ebP{r})\otimes \ebF{r}\\
&\quad-\rho[123]  \sum_{r=1}^{k} ( \ebP{r+1}-\ebP{r})\otimes \ebF{r}\otimes \ebF{r}\ .  \end{split}}
\label{app3pt:5}
\ee

\subsection*{Orthogonal tensors for three pt. functions} 
The expression in \eqref{app3pt:5}  can be proved by demonstrating that the three pt. function should be orthogonal to the following tensors :
\begin{equation}
\begin{split}
 6k^3\ \text{orthogonal tensors} :&\  e_{_{PPP}}^{rsu}\ ,\  e_{_{FPP}}^{rsu}\ ,\  e_{_{PFP}}^{rsu}\ ,\  e_{_{FFP}}^{rsu}\ ,\  e_{_{FPF}}^{rsu}\ ,\  e_{_{FFF}}^{rsu}  \ ,\\
 k^2(k-1)\ \text{orthogonal tensors} :&\ e_{_{PPF}}^{rsu}\ \text{for}\ r\neq s  \ ,\\ 
  k(k-2)\ \text{orthogonal tensors} :&\ e_{_{PPF}}^{rrs}\ \text{for}\ s\neq r,r-1 \ , \\
  k\ \text{orthogonal tensors} :&\  e_{_{PPF}}^{rrr}+ e_{_{PPF}}^{(r+1)(r+1)r} \ ,\\
  k-1\ \text{orthogonal tensors} :&\  e_{_{PPF}}^{rrr}- e_{_{PPF}}^{111} \ ,\\
  k^2(k-1)\ \text{orthogonal tensors} :&\ e_{_{PFF}}^{rsu}\ \text{for}\ s\neq u \ ,\\  
  k(k-2)\ \text{orthogonal tensors} :&\ e_{_{PFF}}^{rss}\ \text{for}\ r\neq s,s+1 \ , \\
  k\ \text{orthogonal tensors} :&\  e_{_{PFF}}^{rrr}+ e_{_{PFF}}^{(r+1)rr} \ ,\\
  k-1\ \text{orthogonal tensors} :&\  e_{_{PFF}}^{rrr}- e_{_{PFF}}^{111} \ .\\
 \end{split}
 \label{app3pt:6}
\end{equation}
This gives in total $8k^3-2$ orthogonal tensors. 

The arguments for the tensors mentioned in \eqref{app3pt:6} being orthogonal to the array of contour correlators are given in tables \ref{tab:3pttrivial-ortho},\ref{tab:3ptppf-ortho} and \ref{tab:3ptpff-ortho}.

\begin{table}
\centering
\begin{tabular}{|c|c|c|}
\hline
 Orthogonal  & Total no. & Argument  \\   
 tensor	&  of tensors  & \\
\hline
$e_{_{PPP}}^{rsu}$ & $k^3$ & $e_{_{PP \textcolor{red}{P}}}^{^{rs\textcolor{red}{u}}} \xrightarrow {\text{P-collapse}} 0$ \\ & & \\
\hline
$e_{_{PFP}}^{rsu}$ & $k^3$ & $e_{_{ PF\textcolor{red}{P}} }^{^{rs\textcolor{red}{u}}} \xrightarrow {\text{P-collapse}} 0$ \\ & &\\
\hline
$e_{_{FPP}}^{rsu}$ & $k^3$ & $e_{_{ \textcolor{red}{F}PP}}^{^{\textcolor{red}{r}su}} \xrightarrow {\text{F-collapse}} 0$ \\& & \\
\hline
$e_{_{FPF}}^{rsu}$ & $k^3$ & $e_{_{ \textcolor{red}{F}PF}}^{^{\textcolor{red}{r}su}} \xrightarrow {\text{F-collapse}} 0$ \\& & \\
\hline
$e_{_{FFP}}^{rsu}$ & $k^3$ & $e_{_{ \textcolor{red}{F}FP}}^{^{\textcolor{red}{r}su}} \xrightarrow {\text{F-collapse}} 0$ \\& & \\
\hline
$e_{_{FFF}}^{rsu}$ & $k^3$ & $e_{_{ \textcolor{red}{F}FF}}^{^{\textcolor{red}{r}su}} \xrightarrow {\text{F-collapse}} 0$ \\& & \\
\hline
\end{tabular}
\caption{Arguments for the tensors trivially orthogonal to the array of 3 point contour correlators \label{tab:3pttrivial-ortho} } 
\end{table}

\begin{table}
\centering
\begin{tabular}{|c|c|c|}
\hline
 Orthogonal  & Total no. & Argument  \\   
 tensor	&  of tensors  & \\
\hline
$e_{_{PPF}}^{rsu}$ & $k^2 (k-1)$ & \text{W.L.O.G let us consider} $r>s$.   \\  \text{for} $\ r\neq s$ & & \underline{Case $1:r>(s+1)$}\\ & & If $\ u\neq s,s-1$, then \\ & & $e_{_{ P\textcolor{red}{P}F}}^{^{r\textcolor{red}{s}u}} \xrightarrow {\text{P-collapse}} 0$, otherwise $e_{_{ \textcolor{red}{P}PF}}^{^{\textcolor{red}{r}su}} \xrightarrow {\text{P-collapse}} 0$.\\  & & \underline{Case $2:r=(s+1)$}\\ & & If $\ u= s$, then \\ & & $e_{_{\textcolor{brown}{P} P \textcolor{blue}{F}}}^{^{\textcolor{brown}{(s+1)}s \textcolor{blue}{s}}} \xrightarrow {\text{F-sliding}} -e_{_{P\textcolor{red}{P}F}}^{^{(s+1)\textcolor{red}{s}(s+1)}}\xrightarrow {\text{P-collapse}}0$,\\ & & if $\ u>s$, then $e_{_{P\textcolor{red}{P}F}}^{^{(s+1)\textcolor{red}{s}u}}\xrightarrow {\text{P-collapse}}0$,\\ & & and if $\ u<s$, then $e_{_{\textcolor{red}{P}PF}}^{^{\textcolor{red}{(s+1)}su}}\xrightarrow {\text{P-collapse}}0$.\\
\hline
$e_{_{PPF}}^{rrs}$ & $k(k-2)$ & $e_{_{ PP\textcolor{red}{F}}}^{^{rr\textcolor{red}{s}}} \xrightarrow {\text{F-collapse}} 0$ \\  \text{for}$\ s\neq r,r-1$ & & \\
\hline
$e_{_{PPF}}^{rrr}+ e_{_{PPF}}^{(r+1)(r+1)r}$ & $k$ & $e_{_{PPF}}^{(r+1)(r+1)r}\xrightarrow {\text{C-shift}}e_{_{ P\textcolor{brown}{P} \textcolor{blue}{F}} }^{^{r\textcolor{brown}{r}\textcolor{blue}{(r-1)}}} \xrightarrow {\text{F-sliding}} -e_{_{ PPF} }^{rrr} $ \\ & &\\
\hline
$e_{_{PPF}}^{rrr}- e_{_{PPF}}^{111}$ & $k-1$ & $e_{_{PPF}}^{rrr}\xrightarrow {\text{C-shift}} e_{_{PPF}}^{111} $ \\ & &\\
\hline
\end{tabular}
\caption{Arguments for the tensors in the PPF sector orthogonal to the array of 3 point contour correlators \label{tab:3ptppf-ortho}} 
\end{table}

\begin{table}
\centering
\begin{tabular}{|c|c|c|}
\hline
 Orthogonal  & Total no. & Argument  \\   
 tensor	&  of tensors  & \\
\hline
$e_{_{PFF}}^{rsu}$ & $k^2 (k-1)$ & \text{W.L.O.G let us consider} $s>u$.   \\  \text{for} $\ s\neq u$ & & \underline{Case $1:s>(u+1)$}\\ & & If $\ r\neq s,s+1$, then \\ & & $e_{_{ P\textcolor{red}{F}F}}^{^{r\textcolor{red}{s}u}} \xrightarrow {\text{F-collapse}} 0$, otherwise $e_{_{ PF\textcolor{red}{F}}}^{^{rs\textcolor{red}{u}}} \xrightarrow {\text{F-collapse}} 0$.\\  & & \underline{Case $2:s=(u+1)$}\\ & & If $\ r= u+1$, then \\ & & $e_{_{ \textcolor{blue}{P}\textcolor{brown}{F}F}}^{^{\textcolor{blue}{(u+1)}\textcolor{brown}{(u+1)}u}} \xrightarrow {\text{P-sliding}} -e_{_{PF\textcolor{red}{F}}}^{^{(u+2)(u+1)\textcolor{red}{u}}}\xrightarrow {\text{F-collapse}}0$,\\ & & if $\ r>(u+1)$, then $e_{_{PF\textcolor{red}{F}}}^{^{r(u+1)\textcolor{red}{u}}}\xrightarrow {\text{F-collapse}}0$,\\ & & and if $\ r<(u+1)$, then $e_{_{P\textcolor{red}{F}F}}^{^{r\textcolor{red}{(u+1)}u}}\xrightarrow {\text{F-collapse}}0$.\\
\hline
$e_{_{PFF}}^{rss}$ & $k(k-2)$ & $e_{_{ P\textcolor{red}{F}F}}^{^{r\textcolor{red}{s}s}} \xrightarrow {\text{F-collapse}} 0$ \\  \text{for}$\ r\neq s,s+1$ & & \\
\hline
$e_{_{PFF}}^{rrr}+ e_{_{PFF}}^{(r+1)rr}$ & $k$ & $e_{_{ \textcolor{blue}{P} \textcolor{brown}{F}F} }^{^{\textcolor{blue}{(r+1)}\textcolor{brown}{r}r }} \xrightarrow {\text{P-sliding}} -e_{_{ PFF} }^{rrr} $ \\ & &\\
\hline
$e_{_{PFF}}^{rrr}- e_{_{PFF}}^{111}$ & $k-1$ & $e_{_{PFF}}^{rrr}\xrightarrow {\text{C-shift}} e_{_{ PFF} }^{111} $ \\ & &\\
\hline
\end{tabular}
\caption{Arguments for the tensors in the PFF sector orthogonal to the array of 3 point contour correlators \label{tab:3ptpff-ortho}} 
\end{table}

As before, in tables \ref{tab:3pttrivial-ortho}, \ref{tab:3ptppf-ortho} and \ref{tab:3ptpff-ortho} we have indicated the $e_{_P}$($e_{_F}$) which leads to a P(F)-collapse by red colour, and in case of a P/F-sliding we have indicated the $e_{_P}/e_{_F}$ that slides by blue colour and the corresponding $e_{_F}/e_{_P}$ that acts as the anchor by brown colour.

The array that is orthogonal to all the tensors mentioned above must have the form
\begin{equation}
\begin{split}
&\matMt{k,3}(\text{3-Pt})\\&
= \alpha_1^{(3)}\Big ( \sum_{r=1}^{k} ( \ebP{r+1} (\omega_1)\otimes \ebP{r+1}(\omega_2)\otimes \ebF{r}(\omega_3)-\ebP{r} (\omega_1)\otimes \ebP{r}(\omega_2)\otimes \ebF{r}(\omega_3)) \Big )\\
&+ \alpha_2^{(3)} \Big ( \sum_{r=1}^{k} ( \ebP{r+1} (\omega_1)-\ebP{r}(\omega_1))\otimes \ebF{r}(\omega_2)\otimes \ebF{r}(\omega_3) \Big ) ~.
\end{split}
\label{app3pt:7}
\end{equation}
Now, the coefficient $ \alpha_1^{(3)}$ is given by the contraction with the tensor $e_{PPF}^{(r+1) (r+1) r}$, an example of which is the case $r=1$. Then this coefficient is given by 
\begin{equation}
\begin{split}
 \alpha_1^{(3)} &=\matMt{k,3}(\text{3-Pt})\cdot e_{PPF}^{\ 221} =\rho[321]\ .
\end{split}
 \label{app3pt:12}
\end{equation}
The coefficient $ \alpha_2^{(3)}$ is given by the contraction with the tensor $e_{PFF}^{(r+1) rr}$ , an example of which is the case $r=1$. Then this coefficient is given by 
\begin{equation}
\begin{split}
 \alpha_2^{(3)} &=\matMt{k,3}(\text{3-Pt})\cdot e_{PFF}^{\ 211}=-\rho[123]\ .
\end{split}
 \label{app3pt:17}
\end{equation}
Substituting the values of $\alpha_1^{(3)}$ and $\alpha_2^{(3)}$ that were obtained in \eqref{app3pt:12} and \eqref{app3pt:17} into the equation   \eqref{app3pt:7} we get the expansion that was mentioned in \eqref{app3pt:5}.

\subsection{Column Vector Representation for $4$ pt. functions}
\label{app4pt}

Finally, let us discuss the  column vector representation of the array of 4-point correlators on a k-fold contour. We take  $\matMt{k,4}(\text{4-Pt})$
to be the Wightman array in the  Fourier domain. Its components in column vector basis  satisfy the rules mentioned in \ref{rulesummary}. Using these rules, one can show that 
\be
\boxed{\begin{split}
\matMt{k,4}(\text{4-Pt})&=\matMt{k,4}_{PPPF}+\matMt{k,4}_{PFFF}+\matMt{k,4}_{PFPF}+\matMt{k,4}_{PPFF}
\end{split}}
\label{app4pt:5}
\ee
where
\be
\boxed{\begin{split}
\matMt{k,4}_{PPPF}&=-\rho[4321]
\sum_{r=1}^k \Big (\ebP{r+1} \otimes \ebP{r+1}\otimes \ebP{r+1} \otimes \ebF{r}-\ebP{r} \otimes \ebP{r} \otimes \ebP{r} \otimes \ebF{r}\Big )\ , 
\end{split}}
\label{app4pt:6}
\ee
\be
\boxed{\begin{split}
\matMt{k,4}_{PFFF}&=\rho[1234]
\sum_{r=1}^k \Big (\ebP{r+1}-\ebP{r}\Big ) \otimes \ebF{r} \otimes \ebF{r}\otimes \ebF{r}\ ,
\end{split}}
\label{app4pt:7}
\ee
\be
\boxed{
\begin{split}
\matMt{k,4}_{PFPF}&= \sum_{r,s=1}^k  \Big(\theta_{r>s}\ \rho[12][34]+\theta_{r \leq s}\ \rho[34][12]\Big)\Big (\ebP{r+1}-\ebP{r}\Big ) \otimes \ebF{r} \otimes \ebP{s+1} \otimes \ebF{s}\\
& -\sum_{r,s=1}^k   \Big(\theta_{r\geq s} \rho[12][34] +\theta_{r < s}\ \rho[34][12]\Big)  \Big(\ebP{r+1}-\ebP{r}\Big) \otimes \ebF{r}\otimes \ebP{s} \otimes \ebF{s}\ ,
\end{split}}
\label{app4pt:8}
\ee
\be
\boxed{\begin{split}
\matMt{k,4}_{PPFF}= \sum_{r,s=1}^k &  \Big(\theta_{r>s}\ \rho[13][24]
+\theta_{r \leq s}\ \rho[24][13] \Big)\Big (\ebP{r+1}-\ebP{r}\Big ) \otimes \ebP{s+1} \otimes \ebF{r} \otimes \ebF{s}\\
& -\sum_{r,s=1}^k \Big(\theta_{r\geq s}\ \rho[13][24]+\theta_{r < s}\ \rho[24][13] \Big)  \Big (\ebP{r+1}-\ebP{r}\Big ) \otimes \ebP{s} \otimes \ebF{r} \otimes \ebF{s}\\
& +\sum_{r,s=1}^k \Big(\theta_{r\geq s}\ \rho[14][23] +\theta_{r < s}\ \rho[23][14] \Big)  \ebP{r+1}\otimes \Big (\ebP{s+1}-\ebP{s}\Big )\otimes \ebF{s} \otimes \ebF{r}\\
&-\sum_{r,s=1}^k   \Big(\theta_{r> s}\ \rho[14][23]+\theta_{r \leq s}\ \rho[23][14] \Big)  \ebP{r}\otimes \Big (\ebP{s+1}-\ebP{s}\Big )\otimes \ebF{s} \otimes \ebF{r}\\
&+\rho[2314]\ \sum_{r=1}^k \Big (\ebP{r+1}\otimes \ebP{r+1}-\ebP{r}\otimes \ebP{r}\Big) \otimes \ebF{r} \otimes \ebF{r}\ .
\end{split}}
\label{app4pt:9}
\ee

\subsection*{Orthogonal tensors for four pt. functions:} 
The results in \eqref{app4pt:5},\eqref{app4pt:6},\eqref{app4pt:7},\eqref{app4pt:8} and \eqref{app4pt:9} can be proved by demonstrating that the four pt. function should be orthogonal to the following tensors.

\paragraph{Trivial orthogonal tensors}
\begin{equation}
\begin{split}
 12k^4\ \text{orthogonal tensors} :&\  e_{_{PPPP}}^{rsuv}\ ,\  e_{_{PPFP}}^{rsuv}\ ,\  e_{_{PFPP}}^{rsuv}\ ,\  e_{_{PFFP}}^{rsuv}\ ,\  e_{_{FPPP}}^{rsuv}\ ,\  e_{_{FPPF}}^{rsuv},  \\
   &\  e_{_{FPFP}}^{rsuv}\ ,\  e_{_{FPFF}}^{rsuv}\ ,\  e_{_{FFPP}}^{rsuv}\ ,\  e_{_{FFPF}}^{rsuv}\ ,\  e_{_{FFFP}}^{rsuv}\ ,\  e_{_{FFFF}}^{rsuv}  \\
  \end{split}
 \label{app4pt:10}
\end{equation}

\begin{table}
\centering
\begin{tabular}{|c|c|c|}
\hline
 Orthogonal  & Total no. & Argument  \\   
 tensor	&  of tensors  & \\
\hline
$e_{_{PPPP}}^{rsuv}$ & $k^4$ & $e_{_{PPP \textcolor{red}{P}}}^{^{rsu\textcolor{red}{v}}} \xrightarrow {\text{P-collapse}} 0$ \\ & & \\
\hline
$e_{_{PPFP}}^{rsuv}$ & $k^4$ & $e_{_{ PPF\textcolor{red}{P}} }^{^{rsu\textcolor{red}{v}}} \xrightarrow {\text{P-collapse}} 0$ \\ & &\\
\hline
$e_{_{PFPP}}^{rsuv}$ & $k^4$ & $e_{_{ PFP\textcolor{red}{P}} }^{^{rsu\textcolor{red}{v}}} \xrightarrow {\text{P-collapse}} 0$ \\ & &\\
\hline
$e_{_{PFFP}}^{rsuv}$ & $k^4$ & $e_{_{ PFF\textcolor{red}{P}} }^{^{rsu\textcolor{red}{v}}} \xrightarrow {\text{P-collapse}} 0$ \\ & &\\
\hline
$e_{_{FPPP}}^{rsuv}$ & $k^4$ & $e_{_{ \textcolor{red}{F}PPP}}^{^{\textcolor{red}{r}suv}} \xrightarrow {\text{F-collapse}} 0$ \\& & \\
\hline
$e_{_{FPPF}}^{rsuv}$ & $k^4$ & $e_{_{ \textcolor{red}{F}PPF}}^{^{\textcolor{red}{r}suv}} \xrightarrow {\text{F-collapse}} 0$ \\& & \\
\hline
$e_{_{FPFP}}^{rsuv}$ & $k^4$ & $e_{_{ \textcolor{red}{F}PFP}}^{^{\textcolor{red}{r}suv}} \xrightarrow {\text{F-collapse}} 0$ \\& & \\
\hline
$e_{_{FPFF}}^{rsuv}$ & $k^4$ & $e_{_{ \textcolor{red}{F}PFF}}^{^{\textcolor{red}{r}suv}} \xrightarrow {\text{F-collapse}} 0$ \\& & \\
\hline
$e_{_{FFPP}}^{rsuv}$ & $k^4$ & $e_{_{ \textcolor{red}{F}FPP}}^{^{\textcolor{red}{r}suv}} \xrightarrow {\text{F-collapse}} 0$ \\& & \\
\hline
$e_{_{FFPF}}^{rsuv}$ & $k^4$ & $e_{_{ \textcolor{red}{F}FPF}}^{^{\textcolor{red}{r}suv}} \xrightarrow {\text{F-collapse}} 0$ \\& & \\
\hline
$e_{_{FFFP}}^{rsuv}$ & $k^4$ & $e_{_{ \textcolor{red}{F}FFP}}^{^{\textcolor{red}{r}suv}} \xrightarrow {\text{F-collapse}} 0$ \\& & \\
\hline
$e_{_{FFFF}}^{rsuv}$ & $k^4$ & $e_{_{ \textcolor{red}{F}FFF}}^{^{\textcolor{red}{r}suv}} \xrightarrow {\text{F-collapse}} 0$ \\& & \\
\hline
\end{tabular}
\caption{Arguments for the tensors trivially orthogonal to the array of 4 point contour correlators \label{tab:4pttrivial-ortho}} 
\end{table}

\paragraph{Orthogonal tensors in the PPPF sector:}
\begin{equation}
\begin{split}
 k(k^3-k)\ \text{orthogonal tensors} :&\ e_{_{PPPF}}^{rsuv}\ \text{when $r,s$ and $u$ are not all equal}  \ ,\\ 
  k(k-2)\ \text{orthogonal tensors} :&\ e_{_{PPPF}}^{rrrs}\ \text{for}\ r\neq s,s+1 \ ,\\
   2k-1\ \text{orthogonal tensors} :&\ e_{_{PPPF}}^{(r+1)(r+1)(r+1)r}+e_{_{PPPF}}^{rrrr}\ .\\
     \end{split}
 \label{app4pt:11} 
\end{equation}

Total number of orthogonal tensors in this sector $=k^4-1$.

These orthogonal tensors fix $\matMt{k,4}_{PPPF}$ to be of the following form:

\be
\begin{split}
&\matMt{k,4}_{PPPF}\\&=\alpha^{(4)}_1 \sum_{r=1}^k  \Big(\ebP{r+1}(\omega_1)\otimes \ebP{r+1}(\omega_2)\otimes \ebP{r+1}(\omega_3)\otimes\ebF{r}(\omega_4)\\
&\qquad\qquad-\ebP{r}(\omega_1)\otimes \ebP{r}(\omega_2)\otimes \ebP{r}(\omega_3)\otimes\ebF{r}(\omega_4) \Big)~.
\end{split}
\label{app4pt:12}
\ee

The coefficient $ \alpha_1^{(4)}$ is given by the contraction with the tensor $e_{PPPF}^{(r+1) (r+1)(r+1)r}$ , an example of which is the case $r=1$. Then this coefficient is given by 

\begin{equation}
\begin{split}
 \alpha_1^{(4)} &=\matMt{k,4}(\text{4-Pt})\cdot e_{PPPF}^{\ 2221}= \rho[4321]\ .
\end{split}
 \label{app4pt:13}
\end{equation}

\begin{table}[H]
\centering
\begin{tabular}{|c|c|c|}
\hline
 Orthogonal  & Total no. & Argument  \\   
 tensor	&  of tensors  & \\
\hline
$e_{_{PPPF}}^{rsuv}$ & $k (k^3-k)$ & \underline{Case1:$s\neq u$} \\
\text{ when $r,s$ and $u$} & & \text{W.L.O.G let us consider} $s>u$. \\ 
\text{ are not all equal} & &  \underline{Subcase $1:s>(u+1)$}\\ 
& & $e_{_{ PP\textcolor{red}{P}F}}^{^{rs\textcolor{red}{u}v}} \xrightarrow {\text{P-collapse}} 0$, otherwise $e_{_{ P\textcolor{red}{P}PF}}^{^{r\textcolor{red}{s}uv}} \xrightarrow {\text{P-collapse}} 0$.\\  
& & \underline{Subcase $2:s=(u+1)$}\\
& & If $\ v= u$, then \\
 & & $e_{_{P\textcolor{brown}{P} P \textcolor{blue}{F}}}^{^{r\textcolor{brown}{(u+1)}u\textcolor{blue}{u}}} \xrightarrow {\text{F-sliding}} -e_{_{PP\textcolor{red}{P}F}}^{^{r(u+1)\textcolor{red}{u}(u+1)}}\xrightarrow {\text{P-collapse}}0$,\\
 & & if $\ v>u$, then \\
 & & $e_{_{PP\textcolor{red}{P}F}}^{^{r(u+1)\textcolor{red}{u}v}} \xrightarrow {\text{P-collapse}}0$,\\
& & if $\ v<u$, then \\
 & & $e_{_{P\textcolor{red}{P}PF}}^{^{r\textcolor{red}{(u+1)}uv}} \xrightarrow {\text{P-collapse}}0$,\\ 
& & \underline{Case 2: $s=u\neq r$}\\ 
& & W.L.O.G let us consider r>s.\\
& & \underline{Subcase $1:r>(s+1)$}\\
& & If $\ v\neq s,s-1$, then \\& & $e_{_{ PP\textcolor{red}{P}F}}^{^{rs\textcolor{red}{s}v}} \xrightarrow {\text{P-collapse}} 0$, \\& & otherwise $e_{_{ \textcolor{red}{P}PPF}}^{^{\textcolor{red}{r}ssv}} \xrightarrow {\text{P-collapse}} 0$.\\ & & \underline{Subcase $2:r=(s+1)$}\\ & & If $\ v= s$, then \\& & $e_{_{ PP\textcolor{brown}{P}\textcolor{blue}{F}}}^{^{(s+1)s\textcolor{brown}{s}\textcolor{blue}{s}}} \xrightarrow {\text{F-sliding}} -e_{_{\textcolor{red}{P}PPF}}^{^{\textcolor{red}{(s+1)}ss(s-1)}}\xrightarrow {\text{P-collapse}}0$,\\ & & if $\ v>s$, then $e_{_{PP\textcolor{red}{P}F}}^{^{(s+1)s\textcolor{red}{s}v}}\xrightarrow {\text{P-collapse}}0$,\\ & & and if $\ v<s$, then $e_{_{\textcolor{red}{P}PPF}}^{^{\textcolor{red}{(s+1)}ssv}}\xrightarrow {\text{P-collapse}}0$.\\
\hline
$e_{_{PPPF}}^{rrrs}$ & $k(k-2)$ & $e_{_{ PPP\textcolor{red}{F}}}^{^{rrr\textcolor{red}{s}}} \xrightarrow {\text{F-collapse}} 0$ \\  \text{for}$\ s\neq r,r-1$ & & \\
\hline
$e_{_{PPPF}}^{rrrr}+ e_{_{PPPF}}^{(r+1)(r+1)(r+1)r}$ & $2k-1$ & $e_{_{PPPF}}^{(r+1)(r+1)(r+1)r}\xrightarrow {\text{C-shift}}e_{_{ PP\textcolor{brown}{P} \textcolor{blue}{F}} }^{^{rr\textcolor{brown}{r}\textcolor{blue}{(r-1)}}} \xrightarrow {\text{F-sliding}} -e_{_{ PPPF} }^{rrrr} $ \\ & &\\
\hline
\end{tabular}
\caption{Arguments for the tensors in the PPPF sector orthogonal to the array of 4 point contour correlators \label{tab:4ptPPPF-ortho}} 
\end{table}

   \paragraph{Orthogonal tensors in the PFFF sector:}
  \begin{equation}
\begin{split} 
 k(k^3-k)\ \text{orthogonal tensors} :&\ e_{_{PFFF}}^{rsuv}\ \text{when s,u and v are not all equal} \ , \\ 
  k(k-2)\ \text{orthogonal tensors} :&\ e_{_{PFFF}}^{rsss}\ \text{for}\ r\neq s,s+1 \ ,\\
   2k-1\ \text{orthogonal tensors} :&\ e_{_{PFFF}}^{(r+1)rrr}+e_{_{PFFF}}^{rrrr}~.\\
     \end{split}
 \label{app4pt:14}
\end{equation}

Total number of orthogonal tensors in this sector $=k^4-1$.

\begin{table}[H]
\centering
\begin{tabular}{|c|c|c|}
\hline
 Orthogonal  & Total no. & Argument  \\   
 tensor	&  of tensors  & \\
\hline
$e_{_{PFFF}}^{rsuv}$ & $k (k^3-k)$ &  \underline{Case $1:s\neq u$} \\  
\text{ when $s,u$ and $v$}  & & W.L.O.G, let us consider s>u\\
 \text{ are not all equal} & & \underline{Subcase $1:s>(u+1)$}\\
 & & $e_{_{ P\textcolor{red}{F}FF}}^{^{r\textcolor{red}{s}uv}} \xrightarrow {\text{F-collapse}} 0$,  otherwise $e_{_{ PF\textcolor{red}{F}F}}^{^{rs\textcolor{red}{u}v}} \xrightarrow {\text{F-collapse}} 0$.\\
 & &  \underline{Subcase $2:s=(u+1)$}\\
 & &  If $\ r= u+1$, then \\
 & & $e_{_{\textcolor{blue}{P}F\textcolor{brown}{F}  F}}^{^{\textcolor{blue}{(u+1)}(u+1)\textcolor{brown}{u}v}} \xrightarrow {\text{P-sliding}} -e_{_{P\textcolor{red}{P}PF}}^{^{u\textcolor{red}{(u+1)}uv}}\xrightarrow {\text{F-collapse}}0$,\\
 & & if $\ r>u+1$, then \\
 & & $e_{_{PF\textcolor{red}{F}F}}^{^{r(u+1)\textcolor{red}{u}v}} \xrightarrow {\text{F-collapse}}0$,\\
& & if $\ r<u+1$, then \\
 & & $e_{_{P\textcolor{red}{F}FF}}^{^{r\textcolor{red}{(u+1)}uv}} \xrightarrow {\text{F-collapse}}0$.\\ 
& & \underline{Case $2:s=u\neq v$}\\
& & W.L.O.G, let us consider s>v \\   
& & \underline{Subcase $1:s>(v+1)$} \\  & & If $ r\neq s,s+1$, then $e_{_{ P\textcolor{red}{F}FF}}^{^{r\textcolor{red}{s}sv}} \xrightarrow {\text{F-collapse}} 0$, \\& & otherwise $e_{_{ PFF\textcolor{red}{F}}}^{^{rss\textcolor{red}{v}}} \xrightarrow {\text{F-collapse}} 0$.\\ & & \underline{Subcase $2:s=(v+1)$}\\ & & If $ r= v+1$, then \\& & $e_{_{ \textcolor{blue}{P}\textcolor{brown}{F}FF}}^{^{\textcolor{blue}{(v+1)}\textcolor{brown}{(v+1)}(v+1)v}} \xrightarrow {\text{P-sliding}} -e_{_{PFF\textcolor{red}{F}}}^{^{(v+2)(v+1)(v+1)\textcolor{red}{v}}} \xrightarrow {\text{F-collapse}}0$,\\ & & if $\ r>(v+1)$, then $e_{_{PFF\textcolor{red}{F}}}^{^{r(v+1)(v+1)\textcolor{red}{v}}}\xrightarrow {\text{F-collapse}}0$,\\ & & and if $\ r<(v+1)$, then $e_{_{P\textcolor{red}{F}FF}}^{^{r\textcolor{red}{(v+1)}(v+1)v}}\xrightarrow {\text{F-collapse}}0
$\\ 
\hline
$e_{_{PFFF}}^{rsss}$ & $k(k-2)$ & $e_{_{ P\textcolor{red}{F}FF}}^{^{r\textcolor{red}{s}ss}} \xrightarrow {\text{F-collapse}} 0$ \\  \text{for}$\ r\neq s,s+1$ & & \\
\hline
$e_{_{PFFF}}^{rrrr}+ e_{_{PFFF}}^{(r+1)rrr}$ & $2k-1$ & $e_{_{ \textcolor{blue}{P} \textcolor{brown}{F}FF} }^{^{\textcolor{blue}{(r+1)}\textcolor{brown}{r}rr }} \xrightarrow {\text{P-sliding}} -e_{_{ PFFF} }^{rrrr} $ \\ & &\\
\hline
\end{tabular}
\caption{Arguments for the tensors in the PFFF sector orthogonal to the array of 4 point contour correlators \label{tab:4ptPFFF-ortho} } 
\end{table}

These orthogonal tensors fix $\matMt{k,4}_{PFFF}$ to be of the following form:

\be
\begin{split}
&\matMt{k,4}_{PFFF}\\&=\alpha^{(4)}_2 \sum_{r=1}^k  \Big(\ebP{r+1}(\omega_1)-\ebP{r}(\omega_1)\Big)\otimes \ebP{r}(\omega_2)\otimes \ebP{r}(\omega_3)\otimes\ebF{r}(\omega_4) ~.
\end{split}
\label{app4pt:15}
\ee

The coefficient $ \alpha_2^{(4)}$ is given by the contraction with the tensor $e_{PFFF}^{(r+1) rrr}$ , an example of which is the case $r=1$. Then this coefficient is given by 

\begin{equation}
\begin{split}
 \alpha_2^{(4)} &=\matMt{k,4}(\text{4-Pt})\cdot e_{PFFF}^{\ 2111}= \rho[1234]\ .
\end{split}
 \label{app4pt:16}
\end{equation}

\paragraph{Orthogonal tensors in the PFPF sector:}
\begin{equation}
\begin{split}
   k^3 (k-2)\ \text{orthogonal tensors} :& \ e_{_{PFPF}}^{rsuv}\ \text{for}\ r\neq s,s+1 \ ,\\
   k^2 (k-2)\ \text{orthogonal tensors} :& \ e_{_{PFPF}}^{(r+1)ruv}\ \text{for}\ v\neq u,u-1 \ ,\\
    k^2 (k-2)\ \text{orthogonal tensors} :& \ e_{_{PFPF}}^{rruv}\ \text{for}\ v\neq u,u-1 \ ,\\
    k^2\  \text{orthogonal tensors} :& \ e_{_{PFPF}}^{(r+1)ruu}+e_{_{PFPF}}^{rruu} \ ,\\
    k^2\  \text{orthogonal tensors} :& \ e_{_{PFPF}}^{rr(u+1)u}+e_{_{PFPF}}^{rr(u+1)(u+1)} \ ,\\
    k^2\  \text{orthogonal tensors} :& \ e_{_{PFPF}}^{(r+1)r(u+1)u}-e_{_{PFPF}}^{rr(u+1)(u+1)} \ ,\\
     \frac{1}{2}(k^2-k)\  \text{orthogonal tensors} :& \ e_{_{PFPF}}^{(r+1+l)(r+1+l)rr}-e_{_{PFPF}}^{(r+1)(r+1)rr}\ \text{for}\ 1\leq l \leq (k-r) \ ,\\
     \frac{1}{2}(k^2-k)\  \text{orthogonal tensors} :& \ e_{_{PFPF}}^{(r-l)(r-l)(r+1)(r+1)}-e_{_{PFPF}}^{rr(r+1)(r+1)}\ \text{for}\ 1\leq l \leq (r-1) \ ,\\
     k-1\  \text{orthogonal tensors} :& \ e_{_{PFPF}}^{rrrr}-e_{_{PFPF}}^{1111}~.\\ 
      \end{split}
 \label{app4pt:17}
\end{equation}

Total number of orthogonal tensors in this sector $=k^4-1$.

These orthogonal tensors fix $\matMt{k,4}_{PFPF}$ to be of the following form:
\be
\begin{split}
&\matMt{k,4}_{PFPF}\\&=\alpha^{(4)}_3\sum_{r,s=1}^k   \Big(\theta_{r>s}+\theta_{r \leq s}e^{\beta(\omega_3+\omega_4)}\Big)\\& \qquad \qquad \Big (\ebP{r+1}(\omega_1)-\ebP{r}(\omega_1)\Big ) \otimes \ebF{r}(\omega_2) \otimes \ebP{s+1}(\omega_3) \otimes \ebF{s}(\omega_4)\\& -\alpha^{(4)}_3\sum_{r,s=1}^k   \Big(\theta_{r\geq s} +\theta_{r < s}e^{\beta(\omega_3+\omega_4)}\Big)\\&\qquad\qquad  \Big (\ebP{r+1}(\omega_1)-\ebP{r}(\omega_1)\Big ) \otimes \ebF{r}(\omega_2) \otimes \ebP{s}(\omega_3) \otimes \ebF{s}(\omega_4)~.
\end{split}
\label{app4pt:18}
\ee
The coefficient $ \alpha_3^{(4)}$ is given by the contraction with the tensor $e_{PFPF}^{(r+1) r(s+1)s}$ for r>s , an example of which is the case $r=2,s=1$. Then this coefficient is given by 
\begin{equation}
\begin{split}
 \alpha_3^{(4)} &=\matMt{k,4}(\text{4-Pt})\cdot e_{PFPF}^{\ 3221}= \rho[12][34] \ .
\end{split}
 \label{app4pt:19}
\end{equation}

\begin{table}[H]
\centering
\begin{tabular}{|c|c|c|}
\hline
 Orthogonal  & Total no. & Argument  \\   
 tensor	&  of tensors  & \\
\hline
$e_{_{PFPF}}^{rsuv}$ & $k^3 (k-2)$ &  $e_{_{P\textcolor{red}{F}PF}}^{r\textcolor{red}{s}uv} \xrightarrow {\text{F-collapse}}0$ \\  \text{ for} $r \neq s,s+1$ & & \\
\hline
$e_{_{PFPF}}^{(s+1)suv}$ & $k^2 (k-2)$ &  $e_{_{PF\textcolor{red}{P}F}}^{(s+1)s\textcolor{red}{u}v} \xrightarrow {\text{P-collapse}}0$ \\  \text{ for} $v \neq u,u-1$ & & \\
\hline
$e_{_{PFPF}}^{ssuv}$ & $k^2 (k-2)$ &  $e_{_{PF\textcolor{red}{P}F}}^{ss\textcolor{red}{u}v} \xrightarrow {\text{P-collapse}}0$ \\  \text{ for} $v \neq u,u-1$ & & \\
\hline
$e_{_{PFPF}}^{(s+1)suu}+e_{_{PFPF}}^{ssuu}$ & $k^2 $ &  $e_{_{\textcolor{blue}{P}\textcolor{brown}{F}PF}}^{\textcolor{blue}{(s+1)}\textcolor{brown}{s}uu} \xrightarrow {\text{P-sliding}}-e_{_{PFPF}}^{ssuu}$ \\ & & \\
\hline
$e_{_{PFPF}}^{ss(u+1)u}+e_{_{PFPF}}^{ss(u+1)(u+1)}$ & $k^2 $ &  $e_{_{PF\textcolor{brown}{P}\textcolor{blue}{F}}}^{ss\textcolor{brown}{(u+1)}\textcolor{blue}{u}} \xrightarrow {\text{F-sliding}}-e_{_{PFPF}}^{ss(u+1)(u+1)}$ \\ & & \\
\hline
$e_{_{PFPF}}^{(s+1)s(u+1)u}-e_{_{PFPF}}^{ss(u+1)(u+1)}$ & $k^2 $ &  $e_{_{PF\textcolor{brown}{P}\textcolor{blue}{F}}}^{(s+1)s\textcolor{brown}{(u+1)}\textcolor{blue}{u}} \xrightarrow {\text{F-sliding}}-e_{_{\textcolor{blue}{P}\textcolor{brown}{F}PF}}^{\textcolor{blue}{(s+1)}\textcolor{brown}{s}(u+1)(u+1)}$\\ & & $\xrightarrow {\text{P-sliding}} e_{_{PFPF}}^{ss(u+1)(u+1)}$ \\
\hline
$e_{_{PFPF}}^{(s+1+l)(s+1+l)ss}-e_{_{PFPF}}^{(s+1)(s+1)ss}$ & $\frac{1}{2}(k^2-k) $ &  $e_{_{\textcolor{blue}{P}\textcolor{brown}{F}PF}}^{\textcolor{blue}{(s+1)}\textcolor{brown}{(s+1)}ss} \xrightarrow {\text{P-sliding}}-e_{_{\textcolor{brown}{P}\textcolor{blue}{F}PF}}^{\textcolor{brown}{(s+2)}\textcolor{blue}{(s+1)}ss}$ \\ for $1\leq l \leq (k-s)$ & & $\xrightarrow {\text{F-sliding}} e_{_{\textcolor{blue}{P}\textcolor{brown}{F}PF}}^{\textcolor{blue}{(s+2)}\textcolor{brown}{(s+2)}ss} $ \\
& & $\xrightarrow {\text{P-sliding}}-e_{_{\textcolor{brown}{P}\textcolor{blue}{F}PF}}^{\textcolor{brown}{(s+3)}\textcolor{blue}{(s+2)}ss}$\\
& & $\xrightarrow {\text{F-sliding}} e_{_{\textcolor{blue}{P}\textcolor{brown}{F}PF}}^{\textcolor{blue}{(s+3)}\textcolor{brown}{(s+3)}ss} $ \\ & & $\xrightarrow  {\text{P-sliding}} ...  \xrightarrow {\text{F-sliding}} e_{_{PFPF}}^{(s+1+l)(s+1+l)ss}$\\
\hline
$\Big (e_{_{PFPF}}^{(s-l)(s-l)(s+1)(s+1)}$ & $\frac{1}{2}(k^2-k) $ &  $e_{_{\textcolor{brown}{P}\textcolor{blue}{F}PF}}^{\textcolor{brown}{s}\textcolor{blue}{s}(s+1)(s+1)} \xrightarrow {\text{F-sliding}}-e_{_{\textcolor{blue}{P}\textcolor{brown}{F}PF}}^{\textcolor{blue}{s}\textcolor{brown}{(s-1)}(s+1)(s+1)} $ \\ $-e_{_{PFPF}}^{ss(s+1)(s+1)}\Big )$ & & $\xrightarrow {\text{P-sliding}} e_{_{\textcolor{brown}{P}\textcolor{blue}{F}PF}}^{\textcolor{brown}{(s-1)}\textcolor{blue}{(s-1)}(s+1)(s+1)}$ \\ for $1\leq l \leq (s-1)$ & & $\xrightarrow {\text{F-sliding}}-e_{_{\textcolor{blue}{P}\textcolor{brown}{F}PF}}^{\textcolor{brown}{(s-1)}\textcolor{blue}{(s-2)}(s+1)(s+1)} $\\ & & $\xrightarrow {\text{P-sliding}} e_{_{\textcolor{brown}{P}\textcolor{blue}{F}PF}}^{\textcolor{brown}{(s-2)}\textcolor{blue}{(s-2)}(s+1)(s+1)}$\\ & & $\xrightarrow  {\text{F-sliding}} ...  \xrightarrow {\text{P-sliding}} e_{_{PFPF}}^{(s-l)(s-l)(s+1)(s+1)}$\\
\hline
$e_{_{PFPF}}^{ssss}-e_{_{PFPF}}^{1111}$ & $k-1$ &  $e_{_{PFPF}}^{ssss}\xrightarrow {\text{C-shift}}e_{_{PFPF}}^{1111}$ \\ & & \\
\hline
\end{tabular}
\caption{Arguments for the tensors in the PFPF sector orthogonal to the array of 4 point contour correlators \label{tab:4ptPFPF-ortho}} 
\end{table}

\paragraph{Orthogonal tensors in the PPFF sector:}
\begin{equation}
\begin{split}
     k(k-3)(k^2-8)\  \text{orthogonal tensors} :& \ e_{_{PPFF}}^{rsuv}\ \text{for}\ |u-v|>1 \  \text{and} \\& (r,s)\notin \{(u,v),(u,v+1),(u+1,v),(u+1,v+1),\\& \qquad\qquad (v,u),(v+1,u),(v,u+1),(v+1,u+1)\} \ ,\\
      k(k-3)\  \text{orthogonal tensors} :& \ e_{_{PPFF}}^{(u+1)vuv}+e_{_{PPFF}}^{uvuv}\ \text{for}\ |u-v|>1 \ ,\\
      k(k-3)\  \text{orthogonal tensors} :& \ e_{_{PPFF}}^{u(v+1)uv}+e_{_{PPFF}}^{uvuv}\ \text{for}\ |u-v|>1 \ ,\\
       k(k-3)\  \text{orthogonal tensors} :& \ e_{_{PPFF}}^{(u+1)(v+1)uv}-e_{_{PPFF}}^{uvuv}\ \text{for}\ |u-v|>1 \ ,\\
       k(k-3)\  \text{orthogonal tensors} :& \ e_{_{PPFF}}^{v(u+1)uv}+e_{_{PPFF}}^{vuuv}\ \text{for}\ |u-v|>1 \ ,\\
      k(k-3)\  \text{orthogonal tensors} :& \ e_{_{PPFF}}^{(v+1)uuv}+e_{_{PPFF}}^{vuuv}\ \text{for}\ |u-v|>1 \ ,\\
       k(k-3)\  \text{orthogonal tensors} :& \ e_{_{PPFF}}^{(v+1)(u+1)uv}-e_{_{PPFF}}^{vuuv}\ \text{for}\ |u-v|>1~.\\
\end{split}
\label{app4pt:20}
\end{equation}

\begin{equation}
\begin{split}
       k(k^2-7)\  \text{orthogonal tensors} :& \ e_{_{PPFF}}^{rsu(u+1)}\\& \text{ for} (r,s)\notin \{(u,u+1),(u,u+2), (u+1,u+1),\\ & \qquad\qquad\qquad (u+1,u+2),(u+1,u),(u+2,u),\\&  \qquad\qquad\qquad (u+2,u+1)\} \ , \\
       k(k^2-7)\  \text{orthogonal tensors} :& \ e_{_{PPFF}}^{rs(v+1)v}\\& \text{ for} (r,s)\notin \{(v+1,v),(v+2,v), (v+1,v+1),\\ & \qquad\qquad\qquad (v+2,v+1),(v,v+1),(v,v+2),\\&  \qquad\qquad\qquad (v+1,v+2)\} \ , \\
       k\  \text{orthogonal tensors} :& \  e_{_{PPFF}}^{u(u+2)u(u+1)}+e_{_{PPFF}}^{u(u+1)u(u+1)} \ ,\\
        k\  \text{orthogonal tensors} :& \  e_{_{PPFF}}^{(u+1)(u+2)u(u+1)}-e_{_{PPFF}}^{u(u+1)u(u+1)} \ ,\\
        k\  \text{orthogonal tensors} :& \  e_{_{PPFF}}^{(u+2)uu(u+1)}+e_{_{PPFF}}^{(u+1)uu(u+1)} \ ,\\
        k\  \text{orthogonal tensors} :& \  e_{_{PPFF}}^{(u+2)(u+1)u(u+1)}-e_{_{PPFF}}^{(u+1)uu(u+1)} \ ,\\
          k\  \text{orthogonal tensors} :& \  e_{_{PPFF}}^{v(v+2)(v+1)v}+e_{_{PPFF}}^{v(v+1)(v+1)v} \ ,\\
        k\  \text{orthogonal tensors} :& \  e_{_{PPFF}}^{(v+1)(v+2)(v+1)v}-e_{_{PPFF}}^{v(v+1)(v+1)v} \ ,\\
        k\  \text{orthogonal tensors} :& \  e_{_{PPFF}}^{(v+2)v(v+1)v}+e_{_{PPFF}}^{(v+1)v(v+1)v} \ ,\\
        k\  \text{orthogonal tensors} :& \  e_{_{PPFF}}^{(v+2)(v+1)(v+1)v}-e_{_{PPFF}}^{(v+1)v(v+1)v} \ ,\\
	k-1\  \text{orthogonal tensors} :& \  e_{_{PPFF}}^{(u+1)(u+1)u(u+1)}-e_{_{PPFF}}^{2212} \ ,\\
	k-1\  \text{orthogonal tensors} :& \  e_{_{PPFF}}^{(v+1)(v+1)(v+1)v}-e_{_{PPFF}}^{2221}~.\\
\end{split}
\label{app4pt:21}
\end{equation}

\begin{equation}
\begin{split}
       \frac{1}{2}k(k-1)-1\  \text{orthogonal tensors} :& \ e_{_{PPFF}}^{uvuv}-e_{_{PPFF}}^{2121}  \text{for}\ u>v\ ,\\
        \frac{1}{2}k(k-1)-1\  \text{orthogonal tensors} :& \ e_{_{PPFF}}^{uvuv}-e_{_{PPFF}}^{1212}  \text{for}\ u<v\ ,\\
        \frac{1}{2}k(k-1)-1\  \text{orthogonal tensors} :& \ e_{_{PPFF}}^{vuuv}-e_{_{PPFF}}^{1221}  \text{for}\ u>v \ ,\\
        \frac{1}{2}k(k-1)-1\  \text{orthogonal tensors} :& \ e_{_{PPFF}}^{vuuv}-e_{_{PPFF}}^{2112}  \text{for}\ u<v\ ,\\
        1\  \text{orthogonal tensor} :& \ e_{_{PPFF}}^{1212}-e^{\beta(\omega_2+\omega_4)}e_{_{PPFF}}^{2121}\ ,\\
        1\  \text{orthogonal tensor} :& \ e_{_{PPFF}}^{2112}-e^{\beta(\omega_1+\omega_4)}e_{_{PPFF}}^{1221}\ .\\
      \end{split}
\label{app4pt:22}
\end{equation}

\begin{equation}
\begin{split}
k(k^2-4)\  \text{orthogonal tensors} :& \ e_{_{PPFF}}^{rsuu} \text{for} (r,s)\notin \{(u,u),(u+1,u),(u,u+1),(u+1,u+1)\}\ , \\
k-1\  \text{orthogonal tensors} :&\ e_{_{PPFF}}^{uuuu}-e_{_{PPFF}}^{1111}\ ,\\
k-1\  \text{orthogonal tensors} :&\ e_{_{PPFF}}^{(u+1)(u+1)uu}-e_{_{PPFF}}^{2211}\ ,\\
k-1\  \text{orthogonal tensors} :&\ e_{_{PPFF}}^{(u+1)uuu}-e_{_{PPFF}}^{2111}\ ,\\
k-1\  \text{orthogonal tensors} :&\ e_{_{PPFF}}^{u(u+1)uu}-e_{_{PPFF}}^{1211}\ ,\\ 
1\  \text{orthogonal tensor} :&\ e_{_{PPFF}}^{2111}-e_{_{PPFF}}^{3121}-e_{_{PPFF}}^{3112}\ ,\\ 
1\  \text{orthogonal tensor} :&\ e_{_{PPFF}}^{1211}-e_{_{PPFF}}^{1321}-e_{_{PPFF}}^{1312}\ ,\\        
1\  \text{orthogonal tensor} :&\ e_{_{PPFF}}^{2221}-e_{_{PPFF}}^{2331}-e_{_{PPFF}}^{3231}\ ,\\        
1\  \text{orthogonal tensor} :&\ e_{_{PPFF}}^{2212}-e_{_{PPFF}}^{2313}-e_{_{PPFF}}^{3213}\ ,\\        
1\  \text{orthogonal tensor} :&\ e_{_{PPFF}}^{1111}+e_{_{PPFF}}^{2111}+e_{_{PPFF}}^{1211}+e_{_{PPFF}}^{2211}~.\\                    
\end{split}
\label{app4pt:23}
\end{equation}

Total number of orthogonal tensors in this sector $=k^4-3$.

\begin{table}[H]
\centering
\begin{tabular}{|c|c|c|}
\hline
 Orthogonal  & Total no. & Argument  \\   
 tensor	&  of tensors  & \\
\hline
$e_{_{PPFF}}^{rsuv}$ & $k (k-3)$ & \text{If} $r\neq u,u+1$ \text{ and } $s\neq u,u+1$ , \\  
\text{ for} $|u-v|>1$ \text{and} & $(k^2-8)$ &  \text{ then} $e_{_{PP\textcolor{red}{F}F}}^{rs\textcolor{red}{u}v}\xrightarrow{\text{F-collapse}} 0$ \\ 
$(r,s)\notin \{(u,v),(u,v+1),$ & &  \text{If} $r\neq v,v+1$ \text{ and } $s\neq v,v+1$\\
$(u+1,v),(u+1,v+1),$  & &  \text{ then} $e_{_{PPF\textcolor{red}{F}}}^{rsu\textcolor{red}{v}}\xrightarrow{\text{F-collapse}} 0$\\
$(v,u),(v+1,u),$ & &\\
$(v,u+1),(v+1,u+1)\}$ & &\\
\hline
$e_{_{PPFF}}^{(u+1)vuv}+e_{_{PPFF}}^{uvuv}$ & $k(k-3) $ & $e_{_{\textcolor{blue}{P}P\textcolor{brown}{F}F}}^{\textcolor{blue}{(u+1)}v\textcolor{brown}{u}v}\xrightarrow{\text{P-sliding}}-e_{_{PPFF}}^{uvuv}$\\ 
\text{for} $|u-v|>1$  &  &\\  
\hline
$e_{_{PPFF}}^{u(v+1)uv}+e_{_{PPFF}}^{uvuv}$ & $k(k-3) $ & $e_{_{P\textcolor{blue}{P}F\textcolor{brown}{F}}}^{u\textcolor{blue}{(v+1)}u\textcolor{brown}{v}}\xrightarrow{\text{P-sliding}}-e_{_{PPFF}}^{uvuv}$\\ 
\text{for} $|u-v|>1$  &  &\\  
\hline
$e_{_{PPFF}}^{(u+1)(v+1)uv}-e_{_{PPFF}}^{uvuv}$ & $k(k-3) $ & $e_{_{P\textcolor{blue}{P}F\textcolor{brown}{F}}}^{(u+1)\textcolor{blue}{(v+1)}u\textcolor{brown}{v}}\xrightarrow{\text{P-sliding}}-e_{_{\textcolor{blue}{P}P\textcolor{brown}{F}F}}^{\textcolor{blue}{(u+1)}v\textcolor{brown}{u}v}$\\ 
\text{for} $|u-v|>1$  &  & $\xrightarrow{\text{P-sliding}}e_{_{PPFF}}^{uvuv}$\\  
\hline
$e_{_{PPFF}}^{v(u+1)uv}+e_{_{PPFF}}^{vuuv}$ & $k(k-3) $ & $e_{_{P\textcolor{blue}{P}\textcolor{brown}{F}F}}^{v\textcolor{blue}{(u+1)}\textcolor{brown}{u}v}\xrightarrow{\text{P-sliding}}-e_{_{PPFF}}^{vuuv}$\\ 
\text{for} $|u-v|>1$  &  &\\  
\hline
$e_{_{PPFF}}^{(v+1)uuv}+e_{_{PPFF}}^{vuuv}$ & $k(k-3) $ & $e_{_{\textcolor{blue}{P}PF\textcolor{brown}{F}}}^{\textcolor{blue}{(v+1)}uu\textcolor{brown}{v}}\xrightarrow{\text{P-sliding}}-e_{_{PPFF}}^{vuuv}$\\ 
\text{for} $|u-v|>1$  &  &\\  
\hline
$e_{_{PPFF}}^{(v+1)(u+1)uv}-e_{_{PPFF}}^{uvuv}$ & $k(k-3) $ & $e_{_{P\textcolor{blue}{P}F\textcolor{brown}{F}}}^{\textcolor{blue}{(v+1)}(u+1)u\textcolor{brown}{v}}\xrightarrow{\text{P-sliding}}-e_{_{P\textcolor{blue}{P}\textcolor{brown}{F}F}}^{v\textcolor{blue}{(u+1)}\textcolor{brown}{u}v}$\\ 
\text{for} $|u-v|>1$  &  & $\xrightarrow{\text{P-sliding}}e_{_{PPFF}}^{vuuv}$\\  
\hline
\end{tabular}
\caption{Arguments for the tensors in the PPFF sector orthogonal to the array of 4 point contour correlators \label{tab:4ptPPFF1-ortho} } 
\end{table}

\begin{table}[H]
\centering
\begin{tabular}{|c|c|c|}
\hline
 Orthogonal  & Total no. & Argument  \\   
 tensor	&  of tensors  & \\
\hline
$e_{_{PPFF}}^{rsu(u+1)}$ & $k(k^2-7) $ & \text{If} $r\neq u,u+1$ \text{ and } $s\neq u,u+1$ , \\ 
\text{ for} $(r,s)\notin \{(u,u+1),(u,u+2),$ &  &\text{ then} $e_{_{PP\textcolor{red}{F}F}}^{rs\textcolor{red}{u}(u+1)}\xrightarrow{\text{F-collapse}} 0$.\\  
$(u+1,u+1),(u+1,u+2),$  &  & \text{If} $r\neq u+1,u+2$ \\
$(u+1,u),(u+2,u),(u+2,u+1)\}$ &  &\text{ and } $s\neq u+1,u+2$, \\  
& & \text{ then} $e_{_{PPF\textcolor{red}{F}}}^{rsu\textcolor{red}{(u+1)}}\xrightarrow{\text{F-collapse}} 0$.\\
\hline
$e_{_{PPFF}}^{rs(v+1)v}$ & $k(k^2-7) $ & \text{If} $r\neq v,v+1$ \text{ and } $s\neq v,v+1$ , \\ 
\text{ for} $(r,s)\notin \{(v+1,v),(v+2,v),$ &  &\text{ then} $e_{_{PPF\textcolor{red}{F}}}^{rs(v+1)\textcolor{red}{v}}\xrightarrow{\text{F-collapse}} 0$.\\  
$(v+1,v+1),(v+2,v+1),$  &  & \text{If} $r\neq v+1,v+2$ \\
$(v,v+1),(v,v+2),(v+1,v+2)\}$ &  &\text{ and } $s\neq v+1,v+2$ , \\   
& & \text{ then} $e_{_{PP\textcolor{red}{F}F}}^{rs\textcolor{red}{(v+1)}v}\xrightarrow{\text{F-collapse}} 0$.\\   
\hline
$e_{_{PPFF}}^{u(u+2)u(u+1)}+e_{_{PPFF}}^{u(u+1)u(u+1)}$ & $k $ & $e_{_{P\textcolor{blue}{P}F\textcolor{brown}{F}}}^{u\textcolor{blue}{(u+2)}u\textcolor{brown}{(u+1)}}\xrightarrow{\text{P-sliding}}-e_{_{PPFF}}^{u(u+1)u(u+1)}$\\ 
\hline
$e_{_{PPFF}}^{(u+1)(u+2)u(u+1)}-e_{_{PPFF}}^{u(u+1)u(u+1)}$ & $k $ & $e_{_{\textcolor{blue}{P}P\textcolor{brown}{F}F}}^{\textcolor{blue}{(u+1)}(u+2)\textcolor{brown}{u}(u+1)}\xrightarrow{\text{P-sliding}}-e_{_{P\textcolor{blue}{P}F\textcolor{brown}{F}}}^{u\textcolor{blue}{(u+2)}u\textcolor{brown}{(u+1)}}$\\ 
& & $\xrightarrow{\text{P-sliding}} e_{_{PPFF}}^{u(u+1)u(u+1)}$\\
\hline
$e_{_{PPFF}}^{(u+2)uu(u+1)}+e_{_{PPFF}}^{(u+1)uu(u+1)}$ & $k $ & $e_{_{\textcolor{blue}{P}PF\textcolor{brown}{F}}}^{\textcolor{blue}{(u+2)}uu\textcolor{brown}{(u+1)}}\xrightarrow{\text{P-sliding}}-e_{_{PPFF}}^{(u+1)uu(u+1)}$\\ 
\hline
$e_{_{PPFF}}^{(u+2)(u+1)u(u+1)}-e_{_{PPFF}}^{(u+1)uu(u+1)}$ & $k $ & $e_{_{P\textcolor{blue}{P}\textcolor{brown}{F}F}}^{(u+2)\textcolor{blue}{(u+1)}\textcolor{brown}{u}(u+1)}\xrightarrow{\text{P-sliding}}-e_{_{P\textcolor{blue}{P}FF}}^{\textcolor{blue}{(u+2)}uu\textcolor{brown}{(u+1)}}$\\ 
& & $\xrightarrow{\text{P-sliding}} e_{_{PPFF}}^{(u+1)uu(u+1)}$\\
\hline
$e_{_{PPFF}}^{v(v+2)(v+1)v}+e_{_{PPFF}}^{v(v+1)(v+1)v}$ & $k $ & $e_{_{P\textcolor{blue}{P}\textcolor{brown}{F}F}}^{v\textcolor{blue}{(v+2)}\textcolor{brown}{(v+1)}v}\xrightarrow{\text{P-sliding}}-e_{_{PPFF}}^{v(v+1)(v+1)v}$\\ 
\hline
$e_{_{PPFF}}^{(v+1)(v+2)(v+1)v}-e_{_{PPFF}}^{v(v+1)(v+1)v}$ & $k $ & $e_{_{P\textcolor{blue}{P}\textcolor{brown}{F}F}}^{\textcolor{blue}{(v+1)}(v+2)(v+1)\textcolor{brown}{v}}\xrightarrow{\text{P-sliding}}-e_{_{P\textcolor{blue}{P}\textcolor{brown}{F}F}}^{v\textcolor{blue}{(v+2)}\textcolor{brown}{(v+1)}v}$\\ 
& & $\xrightarrow{\text{P-sliding}} e_{_{PPFF}}^{v(v+1)(v+1)v}$\\
\hline
$e_{_{PPFF}}^{(v+2)v(v+1)v}+e_{_{PPFF}}^{(v+1)v(v+1)v}$ & $k $ & $e_{_{\textcolor{blue}{P}P\textcolor{brown}{F}F}}^{\textcolor{blue}{(v+2)}v\textcolor{brown}{(v+1)}v}\xrightarrow{\text{P-sliding}}-e_{_{PPFF}}^{(v+1)v(v+1)v}$\\ 
\hline
$e_{_{PPFF}}^{(v+2)(v+1)(v+1)v}-e_{_{PPFF}}^{(v+1)v(v+1)v}$ & $k $ & $e_{_{P\textcolor{blue}{P}F\textcolor{brown}{F}}}^{(v+2)\textcolor{blue}{(v+1)}(v+1)\textcolor{brown}{v}}\xrightarrow{\text{P-sliding}}-e_{_{\textcolor{blue}{P}P\textcolor{brown}{F}F}}^{\textcolor{blue}{(v+2)}v\textcolor{brown}{(v+1)}v}$\\ 
& & $\xrightarrow{\text{P-sliding}} e_{_{PPFF}}^{(v+1)v(v+1)v}$\\
\hline
$e_{_{PPFF}}^{(u+1)(u+1)u(u+1)}-e_{_{PPFF}}^{2212}$ & $k-1 $ & $e_{_{PPFF}}^{2212}\xrightarrow{\text{C-shift}}e_{_{PPFF}}^{(u+1)(u+1)u(u+1)}$\\
\hline
$e_{_{PPFF}}^{(v+1)(v+1)(v+1)v}-e_{_{PPFF}}^{2221}$ & $k-1 $ & $e_{_{PPFF}}^{2221}\xrightarrow{\text{C-shift}}e_{_{PPFF}}^{(v+1)(v+1)(v+1)v}$\\
\hline
\end{tabular}
\caption{Arguments for the tensors in the PPFF sector orthogonal to the array of 4 point contour correlators \label{tab:4ptPPFF2-ortho} } 
\end{table}

\begin{table}[H]
\centering
\begin{tabular}{|c|c|c|}
\hline
 Orthogonal  & Total no. & Argument  \\   
 tensor	&  of tensors  & \\
\hline
$e_{_{PPFF}}^{uvuv}-e_{_{PPFF}}^{2121}$ & $\frac{1}{2}k(k-1)-1$ & \underline{Case 1:\ u=v+1}\\
for u>v &  & $e_{_{PPFF}}^{2121}\xrightarrow{\text{C-shift}} e_{_{PPFF}}^{(v+1)v(v+1)v}$ \\
& & \underline{Case 2:\ u=(v+m) where m>1}\\
& & $e_{_{\textcolor{blue}{P}P\textcolor{brown}{F}F}}^{\textcolor{blue}{2}1\textcolor{brown}{2}1}\xrightarrow{\text{P-sliding}} e_{_{\textcolor{brown}{P}P\textcolor{blue}{F}F}}^{\textcolor{brown}{3}1\textcolor{blue}{2}1}\xrightarrow{\text{F-sliding}} e_{_{\textcolor{blue}{P}P\textcolor{brown}{F}F}}^{\textcolor{blue}{3}1\textcolor{brown}{3}1}$\\
& & $\xrightarrow{\text{P-sliding}} \cdots \xrightarrow{\text{F-sliding}}e_{_{PPFF}}^{(1+m)1(1+m)1}$\\
& & $\xrightarrow{\text{C-shift}}e_{_{PPFF}}^{(v+m)v(v+m)v}$\\
\hline
$e_{_{PPFF}}^{uvuv}-e_{_{PPFF}}^{1212}$ & $\frac{1}{2}k(k-1)-1$ & \underline{Case 1:\ v=u+1}\\
for u<v &  & $e_{_{PPFF}}^{1212}\xrightarrow{\text{C-shift}} e_{_{PPFF}}^{u(u+1)u(u+1)}$ \\
& & \underline{Case 2:\ v=(u+m) where m>1}\\
& & $e_{_{P\textcolor{blue}{P}F\textcolor{brown}{F}}}^{1\textcolor{blue}{2}1\textcolor{brown}{2}}\xrightarrow{\text{P-sliding}} e_{_{P\textcolor{brown}{P}F\textcolor{blue}F}}^{1\textcolor{brown}{3}1\textcolor{blue}{2}}\xrightarrow{\text{F-sliding}} e_{_{P\textcolor{blue}{P}F\textcolor{brown}{F}}}^{1\textcolor{blue}{3}1\textcolor{brown}{3}}$\\
& & $\xrightarrow{\text{P-sliding}} \cdots \xrightarrow{\text{F-sliding}}e_{_{PPFF}}^{1(1+m)1(1+m)}$\\
& & $\xrightarrow{\text{C-shift}}e_{_{PPFF}}^{u(u+m)u(u+m)}$\\
\hline
$e_{_{PPFF}}^{vuuv}-e_{_{PPFF}}^{1221}$ & $\frac{1}{2}k(k-1)-1$ & \underline{Case 1:\ u=v+1}\\
for u>v &  & $e_{_{PPFF}}^{1221}\xrightarrow{\text{C-shift}} e_{_{PPFF}}^{v(v+1)(v+1)v}$ \\
& & \underline{Case 2:\  u=(v+m) where m>1}\\
& & $e_{_{P\textcolor{blue}{P}\textcolor{brown}{F}F}}^{1\textcolor{blue}{2}\textcolor{brown}{2}1}\xrightarrow{\text{P-sliding}} e_{_{P\textcolor{brown}{P}\textcolor{blue}{F}F}}^{1\textcolor{brown}{3}\textcolor{blue}{2}1}\xrightarrow{\text{F-sliding}} e_{_{P\textcolor{blue}{P}\textcolor{brown}{F}F}}^{1\textcolor{blue}{3}\textcolor{brown}{3}1}$\\
& & $\xrightarrow{\text{P-sliding}} \cdots \xrightarrow{\text{F-sliding}}e_{_{PPFF}}^{1(1+m)(1+m)1}$\\
& & $\xrightarrow{\text{C-shift}}e_{_{PPFF}}^{v(v+m)(v+m)v}$\\
\hline
$e_{_{PPFF}}^{vuuv}-e_{_{PPFF}}^{2112}$ & $\frac{1}{2}k(k-1)-1$ & \underline{Case 1:\  v=u+1}\\
for u<v &  & $e_{_{PPFF}}^{2112}\xrightarrow{\text{C-shift}} e_{_{PPFF}}^{(u+1)uu(u+1)}$ \\
& & \underline{Case 2:\ v=(u+m) where m>1}\\
& & $e_{_{\textcolor{blue}{P}PF\textcolor{brown}{F}}}^{\textcolor{blue}{2}11\textcolor{brown}{2}}\xrightarrow{\text{P-sliding}} e_{_{\textcolor{brown}{P}PF\textcolor{blue}F}}^{\textcolor{brown}{3}11\textcolor{blue}{2}}\xrightarrow{\text{F-sliding}} e_{_{\textcolor{blue}{P}PF\textcolor{brown}{F}}}^{\textcolor{blue}{3}11\textcolor{brown}{3}}$\\
& & $\xrightarrow{\text{P-sliding}} \cdots \xrightarrow{\text{F-sliding}}e_{_{PPFF}}^{(1+m)11(1+m)}$\\
& & $\xrightarrow{\text{C-shift}}e_{_{PPFF}}^{(u+m)uu(u+m)}$\\
\hline
$e_{_{PPFF}}^{1212}-e^{\beta(\omega_2+\omega_4)}e_{_{PPFF}}^{2121}$ & 1 & $e_{_{PPFF}}^{1212}\xrightarrow{\text{C-shift}}e_{_{PPFF}}^{k(k+1)k(k+1)}\xrightarrow{}e^{\beta(\omega_2+\omega_4)}e_{_{PPFF}}^{k1k1}$\\
& & \text{We have already shown that}  $e_{_{PPFF}}^{k1k1}\xrightarrow{}e_{_{PPFF}}^{2121}$.\\
& & \text{Therefore, }$e_{_{PPFF}}^{1212}\xrightarrow{}e^{\beta(\omega_2+\omega_4)}e_{_{PPFF}}^{2121}$.\\
\hline
$e_{_{PPFF}}^{2112}-e^{\beta(\omega_1+\omega_4)}e_{_{PPFF}}^{1221}$ & 1 & $e_{_{PPFF}}^{2112}\xrightarrow{\text{C-shift}}e_{_{PPFF}}^{(k+1)kk(k+1)}\xrightarrow{}e^{\beta(\omega_1+\omega_4)}e_{_{PPFF}}^{1kk1}$\\
& & \text{We have already shown that}  $e_{_{PPFF}}^{1kk1}\xrightarrow{}e_{_{PPFF}}^{1221}$.\\
& & \text{Therefore, }$e_{_{PPFF}}^{2112}\xrightarrow{}e^{\beta(\omega_2+\omega_4)}e_{_{PPFF}}^{1221}$.\\
\hline
\end{tabular}
\caption{Arguments for the tensors in the PPFF sector orthogonal to the array of 4 point contour correlators \label{tab:4ptPPFF3-ortho}} 
\end{table}

\begin{table}[H]\
\centering
\begin{tabular}{|c|c|c|}
\hline
 Orthogonal  & Total no. & Argument  \\   
 tensor	&  of tensors  & \\
\hline
$e_{_{PPFF}}^{rsuu}$ & $k(k^2-4)$ & \text{If} $r\neq u,u+1$ \text{then}, \\
\text{for} $(r,s)\notin \{(u,u),(u+1,u),$ &  & $e_{_{\textcolor{red}{P}PFF}}^{\textcolor{red}{r}suu}\xrightarrow{\text{P-collapse}}0$. \\
$(u,u+1),(u+1,u+1)\}$ & & \text{If} $s\neq u,u+1$ \text{then}, \\
& & $e_{_{P\textcolor{red}{P}FF}}^{r\textcolor{red}{s}uu}\xrightarrow{\text{P-collapse}}0$. \\
\hline
$e_{_{PPFF}}^{uuuu}-e_{_{PPFF}}^{1111}$ & $k-1$ & $e_{_{PPFF}}^{1111}\xrightarrow{\text{C-shift}}e_{_{PPFF}}^{uuuu}$\\
\hline
$e_{_{PPFF}}^{(u+1)(u+1)uu}-e_{_{PPFF}}^{2211}$ & $k-1$ & $e_{_{PPFF}}^{2211}\xrightarrow{\text{C-shift}}e_{_{PPFF}}^{(u+1)(u+1)uu}$\\
\hline
$e_{_{PPFF}}^{(u+1)uuu}-e_{_{PPFF}}^{2111}$ & $k-1$ & $e_{_{PPFF}}^{2111}\xrightarrow{\text{C-shift}}e_{_{PPFF}}^{(u+1)uuu}$\\
\hline
$e_{_{PPFF}}^{u(u+1)uu}-e_{_{PPFF}}^{1211}$ & $k-1$ & $e_{_{PPFF}}^{1211}\xrightarrow{\text{C-shift}}e_{_{PPFF}}^{u(u+1)uu}$\\
\hline
$e_{_{PPFF}}^{2111}-e_{_{PPFF}}^{3121}-e_{_{PPFF}}^{3112}$ & $1$ & $e_{_{PP\textcolor{green}{FF}}}^{21\textcolor{green}{11}}\xrightarrow{\text{F-fragmentation}}e_{_{\textcolor{red}{P}PFF}}^{\textcolor{red}{3}111}+e_{_{PPFF}}^{3121}$\\
& & $\qquad \qquad \qquad \qquad+e_{_{PPFF}}^{3112}+e_{_{P\textcolor{red}{P}FF}}^{3\textcolor{red}{1}22}$\\
& & $\xrightarrow{\text{P-collapse}}e_{_{PPFF}}^{3121}+e_{_{PPFF}}^{3112}$\\
\hline
$e_{_{PPFF}}^{1211}-e_{_{PPFF}}^{1321}-e_{_{PPFF}}^{1312}$ & $1$ & $e_{_{PP\textcolor{green}{FF}}}^{12\textcolor{green}{11}}\xrightarrow{\text{F-fragmentation}}e_{_{P\textcolor{red}{P}FF}}^{1\textcolor{red}{3}11}+e_{_{PPFF}}^{1321}$\\
& & $\qquad \qquad \qquad \qquad+e_{_{PPFF}}^{1312}+e_{_{\textcolor{red}{P}PFF}}^{\textcolor{red}{1}322}$\\
& & $\xrightarrow{\text{P-collapse}}e_{_{PPFF}}^{1321}+e_{_{PPFF}}^{1312}$\\
\hline
$e_{_{PPFF}}^{2221}-e_{_{PPFF}}^{2331}-e_{_{PPFF}}^{3231}$ & $1$ & $e_{_{\textcolor{green}{PP}FF}}^{\textcolor{green}{22}21}\xrightarrow{\text{P-fragmentation}}e_{_{PP\textcolor{red}{F}F}}^{22\textcolor{red}{3}1}+e_{_{PPFF}}^{2331}$\\
& & $\qquad \qquad \qquad \qquad+e_{_{PPFF}}^{3231}+e_{_{PPF\textcolor{red}{F}}}^{333\textcolor{red}{1}}$\\
& & $\xrightarrow{\text{F-collapse}}e_{_{PPFF}}^{2331}+e_{_{PPFF}}^{3231}$\\
\hline
$e_{_{PPFF}}^{2212}-e_{_{PPFF}}^{2313}-e_{_{PPFF}}^{3213}$ & $1$ & $e_{_{\textcolor{green}{PP}FF}}^{\textcolor{green}{22}12}\xrightarrow{\text{P-fragmentation}}e_{_{PPF\textcolor{red}{F}}}^{221\textcolor{red}{3}}+e_{_{PPFF}}^{2313}$\\
& & $\qquad \qquad \qquad \qquad+e_{_{PPFF}}^{3213}+e_{_{PP\textcolor{red}{F}F}}^{33\textcolor{red}{1}3}$\\
& & $\xrightarrow{\text{F-collapse}}e_{_{PPFF}}^{2313}+e_{_{PPFF}}^{3213}$\\
\hline
 & &  \\
$e_{_{PPFF}}^{1111}+e_{_{PPFF}}^{2111}+e_{_{PPFF}}^{1211}+e_{_{PPFF}}^{2211}$ & $1$ & \raisebox{-0.5 cm}{\scalebox{0.4}{\begin{tikzpicture}[scale = 1.5]
\draw[thick,color=red](-1.2, -0) circle (0.5ex);
\draw[thick,color=red](-1.2, -0.5) circle (0.5ex);
\draw[thick,color=red](-0.4, - 0) circle (0.5ex);
\draw[thick,color=red](-0.4, -0.5) circle (0.5ex);
\draw[thick,color=red](0.4, - 0) circle (0.5ex);
\draw[thick,color=red](0.4, -1.5) circle (0.5ex);
\draw[thick,color=red](0.4, - 0.5) circle (0.5ex);
\draw[thick,color=red](0.4, -1.0) circle (0.5ex);
\draw[thick,color=red](1.2, -0) circle (0.5ex);
\draw[thick,color=red](1.2, -1.5) circle (0.5ex);
\draw[thick,color=red](1.2, -0.5) circle (0.5ex);
\draw[thick,color=red](1.2, -1.0) circle (0.5ex);
\draw[thick, color=black]
{
 (-1.2,  - 0) node [above] {\large{{-1}}}
 (-1.2, -0.5) node [above] {\large{{+1}}}
 (-0.4, - 0) node [above] {\large{{-1}}}
 (-0.4, -0.5) node [above] {\large{{+1}}}
 (0.4, - 0) node [above] {\large{{+1}}}
 (0.4, - 1.5) node [above] {\large{{$-e^{-\beta \omega_2}$}}}
(0.4, - 0.5) node [above] {\large{{-1}}}
(0.4, - 1.0) node [above] {\large{{+1}}}
(1.2, - 0) node [above] {\large{{+1}}}
 (1.2, - 1.5) node [above] {\large{{$-e^{-\beta \omega_1}$}}}
(1.2, - 0.5) node [above] {\large{{-1}}}
(1.2, - 1.0) node [above] {\large{{+1}}}
};
\foreach \x in {0,...,-1.5}{
\pgfmathtruncatemacro\z{int(-\x + 1)}
\draw[thick,color=violet,->] (-2,\x cm) -- (2,\x cm) ;
\draw[thick,color=violet,->] (2, \x cm - 0.5 cm) -- (-2, \x cm - 0.5 cm) ;
\draw[thick,color=violet,->] (2, \x cm) arc (90:-90:0.25);
\draw[thick,color=blue] (2 + 0.25, \x cm - 0.25cm) node[right] {\footnotesize{\z}};
}
\foreach \y in {-1,...,-.5}{
\pgfmathtruncatemacro\w{int(-\y + 1)}
\draw[thick,color=violet,->] (-2,\y cm + 0.5 cm) arc (90:270:0.25);
}
\end{tikzpicture}
}}
=\quad \raisebox{-0.5 cm}{\scalebox{0.4}{\begin{tikzpicture}[scale = 1.5]
\draw[thick,color=red](-1.2, -0) circle (0.5ex);
\draw[thick,color=red](-1.2, -0.5) circle (0.5ex);
\draw[thick,color=red](-0.4, - 0) circle (0.5ex);
\draw[thick,color=red](-0.4, -0.5) circle (0.5ex);
\draw[thick,color=red](0.4, -1.5) circle (0.5ex);
\draw[thick,color=red](0.4, -1.0) circle (0.5ex);
\draw[thick,color=red](1.2, -1.5) circle (0.5ex);
\draw[thick,color=red](1.2, -1.0) circle (0.5ex);
\draw[thick, color=black]
{
 (-1.2,  - 0) node [above] {\large{{-1}}}
 (-1.2, -0.5) node [above] {\large{{+1}}}
 (-0.4, - 0) node [above] {\large{{-1}}}
 (-0.4, -0.5) node [above] {\large{{+1}}}
 (0.4, - 1.5) node [above] {\large{{$-e^{-\beta \omega_2}$}}}
(0.4, - 1.0) node [above] {\large{{+1}}}
 (1.2, - 1.5) node [above] {\large{{$-e^{-\beta \omega_1}$}}}
(1.2, - 1.0) node [above] {\large{{+1}}}
};
\foreach \x in {0,...,-1.5}{
\pgfmathtruncatemacro\z{int(-\x + 1)}
\draw[thick,color=violet,->] (-2,\x cm) -- (2,\x cm) ;
\draw[thick,color=violet,->] (2, \x cm - 0.5 cm) -- (-2, \x cm - 0.5 cm) ;
\draw[thick,color=violet,->] (2, \x cm) arc (90:-90:0.25);
\draw[thick,color=blue] (2 + 0.25, \x cm - 0.25cm) node[right] {\footnotesize{\z}};
}
\foreach \y in {-1,...,-.5}{
\pgfmathtruncatemacro\w{int(-\y + 1)}
\draw[thick,color=violet,->] (-2,\y cm + 0.5 cm) arc (90:270:0.25);
}
\end{tikzpicture}
}}= 0\\
 & &  \\
\hline
\end{tabular}
\caption{Arguments for the tensors in the PPFF sector orthogonal to the array of 4 point contour correlators \label{tab:4ptPPFF4-ortho}} 
\end{table}
In table \ref{tab:4ptPPFF4-ortho}, when there is a P/F or fragmentation we indicate the $e_P/e_F$ that fragments by green colour. Let us explain the argument given for the last orthogonal tensor in table  \ref{tab:4ptPPFF4-ortho}. We successively remove the pair of future-most insertions on the first 2 legs whose contributions cancel each other, thus resulting in zero. These orthogonal tensors fix $\matMt{k,4}_{PPFF}$ to be of the following form:

\be
\begin{split}
&\matMt{k,4}_{PPFF}\\&=\alpha^{(4)}_4 \sum_{r,s=1}^k  \Big(\theta_{r>s}+\theta_{r \leq s} e^{\beta(\omega_2+\omega_4)}\Big)\\&\qquad\qquad  \Big (\ebP{r+1}(\omega_1)-\ebP{r}(\omega_1)\Big ) \otimes \ebP{s+1}(\omega_2) \otimes \ebF{r}(\omega_3) \otimes \ebF{s}(\omega_4)\\& -\alpha^{(4)}_4\sum_{r,s=1}^k \Big(\theta_{r\geq s}+\theta_{r < s}e^{\beta(\omega_2+\omega_4)}\Big)\\&\qquad\qquad  \Big (\ebP{r+1}(\omega_1)-\ebP{r}(\omega_1)\Big ) \otimes \ebP{s}(\omega_2) \otimes \ebF{r}(\omega_3) \otimes \ebF{s}(\omega_4)\\
& +\alpha^{(4)}_5\sum_{r,s=1}^k \Big(\theta_{r\geq s} e^{\beta(\omega_1+\omega_4)}+\theta_{r < s}\Big)\\&\qquad\qquad  \ebP{r+1}(\omega_1)\otimes \Big (\ebP{s+1}(\omega_2)-\ebP{s}(\omega_2)\Big )\otimes \ebF{s}(\omega_3) \otimes \ebF{r}(\omega_4)\\&-\alpha^{(4)}_5 \sum_{r,s=1}^k   \Big(\theta_{r> s}e^{\beta(\omega_1+\omega_4)} +\theta_{r \leq s}\Big)\\&\qquad\qquad  \ebP{r}(\omega_1)\otimes \Big (\ebP{s+1}(\omega_2)-\ebP{s}(\omega_2)\Big )\otimes \ebF{s}(\omega_3) \otimes \ebF{r}(\omega_4)\\
&+\alpha^{(4)}_6 \sum_{r=1}^k \Big (\ebP{r+1}(\omega_1)\otimes \ebP{r+1}(\omega_2) \otimes \ebF{r}(\omega_3) \otimes \ebF{r}(\omega_4) \\
&\qquad-\ebP{r}(\omega_1)\otimes \ebP{r}(\omega_2) \otimes \ebF{r}(\omega_3) \otimes \ebF{r}(\omega_4)\Big)~.
\end{split}
\label{app4pt:24}
\ee

The coefficient $ \alpha_4^{(4)}$ is given by the contraction with the tensor $e_{PPFF}^{(r+1) (s+1)rs}$ for $r>s$ , an example of which is the case $r=2,s=1$. Then this coefficient is given by 
\begin{equation}
\begin{split}
 \alpha_4^{(4)} &=\matMt{k,4}(\text{4-Pt})\cdot e_{PPFF}^{\ 3221}= \rho[13][24] \ .
\end{split}
 \label{app4pt:25}
\end{equation}

The coefficient $ \alpha_5^{(4)}$ is given by the contraction with the tensor $e_{PPFF}^{(r+1) (s+1)sr}$ for r<s , an example of which is the case $r=1,s=2$. Then this coefficient is given by 
\begin{equation}
\begin{split}
 \alpha_5^{(4)} &=\matMt{k,4}(\text{4-Pt})\cdot e_{PPFF}^{\ 2321} = \rho[23][14] \ .
\end{split}
 \label{app4pt:26}
\end{equation}

Finally, the contraction with  the tensor $e_{PPFF}^{(r+1) (r+1)rr}$ for any $r\in \{1,\cdots,k\}$ gives \[ \alpha_4^{(4)}e^{\beta(\omega_2+\omega_4)}+\alpha_5^{(4)}e^{\beta(\omega_1+\omega_4)}+\alpha_6^{(4)}\ .\] Let us look at this contraction for $r=1$ which gives
\begin{equation}
\begin{split}
& \alpha_4^{(4)}e^{\beta(\omega_2+\omega_4)}+\alpha_5^{(4)}e^{\beta(\omega_1+\omega_4)}+\alpha_6^{(4)} =\matMt{k,4}(\text{4-Pt})\cdot e_{PPFF}^{\ 2211}\\
&= \rho[2314]+ \rho[24][13]+ \rho[14][23]\ .
\end{split}
 \label{app4pt:27}
\end{equation}

Now, using equations \eqref{app4pt:25} , \eqref{app4pt:26} and the KMS relations, we have 
\begin{equation}
\begin{split}
\alpha_4^{(4)}e^{\beta(\omega_2+\omega_4)}&=e^{\beta(\omega_2+\omega_4)}  \rho[13][24]
=\rho[24][13]\ ,\\
\alpha_5^{(4)}e^{\beta(\omega_1+\omega_4)}&=e^{\beta(\omega_1+\omega_4)} \rho[23][14] 
= \rho[14][23]~.
\end{split}
 \label{app4pt:28}
\end{equation}
Replacing the expressions of $\alpha_4^{(4)}e^{\beta(\omega_2+\omega_4)}$ and $\alpha_5^{(4)}e^{\beta(\omega_1+\omega_4)}$ obtained in \eqref{app4pt:28} 
into \eqref{app4pt:27} we get
\begin{equation}
\begin{split}
& \alpha_6^{(4)} = \rho[2314]\ .
\end{split}
 \label{app4pt:29}
\end{equation}
Next using the values of $\alpha_i^{(4)}$'s obtained in \eqref{app4pt:13}, \eqref{app4pt:16}, \eqref{app4pt:19}, \eqref{app4pt:25},
\eqref{app4pt:26} and \eqref{app4pt:29} respectively, and the following KMS relations:
\begin{equation}
\begin{split}
\rho[34][12]&=e^{\beta(\omega_3+\omega_4)} \rho[12][34]\ , \\
\rho[24][13]&=e^{\beta(\omega_2+\omega_4)} \rho[13][24]\ , \\
 \rho[14][23]&=e^{\beta(\omega_1+\omega_4)} \rho[23][14]\ ,
\end{split}
 \label{app4pt:30}
\end{equation}
we get the expressions in \eqref{app4pt:6},\eqref{app4pt:7},\eqref{app4pt:8} and \eqref{app4pt:9}.
%
%

\bibliographystyle{JHEP}
\bibliography{thermalcvrdraft}

\end{document}